\documentclass{aa}  

\usepackage{graphicx,amsmath}
\usepackage{times}
\usepackage[pdftex,colorlinks,citecolor=blue]{hyperref}
\def\rrrev{}
\def\rrev{}
\def\rev{}
\def\wig#1{\mathrel{\hbox{\hbox to 0pt{%
          \lower.6ex\hbox{$\sim$}\hss}\raise.4ex\hbox{$#1$}}}}
\def\disp{\displaystyle}
\def\bcol{b}

\def\mea{\rm M_\oplus}

\def\rau{r_{\rm AU}}
\def\rhostar{\tilde{\rho}_{*}}
\def\rhos{\tilde{\rho}_{\rm s}}
\def\rhop{\tilde{\rho}_{\rm p}}

\def\hdust{h_{\rm d}}
\def\hp{h_{\rm p}}
\def\hg{h_{\rm g}}

\def\vcol{\Delta v_{\rm col}}
\def\vmeancol{\overline{\Delta v}_{\rm col}}

\def\tauf{\tau_{\rm f}}
\def\taus{\tau_{\rm s}}
\def\tausnu{\tau_{{\rm s},\nu}}
\def\nup{\nu_{\rm p}}
\def\ffocus{f_{\rm focus}}
\def\fei{f_{e,i}}
\def\rmd{{\rm d}}

\def\rmg{{\rm g}}
\def\rms{{\rm s}}
\def\rmp{{\rm p}}
\def\rmf{{\rm f}}
\def\rmK{{\rm K}}
\def\rmH{{\rm H}}
\def\Rp{R_{\rm p}}
\def\mup{\mu_{\rm p}}
\def\AU{{\rm AU}}
\def\FD{F_{\rm D}}
\def\CD{C_{\rm D}}
\def\vth{v_{\rm th}}
\def\vrel{v_{\rm rel}}
\def\min{{\rm min}}
\def\max{{\rm max}}
\def\micron{{\rm\mu m}}
\def\gcc{{\rm g\,cm^{-3}}}
\def\gcms{{\rm g\,cm^{-2}}}
\def\xmmsn{\chi_{_{\rm MMSN}}}
\def\chialpha{\chi_{\alpha,\taus}}

\def\alphadef{\left(\alpha\over 10^{-2}\right)}
\def\rhopdef{\left(\rhop\over 1\,{\rm g/cm^3}\right)}
\def\Rpun{\left(\Rp\over 1\,{\rm km}\right)}
\def\Rpcent{\left(\Rp\over 100\,{\rm km}\right)}
\def\Rpmille{\left(\Rp\over 1000\,{\rm km}\right)}
\def\raudef{\left(r\over 1\,{\rm AU}\right)}

\begin{document}

\title{On the filtering and processing of dust by planetesimals}
\subtitle{1. Derivation of collision probabilities for non-drifting planetesimals}

\author{Tristan Guillot\inst{\ref{inst1},\ref{inst2}} \and Shigeru Ida\inst{\ref{inst2},\ref{inst3}} \and Chris W. Ormel\inst{\ref{inst4}\thanks{Hubble Fellow.}}}


\institute{Laboratoire Lagrange, UMR 7293, Universit\'e de Nice-Sophia Antipolis, CNRS, Observatoire de la C\^ote d'Azur, 
    06304 Nice Cedex 04, France \label{inst1}
\and
    Department of Earth and Planetary Sciences, Tokyo Institute of Technology, Tokyo, Japan\label{inst2}
\and
    Earth-Life Science Institute, Tokyo Institute of Technology, Tokyo 152-8550, Japan\label{inst3}
\and
    Astronomy Department, University of California, Berkeley, CA 94720, USA\label{inst4} 
}

\date{Submitted to A\&A: November 10, 2013; Accepted: September 14, 2014.}

\abstract
{Circumstellar disks are known to contain a significant mass in dust ranging from micron to centimeter size. Meteorites are evidence that individual grains of those sizes were collected and assembled into planetesimals in the young solar system.}
{We assess the efficiency of dust collection of a swarm of {\rev non-drifting} planetesimals {\rev with radii ranging from 1 to $10^3$\,km and beyond.}}
{We calculate the {\rev collision} probability of dust drifting in the disk due to gas drag by planetesimal accounting for several regimes depending on the size of the planetesimal, dust, and orbital distance: the geometric, Safronov, settling, and three-body regimes. We also include a hydrodynamical regime to account for the fact that small grains tend to be carried by the gas flow around planetesimals.}
{We provide expressions for the {\rev collision} probability of dust by planetesimals {\rev and for the filtering efficiency by a swarm of planetesimals}. {\rrev For standard turbulence conditions (i.e., a turbulence parameter $\alpha=10^{-2}$), filtering is found to be inefficient, meaning that when crossing a minimum-mass solar nebula (MMSN) belt of planetesimals extending between 0.1\,AU and 35\,AU most dust particles are eventually accreted by the central star rather than colliding with planetesimals. However, if the disk is weakly turbulent ($\alpha=10^{-4}$) filtering becomes efficient in two regimes: (i) when planetesimals are all smaller than about 10\,km in size, in which case collisions mostly take place in the geometric regime; and (ii) when planetary embryos larger than about 1000\,km in size dominate the distribution, have a scale height smaller than one tenth of the gas scale height, and dust is of millimeter size or larger in which case most collisions take place in the settling regime. These two regimes have very different properties: we find that the local filtering efficiency $x_{\rm filter,MMSN}$ scales with $r^{-7/4}$ (where $r$ is the orbital distance) in the geometric regime, but with $r^{-1/4}$ to $r^{1/4}$ in the settling regime. This implies that the filtering of dust by small planetesimals should occur close to the central star and with a short spread in orbital distances. On the other hand, the filtering by embryos in the settling regime is expected to be more gradual and determined by the extent of the disk of embryos. Dust particles much smaller than millimeter size tend only to be captured by the smallest planetesimals because they otherwise move on gas streamlines and their collisions take place in the hydrodynamical regime. }}
{{\rev Our results hint at an inside-out formation of planetesimals in the infant solar system because small planetesimals in the geometrical limit can filter dust much more efficiently close to the central star. However}, {\rrev even a fully-formed belt of planetesimals such as the MMSN only marginally captures inward-drifting dust and this} seems to imply that dust in the protosolar disk has been filtered by planetesimals even smaller than 1 km (not included in this study) or that it has been assembled into planetesimals by other mechanisms {\rev (e.g., orderly growth, capture into vortexes). Further refinement of our work concerns, among other things: a quantitative description of the transition region between the hydro and settling regimes; an assessment of the role of disk turbulence for collisions, in particular in the hydro regime; and the coupling of our model to a planetesimal formation model.}}



\titlerunning{On the filtering of dust by planetesimals: 1. Derivation}
\authorrunning{T. Guillot, S. Ida \& C.~W. Ormel}

\maketitle
\section{Introduction}

Observations, laboratory experiments, and theoretical studies have shown that dust grows rapidly in protoplanetary disks from submicron to centimeter sizes. Observations show that classical T-Tauri disks present masses in dust that range from about $10^{-5}$ to $10^{-2}\,\rm M_\odot$ and in which the detectable dust grains are between micron and centimeter size \citep[e.g.,][]{Beckwith+1990, AndrewsWilliams2007}. Surprisingly, however, there appears to be no obvious correlation between inferred dust mass, maximum particle sizes, accretion rate onto the star, and stellar age \citep{Ricci+2010a,Ricci+2010b}. To add to the puzzle, theory predicts that grains of millimeter to centimeter sizes should be lost by gas drag and rapid migration onto the central star on timescales of approximately $10^4$\,yrs \citep{Adachi+1976,Weidenschilling1977, Nakagawa+1986}. The formation of non-drifting, km-sized planetesimals appears necessary to keep the dust from being drained away onto the central star but simulations including gas evolution and planetesimal growth have thus far failed to produce disks of dust and planetesimals that are both massive and frequent \citep{StepinskiValageas1996,StepinskiValageas1997,Garaud2007}. 

Planet formation, however, appears to be widespread and efficient. Planets are known to be present around approximately {\rrev $50\%$ of stars} at least \citep{Mayor+2011,Howard+2012}. Some of the giant planets that are observed in transit are very dense and must have collected large amounts of heavy elements, in some case larger than a hundred times the mass of the Earth \citep{Guillot+2006, Moutou+2013}. The solar system itself bears evidence of this efficiency of planet formation: The Sun contains about 5000\,M$_\oplus$ in heavy elements that were for the most part present as solids in the protosolar cloud core \citep[e.g.,][]{Lodders+2009}. The present solar system, not including the Sun itself, contains about 100\,M$_\oplus$ in heavy elements \citep{Guillot+Gautier2014}, to which we can roughly add between 50$\,\mea$ and 100$\,\mea$ which were ejected mostly by Jupiter \citep[e.g.,][]{Tsiganis+2005}. {\rrev This implies an efficiency of planet and planetesimal formation of at least 150/5000=3\%. However,} although a very large fraction of the material that formed the Sun went through a disk phase, the very violent events of the first phases, including gravitational instabilities, FU Orionis events, and a high accretion rate onto the central star \citep[e.g.,][]{HartmannKenyon1996,VorobyovBasu2010} make it difficult to imagine that a large fraction of solids could be retained before it had acquired about 90\% of its mass.  This implies that of the $\approx 500\rm\,M_\oplus$ masses of heavy elements present in the young, $0.1\,\rm M_\odot$ disk, about 30\% to 40\% had to be captured {\rrev into planetesimals and planets in order to account for the solids in the solar system and those lost by dynamical interactions during its formation.}

In parallel, meteorites are evidence that in the inner solar system, individual particles of micron to centimeter size have been collected into much larger, planetesimal-sized objects in a relatively orderly way. Chondrites, the oldest known rocks {\rrev that are} the closest match to the composition of the Sun contain four main kinds of identifiable material: calcium-aluminum inclusions known as CAIs, chondrules, metal grains, and the matrix. The respective proportions of these components, their mean sizes, and their isotopic characteristics vary from one meteorite group to the next, but remain relatively well defined within one group \citep{ScottKrot2005}, while the bulk composition remains close to solar \citep{HezelPalme2010}. The number of presolar grains (in sizes ranging from mere nanometers to $\sim 20\,\mu$m) identified from their anomalous isotopic signatures is tiny \citep{Ott1993,HoppeZinner2000}, indicating that individual grains were efficiently processed in the young solar system and that any later inflow was either not abundant or not captured by planetesimals. Aqueous alterations remain limited indicating that meteorites were not in direct contact with abundant ice grains present in the outer solar system. Altogether, the homogeneity in {\rrev their characteristics indicates that the individual components of each different chondrite group} were assembled together locally rather than from different regions of the solar system. 

Theoretical studies have mostly focused either on grain growth and the formation of planetesimals \citep[e.g.,][]{Weidenschilling1984,Wurm+2004,DullemondDominik2005,Cuzzi+2008,Birnstiel+2011,Okuzumi+2012} or on the growth of a swarm of mutually interacting planetesimals \cite[e.g.,][]{Safronov1972,WetherillStewart1989,Kokubo+Ida1998,Chambers2006,Levison+2010,Johansen+2014}. The interactions of planetesimals and dust {\rev in a gas-rich disk} have been the focus of less attention, apart from some works to which this study will frequently refer \citep{Rafikov2004,OrmelKlahr2010,OrmelKobayashi2012,LambrechtsJohansen2012}. {\rrev The need for an early formation of planetesimals and the prevalence of dust in disks} leads us to consider a situation in which planetesimals are formed in a given region of the disk while dust drifts from the outer region. Calculating the filtering efficiency of this belt of planetesimals, i.e., the fraction of the dust grains which collide with them is required to answer many crucial questions, such as: Could dust particles from the outer solar system reach the inner regions? What was the composition of the material that the Sun accreted? How did planetesimals in the 2-3 AU region collect their chondrules and other components? What prevented ices from reaching the inner solar system? {\rev This work is a first step towards addressing these questions.}

The purpose of the present study is to derive laws of interaction between planetesimals and shear-dominated dust particles in the presence of gas drag in a protoplanetary disk. In the next section, we examine the geometry of the problem and quantify the rate at which dust is lost from the disks. In section~3, we examine how planetesimals and drifting dust interact {\rev in the geometrical circular limit and derive analytical expressions for the collision probability and filtering efficiency. We show that this view is complementary to the usual approach of calculating collision and growth rates. In section~4 we then consider additional effects, namely the possibility of eccentric and/or inclined orbits, gravitational focusing, hydrodynamical effects and the consequence of turbulence in the disk. } {\rrev The resulting collision probabilities are presented in section~5. In section~6, we then apply our results} to the study of filtering by a swarm of planetesimals in the young solar system. {\rrev Appendices~A to E (available as online material) provide scaling relations for the minimum mass solar nebula, further analytical derivations for collisions in the geometric and settling regimes, figures for the weak-turbulence case, and an analysis of how the filtering efficiency depends on the planetesimal scale height.}

Because the material used is diverse and the problem is intrinsically complex, we have chosen to propose a rather long (but homogeneous and hopefully as complete as possible) re-derivation of the equations of the problem. We generally adopt the notations and approach of \cite{OrmelKlahr2010}, on which the main part of the work is based. The reader not interested in the technical details of the derivation of the collision probabilities may {\rrev skip sections~2.4, 2.5, 4, and 5 and} continue on to section~\ref{sec:filtering} where the problem is directly applied to the solar system in the minimum mass solar nebula formalism.

\section{Context} 

\subsection{Geometry of the problem}\label{sec:geometry}

\begin{figure}
\includegraphics[width=\hsize,angle=0]{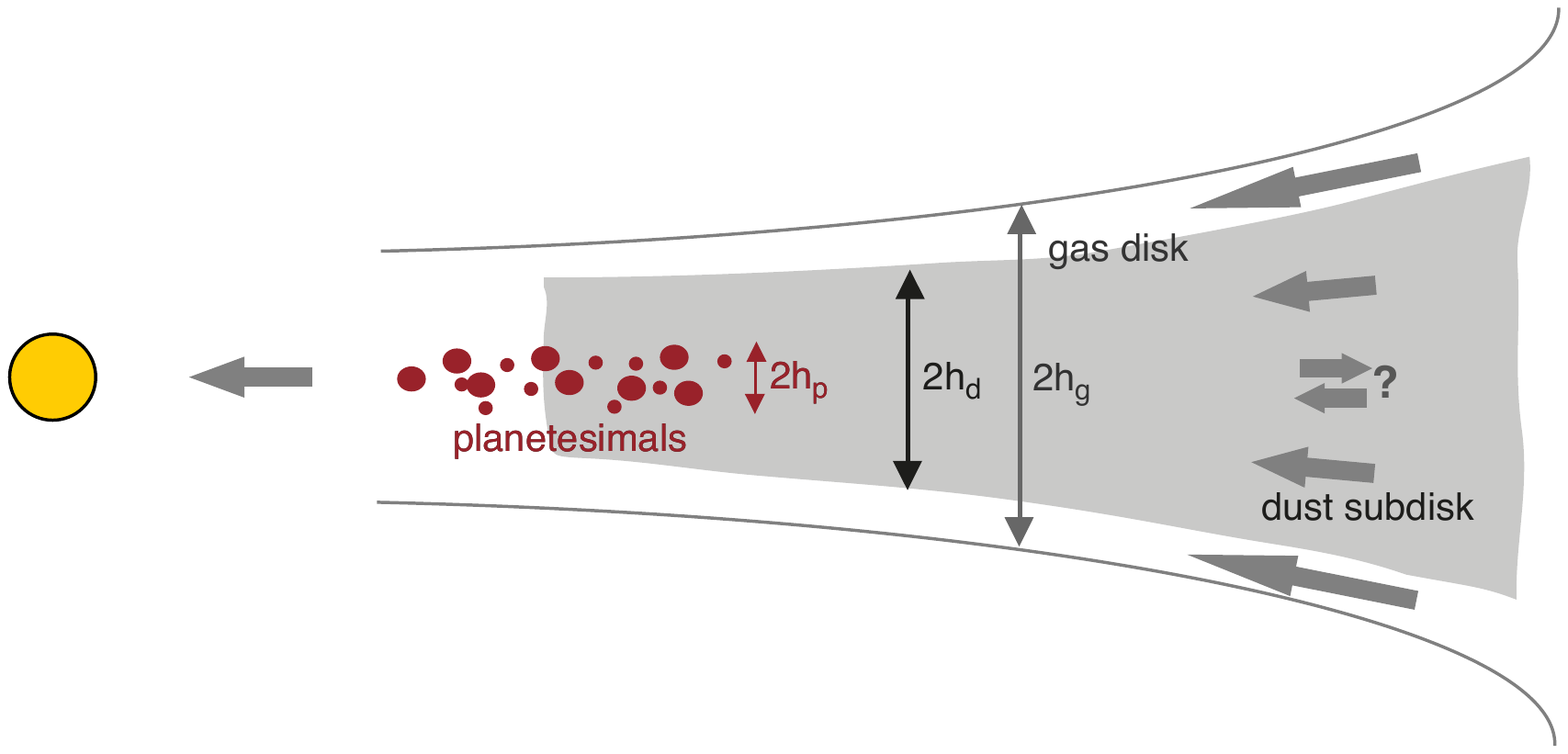}
\caption{Geometry of the problem considered in this study. The gas disk has a characteristic height $h_\rmg$ and is accreting onto the star thus yielding a small but non-negligible inflow velocity. Dust particles grow by mechanisms not modeled in this study. They settle to the mid-plane and drift inward as a result of gas drag, at a pace set both by their size and the thermodynamical conditions in the gas disk. Their vertical extent is $\hdust$. Planetesimals are supposed to have formed preferentially near the star. They have sizes in the kilometer range or much larger and negligible inward drift. Their vertical extent $h_\rmp$ is mainly governed by self-scattering.}
\label{fig:disk-sketch}
\end{figure}

We assume that planetesimals have formed in the inner system by an undefined mechanism, but that a vast reservoir of dust is still present in the outer disk. As described in the previous section, this dust will grow rapidly to a size which is to be defined and drift inward, both as a result of gas drag and the slow inward flow of gas being accreted onto the star. Figure~\ref{fig:disk-sketch} shows the geometry of the problem, with the three main constituents of the disk: the (predominantly) hydrogen and helium gas, the dust, and the planetesimals. The gas forms the thickest disk. Dust tends to settle to the mid-plane as a function of its size but limited by turbulence in the gas. We envision that planetesimals have formed preferentially near the star and will tend to have a smaller vertical extent.  

Circumstellar disks have a structure that is complex and shaped both by the irradiation that they receive from the parent star, viscous heating due to (turbulent) angular momentum transfer, presence or absence of a mechanism to provide this angular momentum transfer, possible accretion from the molecular cloud core, varying composition in dust, etc. A common simplification is to assume that the disk is vertically isothermal and to neglect the disk's gravity over that of the star, in which case the vertical density structure writes \citep[e.g.,][]{HuesoGuillot2005}:
\begin{equation}
\rho_{\rm g,z}=\rho_\rmg e^{-(1/2)(z\Omega_\rmK/c_\rmg)^2}.
\end{equation}
It is natural to define the gas scale height $h_\rmg$ as the one at which the gas density has decreased by a factor $e$ compared to $\rho_\rmg$, that of the mid-plane:
\begin{equation}
h_\rmg=\sqrt{2}{c_\rmg\over \Omega_\rmK}.
\end{equation}
We note that other choices are sometimes made in the literature. With this choice, the relation between the midplane density and the surface density is obtained by a simple vertical integration:
\begin{equation}
\rho_\rmg={\Sigma_\rmg\over \sqrt{\pi} h_\rmg}.
\end{equation}

A further simplification is to assume that the radial structure of the disk is described by power laws. Following \cite{Hayashi1981} and \cite{Nakagawa+1986}, we adopt the following scaling laws for the so-called minimum mass solar nebula (MMSN), 
\begin{eqnarray}
\Sigma_\rmg&=&1.7\times 10^3 \xmmsn \rau^{-3/2}\ \gcms,\\
T&=&280 \rau^{-1/2}\ {\rm K},
\end{eqnarray}
where $\rau\equiv r/1\rm\,AU$, and $\xmmsn$ is a scaling factor on the density. The MMSN is a convenient representation of the planetesimal disk. {\rrev It is based} on the present-day observed planets and can therefore be considered representative of the last stages of the formation of the solar system at the time of the dispersal of the gas disk. {\rrev It does not account for planetesimal-driven migration after the disk disperses. It is also} less clear that is it a good representation of the gas disk (simulations of disk evolution generally yield different power laws both for $\Sigma_\rmg$ and $T$). {\rrev However, the results of this work are easily scalable for any profile other than the MMSN.} 

The values of a number of quantities of relevance to this work based on the MMSN scalings are presented in Appendix~\ref{sec:MMSN}. For the gas disk scale height, the following relation applies:
\begin{equation}
h_\rmg \simeq 4.7 \times 10^{-2} 
\rau^{5/4}
\left(\frac{L_*}{L_{\odot}} \right)^{1/8} 
\left(\frac{M_*}{M_{\odot}} \right)^{-1/2} 
\; {\rm AU}.
\label{eq:scale_height_hayashi}
\end{equation}

The dust grains settle onto the midplane because of friction with the gas and the vertical component of the star's gravitational force, yielding a sedimentation terminal velocity $v_z = -\tau_\rms \Omega_{\rm K} z$ \citep{Nakagawa+1981}. The sedimentation timescale is thus $\tau_{\rm sed} \equiv z/|v_z| \sim (2 \pi \tau_\rms)^{-1} \Omega_\rmK^{-1}$, i.e., it is proportional to the local Keplerian orbital timescale divided by the dimensionless stopping time $\taus$ (see Sect.~\ref{sec:velocities} hereafter for the definition of $\taus$). Except for the smallest particles, we can consider that dust has fully sedimented and has a scale height that is determined by its stopping time and turbulent stirring. \cite {Dubrulle+1995} show that
\begin{equation}
\hdust= \min\left(1,\sqrt{\alpha/\tau_\rms}\right) h_{\rm g}, 
\label{eq:hdust}
\end{equation}
where $\alpha$ is the traditional turbulence parameter\citep{ShakuraSunyaev1973}. 
(We note that \cite{YoudinLithwick2007} extend this relation to any eddy mixing timescale with a slightly more complex dependence on $\taus$ which is minor and not taken into account here). We will use a {\rev fiducial value $\alpha=10^{-2}$} broadly compatible with measured T-Tauri accretion rates and disk observations \citep[e.g.,][]{Hartmann+1998,HuesoGuillot2005} and simulations of magneto-rotational instability in disks \citep[e.g.,][]{HeinemannPapaloizou2009,Flock+2013}. {\rev We will also use a lower value $\alpha=10^{-4}$ more relevant to the turbulence inside }dead zones \citep[e.g.,][]{OkuzumiHirose2011}. For the (compact) grains with sizes ranging from microns to tens of meters considered in this study, values of $\taus$ range between about $10^{-8}$ and $10^{2}$ (see Fig.~\ref{fig:taus} hereafter for the correspondence between grain size and $\taus$ in the MMSN). With this value of the turbulence parameter, $\hdust= h_{\rm g}$ for all grains smaller than about $1\,$cm at $1$\,AU and $0.1$mm at 100\,AU. For larger grains {\rev in the Stokes drag regime} the scale height decreases inversely with the square root of the grain size.   

The scale height of planetesimals is determined directly from their inclinations $i$:
\begin{equation} 
h_\rmp = r <\sin i>.
\end{equation}
Inclinations are determined by the balance between excitation (scattering by planetesimals or density fluctuations in the gas disk) and damping (gas drag,  tidal interactions with the disk and collisions). As discussed in the next section, damping is generally strong and eccentricities small ($e\wig{<}0.05$ for $r<10$\,AU; see Fig.~\ref{fig:ecc_I08}). Eccentricities and inclinations are directly linked. For example, in the case of mutual scattering by planetesimals, $i\sim e/2$ \citep{IdaMakino1992}. On the basis of this approximation, at 1\,AU and for 1000\,km, we can expect $h_\rmp\sim 5\times 10^{-3} r$, about an order of magnitude smaller than $h_\rmg$. 

We can therefore expect that the geometry indicated by Fig.~\ref{fig:disk-sketch}, i.e., $h_\rmp\le\hdust\le h_\rmg$, holds for all grains smaller than about $1$\,m at $1\,$AU and $0.1$\,mm at $100\,AU$. For larger grains the possibility that $h_\rmp >\hdust$ is to be considered.

\subsection{Sizes and eccentricities of planetesimals}\label{sec:sizes}

Planetesimals will have a size distribution that will be affected both by accretion and destruction processes, by mechanisms leading to their formation and by gain and losses due to migration in the protoplanetary disk. In the present work, we will simply assume that their sizes are distributed between $R_{\rm p,min}$ and $R_{\rm p,max}$. For simplicity, we use $R_{\rm p,min}=1\,$km which roughly corresponds to the size below which the drift of the planetesimals must be taken into account. The maximum radius will depend upon accretion processes. Early in the evolution of the disk, streaming instabilities might provide an efficient way of making the first Ceres-sized ($\sim 500\,$km) planetesimals \citep{Johansen+2007}. Subsequent growth must lead to Moon-sized objects and later to planetary mass objects. 

\begin{figure}
\includegraphics[width=\hsize,angle=0]{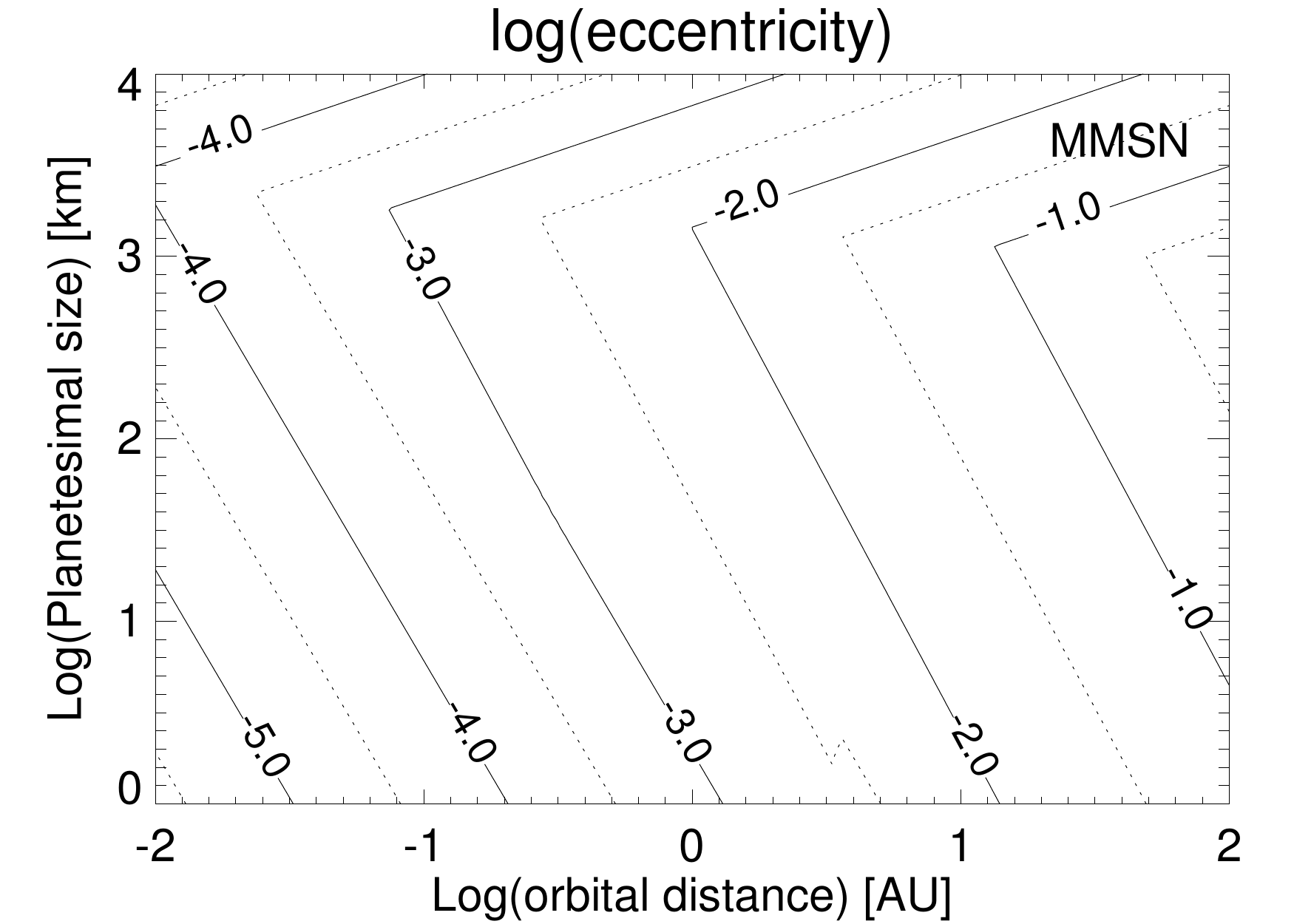}
\caption{Contour plot showing the decimal logarithm of the eccentricity as a function of planetesimal size and orbital distance for a standard MMSN disk model as obtained from an equilibrium between gas drag, tidal interactions, and excitation by a turbulent disk with a turbulence parameter {\rev $\gamma=5\times 10^{-4}$ corresponding to $\alpha=10^{-2}$} \citep[see][]{Ida+2008,OkuzumiOrmel2013}. 
}
\label{fig:ecc_I08}
\end{figure}

The eccentricities of planetesimals are excited by gravitational scattering
by other planetesimals.
As \cite{Ida+2008} and \cite{OkuzumiOrmel2013} suggested, however,
eccentricity excitation by density fluctuations of turbulence of the disk
is comparable to or larger than that by planetesimal scattering,
for usual values of $\alpha = 10^{-3}-10^{-2}$
for magneto-rotational instabilities (MRI).
We here consider the excitation by the  MRI turbulence.
While the excitation by scattering depends on the size distribution of planetesimals,
the excitation by turbulence is independent of the size distribution
of other planetesimals
and also the size of perturbed planetesimals. 
The equilibrium eccentricities are given by a balance between
the excitation and damping due to gas drag and tidal interactions.
\cite{Ida+2008} showed that the equilibrium eccentricities
for small bodies, for which gas drag is dominant, are given by
\begin{equation}
e_{\rm drag} \sim 0.2 \xmmsn^{1/3} \gamma^{2/3}
\left(\frac{R_{\rm p}}{1{\rm km}}\right)^{1/3} 
\left(\frac{\rhop}{3{\rm gcm}^{-2}}\right)^{1/3} 
\rau^{11/12},
\end{equation}
where $R_{\rm p}$ and $\rhop$ are the physical size and density of the perturbed bodies, respectively,
and $\gamma (< 1)$ is a {\rev dimensionless parameter representing the strength of the turbulent stirring and}
is a function of the turbulence parameter $\alpha$.
{\rrev We adopt the relation $\gamma\approx 5\times 10^{-4} (\alpha/10^{-2})^{1/2}$ } provided by \cite{OkuzumiOrmel2013} {\rev for ideal MHD.} 
The equilibrium eccentricities for large bodies, 
for which tidal interactions is dominated, are {\rev then} given by
\begin{equation}
e_{\rm tidal} \sim 20 \xmmsn^{1/2} \gamma
\left(\frac{R_{\rm p}}{1{\rm km}}\right)^{-2/3} 
\left(\frac{\rhop}{3{\rm gcm}^{-2}}\right)^{-1/2} 
\rau^{3/4}.
\end{equation}
The {\rev minimum} of $e_{\rm drag}$ and $e_{\rm tidal}$ is
a value that is actually realized.
{\rev (Given our fiducial value of $\alpha = 10^{-2}$, we use 
$\gamma=5\times 10^{-4}$.)} 
In Fig.~\ref{fig:ecc_I08}, we plot the equilibrium eccentricity
as a function of $R_{\rm p}$ and $r$.
The result shown in Fig.~\ref{fig:ecc_I08} predicts eccentricities that are very small at short orbital distances due to gas drag and higher at large orbital distances. Both small planetesimals and large ones have small eccentricities. This is due to gas drag for small objects and to dynamical friction for the large ones. {\rev We note that Fig.~\ref{fig:ecc_I08} is appropriate for MRI-active zones, but we expect much weaker turbulent stirring {\rrev ($\propto \alpha^{1/2}$)} and therefore smaller eccentricities in dead zones.}

{\rrev As shown by \cite{OrmelKobayashi2012}, the gravitational stirring of planetary embryos starts to dominate over turbulent stirring when their mass becomes larger than $\sim 0.3(\alpha/10^{-2})\,\mea$. However, tidal damping is expected to dominate over self-stirring for the large embryos. The maximum eccentricity of large embryos is therefore expected to occur for the same masses/radii as in Fig.~\ref{fig:ecc_I08}. We estimate a maximum scale height for these of $\hp/r\sim e/2$, i.e., $\hp/r\sim 0.005(\alpha/10^{-2})^{1/2}$ at 1\,AU or $\hp\sim 0.1(\alpha/10^{-2})^{1/2}\hg$. For small values of $\alpha$, viscous stirring sets a floor to the value of $\hp$ \citep{Kokubo+Ida2000}. In any case, the largest objects in the size distribution are expected to have a scale height that is small compared to the gas scale height. In addition, $\hp$ decreases with decreasing $\Rp$. Therefore, we simply adopt $\hp=0.01\hg$ as our standard choice for the scale height of the planetesimals. We experiment with other, more extreme ratios in Appendix E. 

We note that when we consider small planetesimals in the presence large embryos, the former can be excited to greater heights \citep{Kokubo+Ida2002}. Our calculations can easily be extended to the case of a size-dependent planetesimal scale height, but this would require a proper treatment of the evolution of the size distribution of planetesimals, which is beyond the scope of the present work.}



{\rev

\subsection{Rate and geometry of gas accretion}\label{sec:gas}

The flow of gas in a disk is determined by the rate of angular momentum transport and the mass balance between inner and outer regions. In one dimension, the following equation for the mean velocity of the gas can be derived from the equations governing the spreading of a viscous disk \citep{LyndenBellPringle1974},
} 
\begin{equation}
v_\nu={3\over \Sigma_\rmg r^{1/2}}{\partial\over \partial r}\left(\nu\Sigma_\rmg r^{1/2}\right),
\label{eq:vnu}
\end{equation}
where $\nu$ is the turbulent diffusion coefficient in the disk. For a MMSN-type disk with a uniform $\alpha$, $v_\nu=0$ everywhere which is unrealistic (the disk spreads inward and outward at the same rate). When assuming $\Sigma_\rmg\propto r^{-1}$, which is more commonly found in realistic disk evolutions \citep[e.g.,][]{HuesoGuillot2005}, $v_\nu=-(3/2) \nu/r$. The negative sign indicates an inward flow.  

{\rev
When considering the 2D structure of disks within the $\alpha$-turbulence framework, it can be shown \citep[e.g.,][]{TakeuchiLin2002} that for most commonly used power-law relations for the density and temperature radial profiles, a meridional circulation sets in that maintains a strong, inward flow in the upper layers of the disk and a weaker outward flow in the mid-plane. The density-averaged flow still obeys eq.~\eqref{eq:vnu} and is therefore generally inward. This has been advocated as the reason for the presence of chondrules and generally of grains having been formed near the protosun in the outer regions of the solar system \citep{Ciesla2009}.

The presence of such a meridional circulation can lead to the retention in the disk of grains large enough to have settled to the mid-plane so that their outward motion compensates for the inward motion in the upper layers of the disk. Direct numerical simulations of magneto-rotational instability in protoplanetary disks by \cite{Fromang+2011} failed to find such a circulation setting in, but the resulting flow was outward and not inward as was expected for an accretion disk, and the velocity fluctuations were found to be much larger than the mean flow, making the result more tentative. The problem hence still exists. 

For simplicity, and without any pretention of capturing the detailed evolution of real disks} 
we will assume hereafter $v_\nu\sim -\nu/r$. 

{\rev 
\subsection{Particle drift with accreting gas}\label{sec:velocities}

We now rederive drift velocities for dust and planetesimals} using the formalism of \cite{Adachi+1976}. These are widely available in the literature, but the expressions for the radial and azimuthal velocities are derived by assuming that the gas in the disk is at rest. The equation of motion of such a particle in a gas disk under the action of gas drag is 
\begin{equation}
{d^2 {\mathbf r}\over dt^2}+{GM_*\over r^3}{\mathbf r}={\mathbf F}_{\rm D},
\label{eq:motion}
\end{equation}
where ${\mathbf F}_{\rm D}$ is the drag force vector per unit mass which is directed opposite to the velocity vector. Two regimes correspond to the case when the particles are smaller than the mean free path of the gas and the drag can be modeled by accounting for collisions of individual gas molecules (Epstein regime), and when the particles are larger and the gas must be modeled as a fluid. An expression for the amplitude of the drag vector that combines the two regimes \citep[based on][]{Adachi+1976, PeretsMurrayClay2011} is
\begin{equation}
\FD= {\rho_\rmg\over \rhos s} \vth \vrel \min\left[1,{3\over 8}{\vrel\over \vth}\CD(Re)\right],
\label{eq:FD}
\end{equation}
where $\vth=\sqrt{8/\pi}c_\rmg$ is the mean thermal velocity and $Re$ is the Reynolds number measuring the turbulence of the flow with a velocity $\vrel$ around a particle of diameter $2s$ and for a gas {\rev dynamic viscosity $\mu_\rmg$
\begin{equation}
Re={2s\vrel \rho_\rmg / \mu_\rmg},
\label{eq:Re}
\end{equation}
}
and $\CD$ is a dimensionless drag coefficient that is fitted semi-empirically as a function of $Re$ \cite[see][]{PeretsMurrayClay2011}:
\begin{equation}
\CD={24\over Re}(1+0.27 Re)^{0.43}+0.47\left(1-e^{-0.04Re^{0.38}}\right).
\label{eq:CD}
\end{equation}

When $Re\ll 1$ and using {\rev $\mu_\rmg= \rho_\rmg\vth\lambda_\rmg/2$}, it can be shown that eq.~(\ref{eq:FD}) simplifies to the usual Epstein vs. Stokes relations:
\begin{equation}
\lim_{Re\to 0}\FD ={\rho_\rmg\over \rhos s}  \vth \vrel \min\left(1,{9\over 4}{\lambda_\rmg\over s}\right).
\label{eq:limFD}
\end{equation}
Equation~(\ref{eq:motion}) can then be written in $(r,\theta)$ coordinates, assuming planar motions only and accounting for a radial velocity of the gas,
\begin{subequations}
\begin{align}
&{\partial v_r\over \partial t}+v_r{\partial v_r\over \partial r} -{v_\theta^2\over r}=-{GM_*\over r^2}-{\FD\over \vrel}(v_r-v_\nu)\label{eq:dvr_dt} \\ 
&{\partial v_\theta\over \partial t}+v_r{\partial v_\theta\over \partial r}+{v_rv_\theta\over r}=-{\FD\over \vrel}(v_\theta-r\Omega_\rmg), \label{eq:dvtheta_dt}
\end{align}
\end{subequations}
where $r\Omega_\rmg=(1-\eta)v_\rmK$ is the gas azimuthal velocity and
\begin{equation}
\vrel=\left[(v_r-v_\nu)^2+(v_\theta-r\Omega_\rmg)^2\right]^{1/2}.
\end{equation}
The $\eta$ parameter is thus a relative measure of the departure of the gas azimuthal velocity from Keplerian. It can be shown to be directly related to the gas pressure gradient \citep{Adachi+1976}:
\begin{equation}
\eta=-{dP/dr\over 2 r \Omega_{\rm K}^2\rho_{\rm g}}. 
\label{eq:eta}
\end{equation}

Following \cite{Adachi+1976}, we will assume that the azimuthal velocity difference between the gas and the particle is much smaller than the Keplerian azimuthal velocity, i.e., that $ \delta v_\theta \equiv v_\theta-v_\rmK \ll v_\rmK$. Furthermore we assume that the viscous drift velocity is also negligible, i.e., $v_\nu\ll v_\rmK$. Dropping all the second-order terms, eqs.~(\ref{eq:dvtheta_dt})  and (\ref{eq:dvr_dt}) then become 
\begin{subequations}
\begin{align}
v_r&=-{2\over \tau_\rms}(\delta v_\theta+\eta v_\rmK),\\
v_\theta&=v_\rmK+{1\over 2\tau_\rms}(v_r-v_\nu),
\end{align}
\end{subequations}
where we have introduced the stopping time,
\begin{equation}
\tau_\rms\equiv {\vrel\Omega_\rmK\over \FD}.
\label{eq:taus}
\end{equation}

It is then easy to derive the radial velocity and difference between azimuthal and Keplerian velocities of a particle of size $s$ as
\begin{subequations}
\begin{align}
v_r&=-{2\taus\over 1+\taus^2}\left(\eta v_\rmK -{1\over 2\taus}v_\nu\right),\\
v_\theta&=v_\rmK-{1\over 1+\taus^2}\left(\eta v_\rmK -{\taus\over 2}v_\nu\right).
\end{align}
\end{subequations}
A simplification can be made by noticing that the gas velocity in the second equation will always be negligible. For small values of $\taus$, the term is negligible over $\eta v_\rmK$. For large values of $\taus$, the azimuthal velocity becomes Keplerian so that $\delta v_\theta\rightarrow 0$ anyway. We therefore simplify the above system of equations to
\begin{subequations}
\begin{align}
v_r&=-{2(\taus+\tausnu)\over 1+\taus^2}\eta v_\rmK,\label{eq:vrnu}\\
v_\theta&=v_\rmK-{1\over 1+\taus^2}\eta v_\rmK\label{eq:vthetanu},
\end{align}
\end{subequations}
where we have introduced the critical stopping time below which the radial drift of a particle is mostly influenced by gas accretion in the disk:
\begin{equation}
\tausnu \equiv  {1\over 2}{-v_\nu\over \eta v_\rmK}.
\end{equation}

\begin{figure}
\includegraphics[width=\hsize,angle=0]{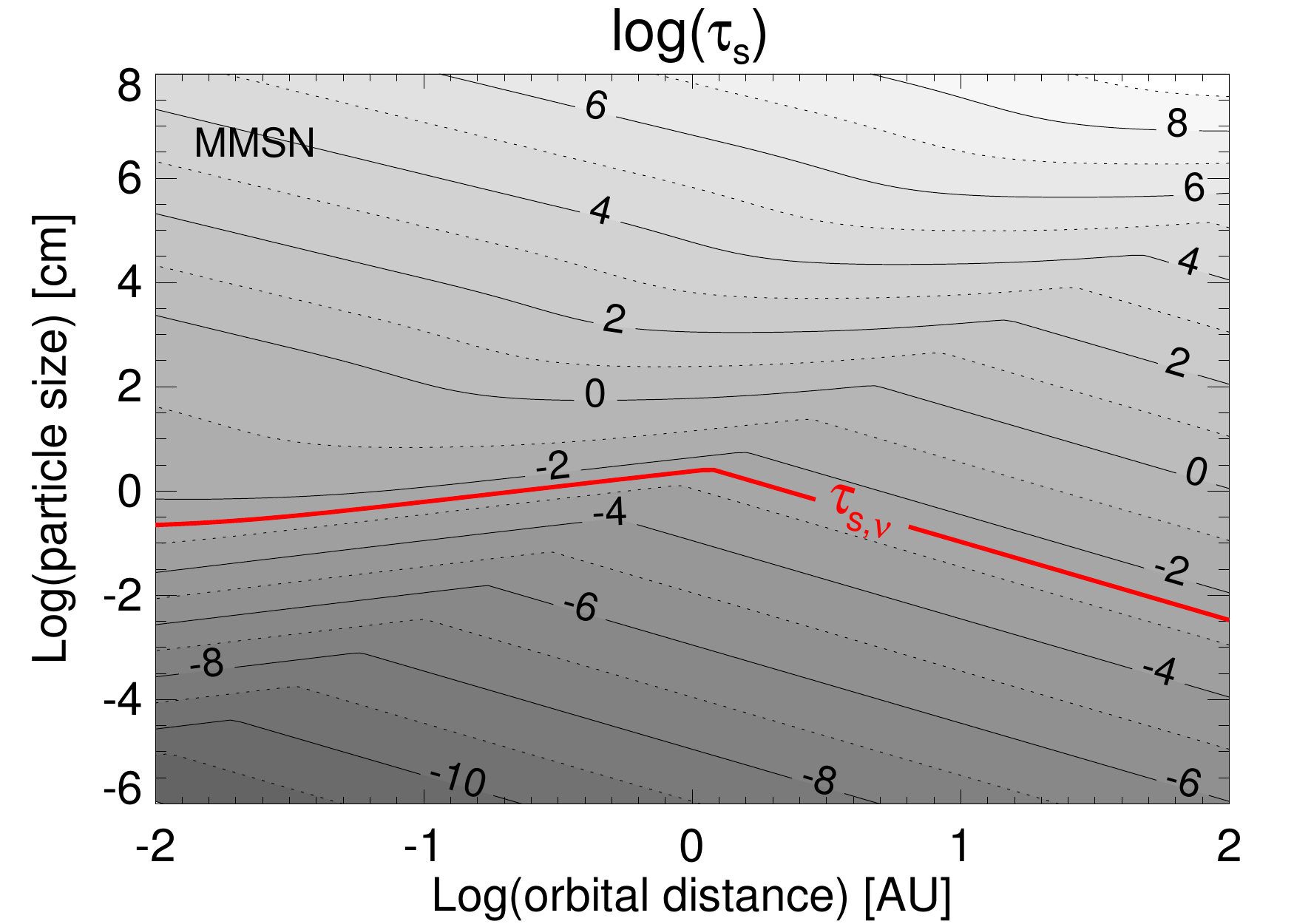}
\caption{Contour plot showing the decimal log of the dimensionless stopping time $\taus$ [see eq.~\protect\eqref{eq:taus}] as a function of particle size and orbital distance for a standard MMSN disk model, $\alpha=10^{-2}$, and assuming a physical density of particles $\rhos=1\,\gcc$. The red contour shows where $\taus=\tausnu$ [see eq. ~\protect\eqref{eq:tausnu}]. Particles below that line will drift mostly with the gas.}
\label{fig:taus}
\end{figure}

Assuming an $\alpha$ prescription for the turbulent viscosity, we can approximate the inward radial velocity of the gas as:
\begin{equation}
v_\nu\sim -\alpha c_\rmg h_\rmg / r,
\end{equation}
so that the critical stopping time becomes
\begin{equation}
\tau_{{\rm s,}\nu}={1\over 2}{\alpha\over \eta}\left(c_\rmg\over v_\rmK\right)^2.
\label{eq:tausnu}
\end{equation}
For the MMSN, $\tau_{{\rm s,}\nu} =0.3 \alpha$, independently of the orbital distance considered. At 1\,AU, this implies that dust particles as large as $0.6\,$cm are {\rev dominated} by the gas inflow for $\alpha=10^{-3}$. Figure~\ref{fig:taus} shows the behavior of both $\taus$ and $\tausnu$ as a function of orbital distance in the MMSN. 

The relative velocity between the particle and the gas is, within the approximations made, independent of the viscous drift:
\begin{equation}
\vrel={\taus\sqrt{4+\taus^2}\over 1+\taus^2}\eta v_\rmK.
\label{eq:vrel}
\end{equation}

Equations~\eqref{eq:FD}, \eqref{eq:Re}, \eqref{eq:CD}, \eqref{eq:taus}, and \eqref{eq:vrel} form a set of non-linear equations that, {\rev except in the  Epstein and $Re<1$ regime,} must be solved iteratively.

\subsection{The filtering length}

In the simplest case, only the dust particles orbiting in the same plane as the planetesimal have a chance to hit it. {\rev We define $2\bcol$ as the linear cross section of planetesimals. (For small planetesimals $\bcol\sim \Rp$.)} If dust particles stayed at the same altitude in the disk, only a maximum fraction $\bcol/\hdust$ of the dust would possibly collide with planetesimals, the remaining $1-\sqrt{\pi}\bcol/2\hdust$ eventually being accreted by the star. {\rev However, particles are moved up and down by turbulence (for small particles) or by their own epicyclic motions (for large particles with $\taus\wig{>}1$). \cite{Ciesla2010} shows that the vertical location of small particles in a disk are governed by an advective-diffusive equation which can be modeled through a Monte Carlo approach,
\begin{equation}
z_i=z_{i-1}-(\alpha+\taus)\Omega_\rmK z \delta t + {Rnd}\left[{2\over {var}}\alpha h_{\rm g}^2\Omega_\rmK \delta t \right]^{1/2},
\end{equation}
where $z_i$ and $z_{i-1}$ are the vertical location of a particle of dimensionless stopping time $\taus\wig{<}1$ at two instants separated by $\delta t$ and $Rnd$ is a random number drawn from a distribution of variance $var$. The advection term (proportional to $(\alpha+\taus)z$) appears as a result of imposing hydrostatic equilibrium in the disk in the vertical direction. It prevents particles from drifting outside the disk. On the basis of simulations of magneto-rotational turbulence in disks, one expects that the particles receive kicks in velocity every $\delta t\sim \Omega_\rmK^{-1}$. Using a Gaussian distribution for $Rnd$ with $var=1$, one can then show numerically that the characteristic time to cross the mid-plane for small particles is within 10\%: $2(\alpha+\taus)^{-1/2}\Omega_\rmK^{-1}$. On the other hand, large particles cross the mid-plane at each orbit \citep[e.g.,][]{YoudinLithwick2007}. Combining the two results yields a mid-plane crossing time
\begin{equation}
t_{\rm cross}\approx \left({2\over \sqrt{\alpha+\taus}}+1\right)\Omega_\rmK^{-1}.
\label{eq:tcross}
\end{equation}
}
The timescale depends on the value of the turbulence parameter only for very small grains that have a scale height equal to the gas scale height. Otherwise, the smaller particle scale height compensates for the added turbulent mixing. Large particles undergo epicyclic motions and their settling is independent of turbulence. 

We can then use eqs.~\eqref{eq:vrnu} and \eqref{eq:tcross} to calculate a filtering length, which is the length after which most grains, regardless of their initial positions, have crossed the mid-plane: 
\begin{equation}
\lambda_{\rm f}\equiv v_r t_{\rm cross} ={2(\taus+\tausnu)\over 1+\taus^2} \left({2\over \sqrt{\alpha+\taus}}+1\right)\eta r ,
\label{eq:lambda_f}
\end{equation}
{\rev
We thus obtain $\lambda_{\rm f}/r \sim \sqrt{\alpha}\eta$ when $\taus\wig{<}\alpha$,  $\lambda_{\rm f}/r \sim 4\sqrt{\taus}\eta$ when $1\wig{>}\taus\wig{>}\alpha$, and $\lambda_{\rm f}/r \sim (2/{\taus})\eta$ when $\taus\wig{>}1$. For the smallest particles, we can expect that $\lambda_{\rm f}/r <10^{-4}$, a very small filtering length. Even for the fastest drifting particles, for which we expect $h_\rmp\sim h_\rmd$ anyway, $\lambda_{\rm f}/r \wig{<} 4\eta\sim 8\times 10^{-3}$. 
}

This implies that even in the case of rapidly drifting particles, turbulence in the disk should ensure that particles always have the opportunity of encountering planetesimals in the mid-plane. 
In other words, if enough planetesimals are present, perfect filtering can occur, i.e., there may be cases for which the star can accrete dust-free gas. 

\section{Filtering in the geometrical limit}

\subsection{2D collision probabilities}\label{sec:2d}

We will now derive the rate of impact of drifting dust onto a non-drifting planetesimal on a circular orbit in the geometrical limit (i.e., neglecting gravitational focusing). In two dimensions, this impact rate is the product of the planetesimal linear cross section $\bcol$ and the encounter velocity $\vcol$:
\begin{equation}
{\cal R}_{\rm 2D}=2 \bcol \vcol.
\label{eq:R}
\end{equation}
The mass accretion of dust by planetesimals is then $\dot{M}_{\rm p}={\cal R} \Sigma_{\rm d}$. 

The velocity difference between the planetesimal and the particle calculated assuming that they do not interact is
\begin{equation}
\Delta v\equiv \sqrt{(v_r-V_r)^2+(v_\theta -V_\theta)^2},
\end{equation} 
where $v_r$ and $v_\theta$ are the particle radial and azimuthal velocities defined by eqs.~\eqref{eq:vrnu} and \eqref{eq:vthetanu} and $V_r$ and $V_\theta$ are those for the planetesimal. {\rrev We assume that $\tausnu\ll 1$}. For non-drifting planetesimals on circular orbits, $V_r=0$ and $V_\theta=v_\rmK$ thus yielding
\begin{equation}
\Delta v_{\rm circ}={\sqrt{1+4\taus^2}\over 1+\taus^2}\eta v_\rmK. 
\label{eq:vcirc}
\end{equation}

In the geometrical circular limit, $\bcol=R_\rmp$ and $\Delta v_{\rm col}=\Delta v_{\rm circ}$, i.e., we neglect any interaction resulting from e.g., gravitational focusing (see below). The impact rate in this limit is then
\begin{equation}
{\cal R}_{\rm 2D,geo,circ}=2R_\rmp\eta v_\rmK {\sqrt{1+4\taus^2}\over 1+\taus^2}.
\label{eq:Rgeocirc}
\end{equation}
When $\taus^2\ll \eta r/\Rp$, this expression is equivalent to eq.~(22) of \cite{OrmelKlahr2010} who derive this impact rate by an analysis of the trajectory of dust particles. 

The 2D probability that a planetesimal will accrete a given dust grain is then given by the ratio between the mass accreted $\dot{M}_{\rm p}$ and the global mass flow of dust grains past the planetesimal $2\pi r v_r \Sigma_{\rm d}$,
\begin{equation}
{\cal P}_{\rm 2D}={{\cal R}_{\rm 2D}\over 2\pi r v_r}.
\label{eq:P2Dbasic}
\end{equation}
In the geometrical limit, this can be shown to yield\footnote{\rev \cite{Kary+1993} obtain a different expression. However, they account for the shear in the disk, which is on the order of $v_{\rm K} \Rp/r$, but not the azimuthal velocity difference between the planetesimal and the particle which is much larger, i.e., on the order of $\eta v_{\rm K}$. Their expression is valid only for large particles, for which three-body effects dominate anyway (see section~\ref{sec:3body}). An expression including both terms is provided by \cite{OrmelKlahr2010} and rederived in Appendix~\ref{sec:shear}.}
\begin{equation}
{\cal P}_{\rm 2D,geo,circ}={R_\rmp\over \pi r}\sqrt{1+\frac{1}{4\left(\taus+\tausnu\right)^2}}.
\label{eq:P2Dcirc}
\end{equation}
The probability thus decreases from a maximum value $\sim \Rp/(2\pi r \tausnu)$ for low values of $\taus$ when the drift is controlled by the gas flow to a {\rev minimum} value of $\Rp/(\pi r)$ for large dust particles. For small particles, the high probability is due to the dust and the planetesimal having very different azimuthal speeds. For large particles, the azimuthal and radial velocity difference between dust and planetesimal is small so that only the dust located at the right azimuth will collide with the planetesimal. We note that the probability remains finite for $\taus\rightarrow \infty$, i.e., for non-drifting dust. Indeed, eq.~\eqref{eq:P2Dcirc} is time-independent: it yields the probability of collision in the limit of infinite time. This view is thus complementary to that obtained by using collision rates to estimate planetesimal growth timescales. 

\subsection{3D collision probability}

\begin{figure}[htb]
\centering\includegraphics[width=\hsize/2,angle=0]{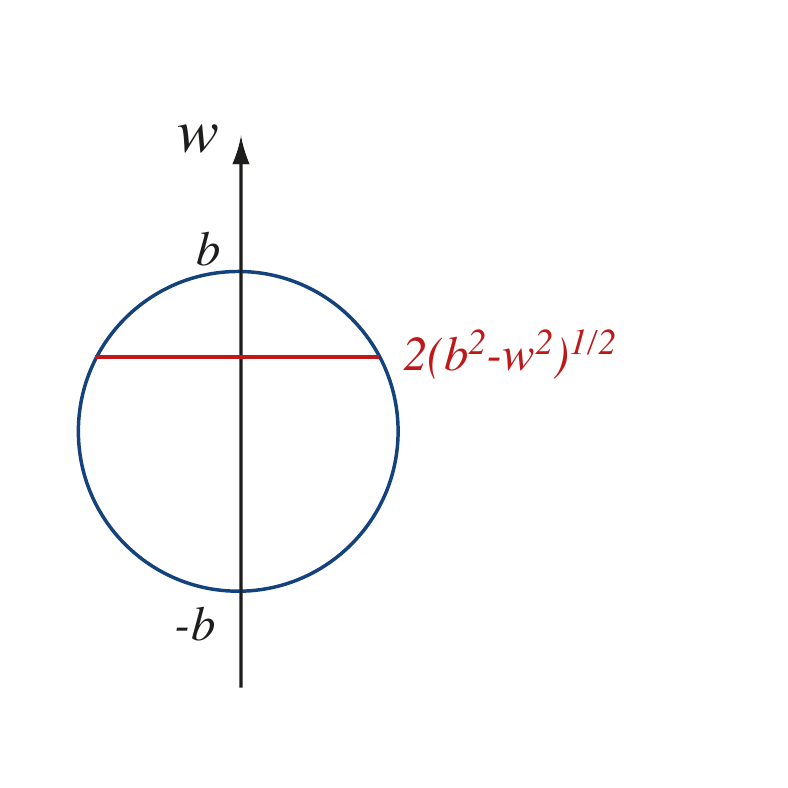}
\caption{Sketch showing how the linear cross section of an assumed spherical planetesimal varies as a function of the impact distance $w$, given a maximum (equatorial) cross section $2b$.}
\label{fig:cross-section}
\end{figure}

 We have so far considered cases when dust and planetesimals orbit in the same plane. In reality, the 3D nature of the disks discussed in section~\ref{sec:geometry} must be considered. As in the 2D case, the collision rate is a product of a linear cross section to a collision velocity, but this time, as shown in Fig.~\ref{fig:cross-section}, the linear cross section depends on $w$, the impact parameter of the dust. The mass accretion rate can thus be calculated by integrating over all impact parameters and all positions of the planetesimal on its trajectory, accounting for variations of the dust density and of the collision velocity,
\begin{equation}
\dot{M}_\rmp =\int_0^{2\pi} \!\int_{-b}^b \! 2\sqrt{\bcol^2-w^2} \vcol \rho_\rmd (r,z+w)dw\,d{\cal M},
\end{equation}
where both $\vcol$, $r$ and $z$ are functions of {\rev the mean anomaly} $\cal M$. 
We simplify the integration by assuming that we can neglect the variations of $\rho_\rmd$ along the trajectory of the planetesimal (i.e., the values of $\cal M$) and that we can calculate a mean collision velocity: 
\begin{equation}
\vmeancol={1\over 2\pi} \int_0^{2\pi}\vcol({\cal M}) d{\cal M}.
\end{equation}
The accretion rate can then be written as an integral over the cross section of the planetesimal:
\begin{equation}
 \dot{M}_\rmp=\int_{-\bcol}^\bcol 2\sqrt{\bcol^2-w^2}\, \vmeancol \rho_{\rm d,0}e^{-w^2/\hdust^2} dw.
\end{equation}
It may seem strange to neglect the variations of $\rho_\rmd$ along the trajectory of the planetesimal and not over its cross section. However, it will enable us to link the 2D solutions obtained in the previous section to the 3D ones. 

It turns out that the above integral may be expressed in terms of a sum of modified Bessel functions {\rev of the first kind $I_0$ and $I_1$}:
\begin{equation}
 \dot{M}_\rmp=\pi \bcol^2 \vmeancol\rho_{\rm d,0} e^{-\bcol^2/2\hdust^2}\left[I_0\left(\bcol^2\over 2\hdust^2\right) +I_1\left(\bcol^2\over 2\hdust^2\right)\right].
\end{equation}
We recover the 2D case when $\bcol\gg \hdust$, in which case 
\[
I_0\left(\bcol^2\over 2\hdust^2\right)\sim I_1\left(\bcol^2\over 2\hdust^2\right)\sim {\hdust\over \bcol} {e^{\bcol^2/2\hdust^2}\over\sqrt{\pi}},
\]
which implies that 
\[
 \dot{M}_\rmp=2\sqrt{\pi} \bcol \vmeancol\rho_{\rm d,0} \hdust.
\]
Given that $\Sigma_\rmd=\sqrt{\pi}\hdust \rho_{\rm d,0}$, we indeed obtain 
\begin{equation}
 \dot{M}_\rmp=2\bcol \vmeancol \Sigma_\rmd \qquad\mbox{when $\bcol\gg \hdust$}.
\end{equation}

In the more natural case when $\bcol\ll \hdust$, then $I_0(\bcol^2/2\hdust^2)\rightarrow 1$ and $I_1(\bcol^2/2\hdust^2)\rightarrow 0$ so that
\begin{equation}
 \dot{M}_\rmp=\pi \bcol^2 \vmeancol\rho_{\rm d,0} \qquad\mbox{when $\bcol\ll \hdust$}.
\end{equation}
We will hereafter assume that this hypothesis is true most of the time. {\rrev When the dust disk is thinner than the planetesimal disk, we will also assume that the relation remains valid but with $\hdust$ replaced by the corresponding scale height for planetesimals, $h_\rmp$.

A simplified relation between the 3D and 2D collisions probabilities is hence: 
\begin{equation}
{\cal P}={\rm min}\left[1,{\sqrt{\pi}\over 2}{\bcol\over {\rm max}\left(\hdust,h_\rmp\right)}\right] {\cal P}_{\rm 2D} . 
\label{eq:P3D}
\end{equation}
}
Given our choice for the expression of $\hg$ and hence $\hdust$, there is a factor $\sqrt{\pi /8}\approx 0.63$ mismatch with the expressions assumed by \cite{OrmelKlahr2010} and \cite{OrmelKobayashi2012}. We note that $\hg$ and $\hdust$ may be defined arbitrarily, so that it is natural that a factor on the order of unity appears next to the $\bcol/\hdust$ ratio. 

{\rrev In the analytical derivations hereafter, we will assume for simplicity that ${\rm max}\left(\hdust,h_\rmp\right) =\hdust$. However in the numerical calculations and plots we will assume $h_\rmp=0.01\,h_{\rm g}$, which will limit the accretion probability of boulders larger than several meters. (We note that in the limit that $\hp<\hdust$, eq.~\eqref{eq:P3D} is independent of $\hp$.)

In the geometrical, circular limit (i.e., $b=\Rp$) and for small-enough dust such that $\hdust >(2/\sqrt{\pi})\Rp$, we obtain} 
\begin{equation}
{\cal P}_{\rm geo,circ}={1\over 2\sqrt{\pi}}{\Rp^2\over \hdust r}\sqrt{1+{1\over 4(\taus+\tausnu)^2}}.
\label{eq:P3Dgeocirc}
\end{equation}
This probability thus depends on size of the planetesimals and properties of the dust (thickness of the dust disk and stopping time). Both the thickness of the dust disk [eq.~\eqref{eq:hdust}] and the quantity $\tausnu$ [eq.~\ref{eq:tausnu}] are directly dependent on the value of the turbulence parameter. We can thus write the probability in terms of the gas (instead of the dust) scale height and further condense the notation,
\begin{equation}
{\cal P}_{\rm geo,circ}={1\over 2\sqrt{\pi}}{\Rp^2\over h_\rmg r}\chialpha,
\label{eq:P3Dgeocirc_chi}
\end{equation}
where 
\begin{equation}
\chialpha \equiv \max\left(1,\sqrt{\taus\over \alpha}\right) \sqrt{1+{1\over 4(\taus+\tausnu)^2}}.
\label{eq:chialpha}
\end{equation}
It is useful to approximate this last quantity depending on the stopping time of dust particles and turbulent viscosity parameter, using the fact that for the MMSN scaling, $\tausnu\approx 0.3\alpha$: 
\begin{equation} 
\chialpha\approx
\begin{cases} 
1.7/\alpha & \mbox{when $\taus\wig{<}\alpha/4$}, \\
1/\sqrt{4\alpha\taus} & \mbox{when $\alpha/4\wig{<}\taus\wig{<}1/2$},\\
\sqrt{\taus/\alpha} & \mbox{when $1/2\wig{<}\taus$}.
\end{cases} 
\label{eq:chi_alphataus}
\end{equation}
{\rev When leaving aside very large particles (governed by different relations), the value of $\chi_{\alpha,\taus}$ ranges from $1.7/\alpha$ to its lowest value $1/\sqrt{2\alpha}$ for $\taus=1/2$ particles. The $1.7/\alpha$ value for small particles is determined entirely by the (mean) gas advection rate. The $1/\sqrt{\alpha}$ dependence, however, relates to the ability of particles to be lofted by turbulence.}

\subsection{Planetesimal size distribution and filtering mass}

We now look for the mass in planetesimals that is required for an efficient filtering of dust grains at any location. If we assume a single size $\Rp$ for all planetesimals (monodisperse size distribution), then we only have to obtain $N_{\Rp}$ the number of planetesimals of that size such that $N_{\Rp} {\cal P}=1$ and obtain the total mass from
\begin{equation}
M_{{\rm filter,}\Rp}(s)={4\pi \over 3}\rhop \Rp^3 {1\over {\cal P}(s,\Rp)},
\end{equation}
where $M_{{\rm filter,}\Rp}(s)$ is thus the mass in planetesimals of size $\Rp$ that is required to efficiently filter dust of size $s$. 
In the geometrical circular regime, this may be approximated as
\begin{equation}
M_{{\rm filter,}\Rp}(s)\sim {8\pi\sqrt{\pi} \over 3}\rhop {h_\rmg  r \Rp\over \chi_{\alpha,\taus}},
\nonumber
\end{equation}
where $\chi_{\alpha,\taus}$ takes the values defined in eq.~\eqref{eq:chi_alphataus} depending on $\alpha$ and $\taus$. In this limit, the filtering mass is therefore linearly proportional to the size of the planetesimals considered. Larger planetesimals have a larger cross section individually, but a smaller collision probability collectively yielding a larger filtering mass.

Of course, planetesimals in the protoplanetary disk will not all be the same size. Instead, we envision that their size distribution is defined by a power law,
\begin{equation}
dN/dR_\rmp = AR_\rmp^{-{q}},
\label{eq:dndmp}
\end{equation}
where $N$ is the number density of planetesimals of size $R_\rmp$ and $A$ is a constant defined by the total number of planetesimals, $N_0=\int A R^{-{q}}dR$. We therefore calculate the mass in planetesimals of all sizes required to capture all incoming dust grains of size $s$ as 
\begin{equation}
M_{\rm filter}(s)={4\pi\over 3}\rhop {\int_{R_{\rm p, min}}^{R_{\rm p, max}} R^{3-{q}} dR  \over \int_{R_{\rm p, min}}^{R_{\rm p, max}} {\cal P}(s,R)\,R^{-{q}} dR },
\label{eq:Mfilter}
\end{equation}
where ${\cal P}(s,R)$ is given by eq.~(\ref{eq:P3D}). 

{\rrev A collisional cascade yields a size distribution characterized by} ${q}\sim 7/2$ \citep{Tanaka+1996}. (This corresponds to a mass distribution $dN/dM_\rmp\propto M_\rmp^{-11/6}$). For comparison, the size distribution in the asteroid belt is characterized by ${q}\approx 4$ for asteroids with radii above 70\,km, and ${q}\approx 2.5$ for smaller asteroids \citep[e.g.,][]{Morbidelli+2009}. {\rrev For simplicity, we adopt $q=7/2$ by default.}
{\rev In the geometrical circular limit, we then obtain}
\begin{equation}
M_{{\rm filter,geo,circ}}(s)\sim {8\pi\sqrt{\pi} \over 3}\rhop {h_\rmg  r R_{\rm p,mean}\over \chi_{\alpha,\taus}},
\label{eq:Mfilter,geo,circ}
\end{equation}
where 
\begin{equation}
R_{\rm p,mean}={\int_{R_{\rm p, min}}^{R_{\rm p, max}} R^{3-{q}} dR\over \int_{R_{\rm p, min}}^{R_{\rm p, max}} R^{2-{q}} dR}.
\end{equation} 
It is easy to show that $R_{\rm p,mean}=R_{\rm p,min}$ if ${q}\gg 4$, $R_{\rm p,mean}=R_{\rm p,max}$ if ${q}\ll 3$ and for ${q}=7/2$, $R_{\rm p,mean}=\sqrt{R_{\rm p, min}R_{\rm p, max}}$. 

For the MMSN, this translates into
\begin{equation}
M_{{\rm filter,geo,circ,MMSN}}(s)=\rau^{9/4} {\rhop\over 1\,\gcc} {R_{\rm p,mean}\over 1\,{\rm km}}M_0
\end{equation}
with 
\begin{equation}
M_0=
\begin{cases}
(\alpha/10^{-3})\times 1.86\,{\rm M_\oplus} & \mbox{if $\taus\wig{<}\alpha/4$,} \\
\sqrt{\alpha/10^{-3}}\sqrt{2\taus}\times 83\,{\rm M_\oplus} & \mbox{if $\alpha/4\wig{<}\taus\wig{<}1/2$,} \\
\sqrt{\alpha/10^{-3}}/\sqrt{2\taus}\times 83\,{\rm M_\oplus} & \mbox{if $1/2\wig{<}\taus$.}
\end{cases}
\end{equation}
These estimates show that even with very small planetesimals, efficient filtering requires very large masses in the geometrical limit. Small grains appear to be easier to filter, but this does not include the hydrodynamical effects which tend to drastically suppress accretion. Conversely, the estimates do not include possible focusing for large mass planetesimals and/or dust. 

\subsection{Filtering efficiency}\label{sec:xfilter}


We now want to estimate the fraction of dust that is effectively filtered by the planetesimals and its complement,  the fraction of dust that passes through the planetesimal belt and is accreted by the star. For simplicity we will assume that collisions effectively lead to accretion, although the reality may be different. We consider $N_\rmp(\Rp,r)d\Rp dr$, the number of planetesimals with sizes between $\Rp$ and $\Rp+d\Rp$, and at orbital distances between $r$ and $r+dr$. We define the disk-integrated filtering efficiency for dust of size $s$ as
\begin{equation}
X_{\rm filter}(s)\equiv \int_{r_{\rm in}}^{r_{\rm out}} {d\over dr}\left( \int_{R_{\rm p,min}}^{R_{\rm p,max}} {\cal P}{dN_\rmp\over d\Rp} d\Rp \right) dr.
\label{eq:Xfilter_def}
\end{equation}
Using the distribution of planetesimal sizes defined by eq.~\eqref{eq:dndmp} and the surface density of planetesimals $\Sigma_\rmp(r)$, the disk-integrated filtering efficiency can be written in the form
\begin{equation}
X_{\rm filter}(s)=\int_{r_{\rm in}}^{r_{\rm out}} 2 x_{\rm filter}(s,r) {dr\over r},
\end{equation}
where $x_{\rm filter}(s,r)$ is the characteristic filtering efficiency for dust of size $s$ at orbital distance $r$,
\begin{equation}
x_{\rm filter}(s,r)={\pi r^2 \Sigma_\rmp(r) \over M_{\rm filter}(s,r)},
\label{eq:xfilter}
\end{equation} 
where $M_{\rm filter}$ is defined by eq.~\eqref{eq:Mfilter}. The quantity $x_{\rm filter}$ thus measures the filtering efficiency at each orbital distance, allowing us to estimate where the dust is mostly absorbed. 


In the geometrical, circular limit, eq.~\eqref{eq:Mfilter,geo,circ} leads us to
\begin{equation}
x_{\rm filter,geo,circ}={3\over 8\sqrt{\pi}}{\Sigma_\rmp\over \rhop R_{\rm p,mean}} {r\over h_\rmg} \chi_{\alpha,\taus}.
\label{eq:xfilter_geo}
\end{equation}
Replacing $\Sigma_\rmp$ and $h_\rmg$ by their MMSN scaling relations yields the filtering efficiency of a swarm of planetesimals with the same total mass as the present solar system:
\begin{equation} 
\begin{split}
&x_{\rm filter,geo,circ,MMSN}=1.1 \left({\chi_{\taus,\alpha}\over 10^{3}}\right) \times \\
&\qquad \left({\rhop \over  1\,\gcc}\right)^{-1} \left({R_{\rm p,mean}\over 1\,{\rm km}}\right)^{-1} \rau^{-7/4}. 
\end{split}
\label{eq:xfilter_geo_MMSN}
\end{equation}
{\rev Integrating over orbital distances, and assuming that $r_{\rm in}\ll r_{\rm out}$ yields
\begin{equation} 
\begin{split}
&X_{\rm filter,geo,circ,MMSN}=1.2 \left({\chi_{\taus,\alpha}\over 10^{3}}\right) \times \\
&\qquad\qquad\qquad \left({\rhop \over  1\,\gcc}\right)^{-1} \left({R_{\rm p,mean}\over 1\,{\rm km}}\right)^{-1} r_{\rm in,AU}^{-7/4}. 
\end{split}
\label{eq:X_filter_geo}
\end{equation}

These equations show that, in the geometrical circular limit, perfect filtering (i.e., $X_{\rm filter,geo,circ,MMSN}\sim 1$) by a MMSN disk of solids is only possible close to the star, for small particles, and/or weak turbulence (keeping in mind that $\chi_{\taus,\alpha}$ is between $1/\alpha$ for small particles and $1/\sqrt(2\alpha)$ for larger ones), and by a swarm of {\em very} small planetesimals dominated by km-sized planetesimals. This is shown in Fig.~\ref{fig:xfilter_geo} for two values of $\alpha$: the filtering efficiency in the geometrical circular limit is minimum for $\taus\sim 1$, and increases both for larger and smaller particle sizes. As shown by the dashed curves in Fig.~\ref{fig:xfilter_geo}, if gas advection by the accreting disk could be neglected ($\chi_{\taus,\alpha}\sim 0$), the filtering efficiency would increase steadily for smaller dust particles because of their very slow migration which increases the probability of hitting a planetesimal. However, gas advection forces small particles to move at a rate which prevents this increase: the filtering efficiency then becomes independent of the size for particles such that $\taus\wig{<}\alpha$.     

\begin{figure}
\includegraphics[width=\hsize,angle=0]{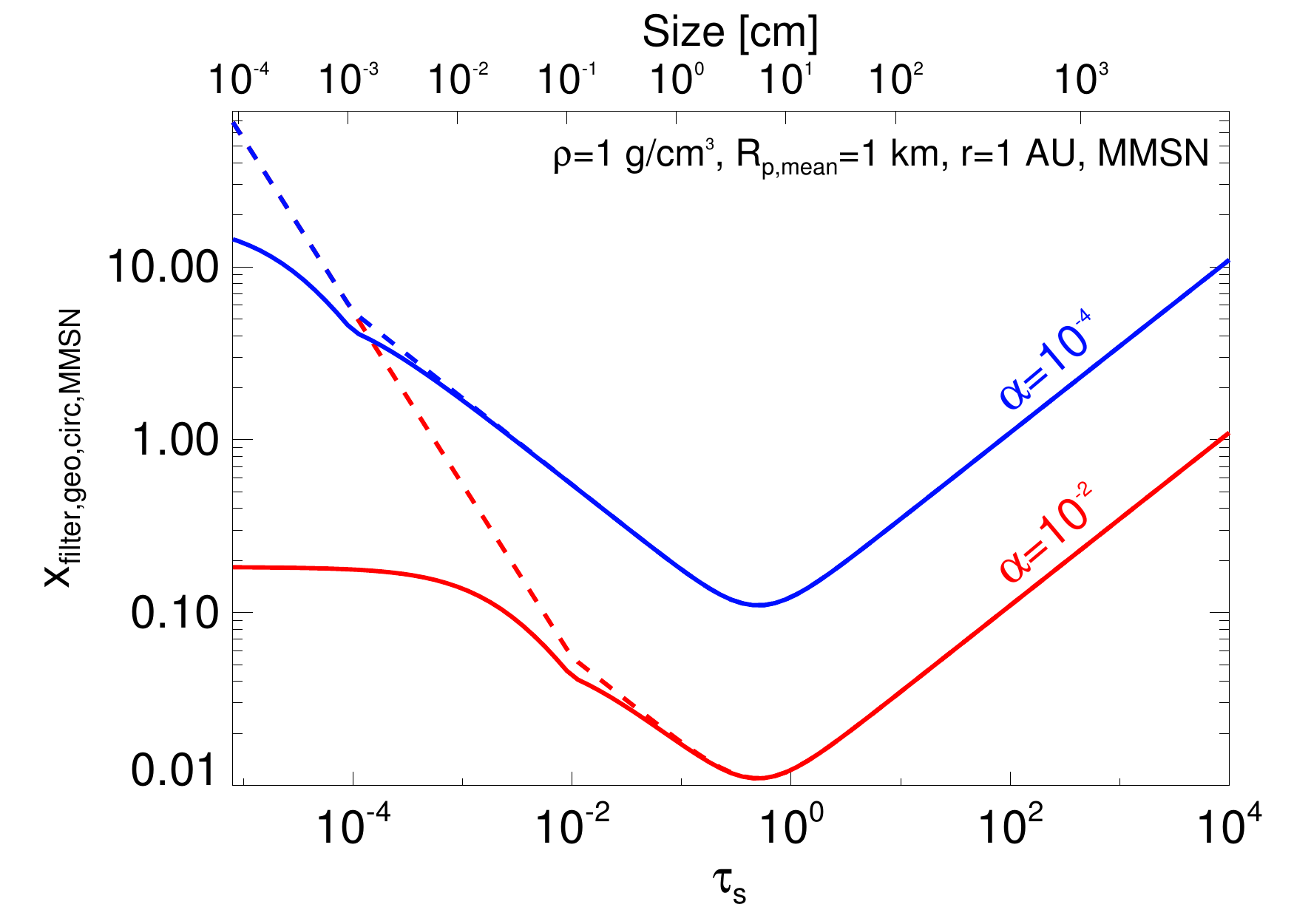}
\caption{{\rev Filtering factor by an MMSN disk of planetesimals of size $R_{\rm p, mean}=1\,$km and density $\rhop=1\,\gcc$ in a MMSN gas disk at 1\,AU, as a function of the dust stopping time $\taus$ for two values of the turbulence parameter $\alpha=10^{-4}$ (blue curves) and $\alpha=10^{-2}$ (red curves), respectively. The corresponding physical sizes of dust particles (top axis) assume a density of $1\,\gcc$ and the MMSN conditions at 1\,AU. The solid lines are obtained from eq.~\protect\eqref{eq:xfilter_geo_MMSN}. The dashed lines indicate the results obtained when neglecting gas advection ($\chi_{\taus,\alpha}$=0). Small values of $\taus<\alpha$ correspond to a situation in which the dust particles are mixed with the gas vertically and have a radial migration due to gas drag that is slower than advection by the gas. We note that the filtering factor is inversely proportional to $R_{\rm p, mean}$, implying that for size distributions equivalent to that of the present-day asteroid belt ($R_{\rm p,mean}\sim 70\,\rm km$), filtering would be very inefficient.}} 
\label{fig:xfilter_geo}
\end{figure}

This} demonstrates that an efficient filtering of dust particles is difficult to achieve at least in the geometrical, circular limit, even with a mass of planetesimals equivalent to the MMSN, which likely represents the maximum available limit, at least for the solar system. Importantly, the filtering efficiency is inversely proportional to planetesimal density and size, implying that it is most effectively done by small and/or porous planetesimals. It is also proportional to the surface density of planetesimals which is generally thought to be a strong function of orbital distance, implying that filtering is most effective close to the star. 

{\rev However, we must consider additional processes arising in real disks.}

\section{Filtering: Additional effects}

\subsection{Accounting for eccentric orbits}\label{sec:eccentricity}

Following \cite{OrmelKlahr2010}, we have so far only considered planetesimals with circular orbits. In this case, the relative motion between the planetesimal and the dust results from the combination of the difference in azimuthal velocity of the planetesimal (Keplerian) and of the dust (sub-Keplerian) and the small radial drift velocity of the dust. For the more general case of planetesimals on eccentric orbits, the encounter velocities are generally much larger because of the non-zero radial velocity of the planetesimals themselves. {\rev A detailed solution of the problem is complex \citep[see][]{Kary+Lissauer1995}. We provide instead a simplified treatment.} 

We now consider the encounter of a dust grain at orbital distance $r$ with a planetesimal with eccentricity $e$ and true anomaly $\nu_\rmp$. Its semi-major axis $a$ is thus such that $a(1-e^2)=r(1+e\cos\nu_\rmp)$, and its radial and azimuthal velocities are, respectively,
\begin{subequations}
\begin{align}
V_r &= {e\sin\nu_\rmp\over (1+e\cos\nu_\rmp)^{1/2}} v_\rmK ,\label{eq:Vr}\\ 
V_\theta &= (1+e\cos\nu_\rmp)^{1/2} v_\rmK , \label{eq:Vtheta}
\end{align}
\end{subequations}
where $v_\rmK$ is the Keplerian velocity at $r$. 

The encounter velocity between the dust grain and the eccentric planetesimal is now obtained by calculating $\Delta v_r=v_r- V_r$ and $\Delta v_\theta=v_\theta - V_\theta$,
\begin{subequations}
\begin{align}
{\Delta v_r\over v_\rmK} &=-{2(\taus+\tausnu)\over 1+\taus^2}\eta - {e\sin\nu_\rmp\over (1+e\cos\nu_\rmp)^{1/2}} ,\label{eq:Deltavr}\\
{\Delta v_\theta\over v_\rmK} & = -{1\over 1+\tau_\rms^2} \eta -\left[(1+e\cos\nu_\rmp)^{1/2}-1\right] , \label{eq:Deltavtheta}
\end{align}
\end{subequations}
where we have neglected variations of $\taus$ and $\eta$ with $r$. The instantaneous accretion rate is still calculated as eq.~\eqref{eq:Rgeocirc}, but because it is now time dependent, we integrate over the planetesimal's orbit,
\begin{equation}
<{\cal R}_{\rm 2D,geo}>=2\Rp{1\over 2\pi}\int_0^{2\pi}\sqrt{(\Delta v_r)^2+(\Delta v_\theta)^2}d{\cal M},
\end{equation}
where the variations of $\Sigma_\rmd$ over the planetesimal's orbit have been neglected and the integral is performed over all mean anomalies ${\cal M}$ ($\nu_\rmp$ is calculated as a function of ${\cal M}$ using Kepler's equation).

\begin{figure}
\includegraphics[width=\hsize,angle=0]{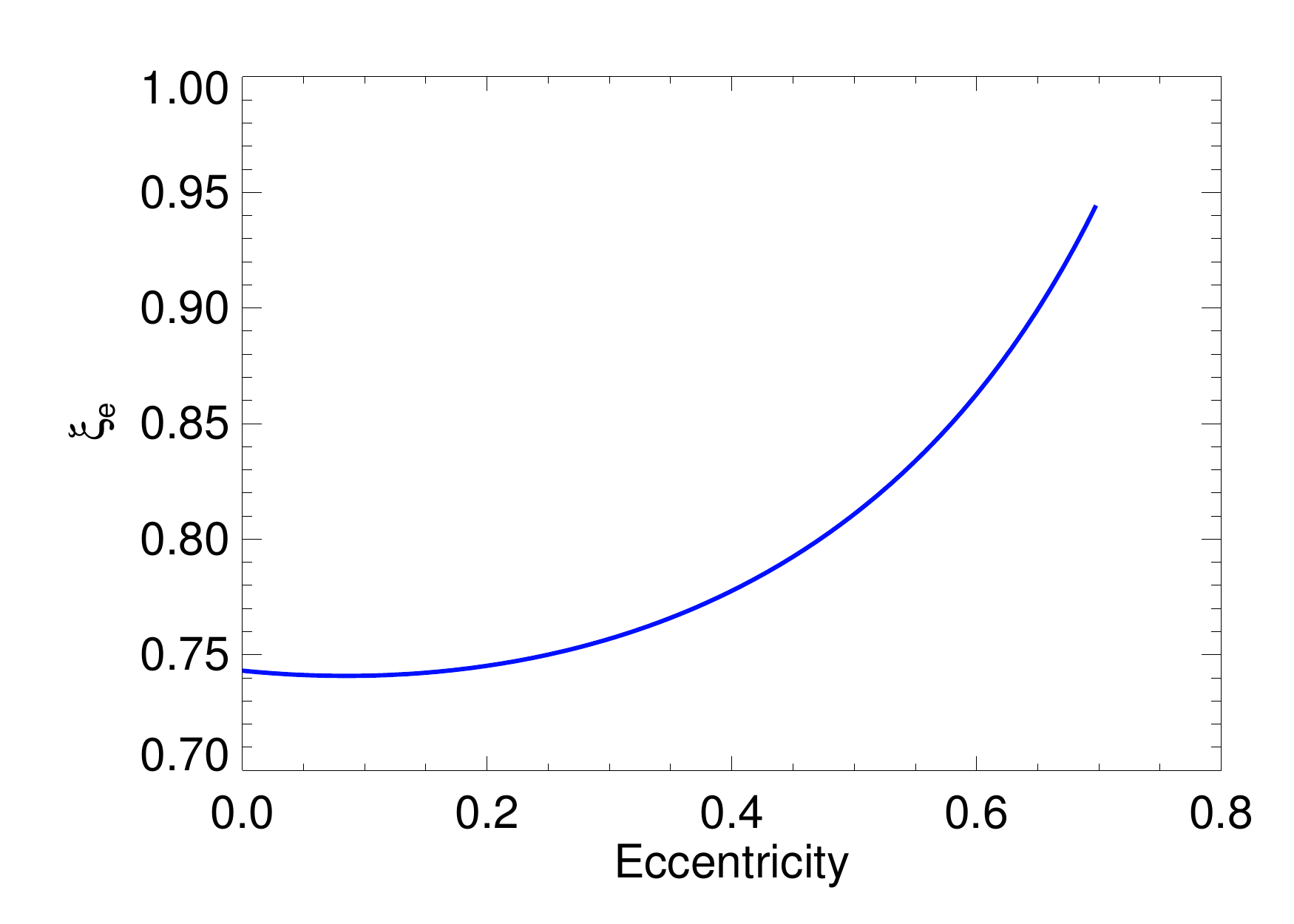}
\caption{Eccentricity factor $\xi_e$ [see eq.~\protect\eqref{eq:xie}] as a function of the eccentricity of the planetesimal.}
\label{fig:xie}
\end{figure}

When performing the integral, $\Delta v_r$ and $\Delta v_\theta$ both contain a constant part and a variable part proportional to the planetesimal's eccentricity. Because this variable part is proportional to $e\sin\nu_\rmp$ and $e\cos\nu_\rmp$, respectively, its contribution is either positive or negative. If its amplitude is smaller than that of the constant part, we can expect it to average out to a negligible amount. If, however, its amplitude is larger than the constant term, because of the square function, it will eventually become dominant. This will occur when
\begin{equation}
\begin{cases} 
e> {2(\taus+\tausnu)\over 1+\taus^2}\eta & \mbox{for $V_r^2>v_r^2$}\\
e> {1\over 1+\taus^2}\eta & \mbox{for $V_\theta^2>v_\theta^2$}.
\end{cases} 
\nonumber
\end{equation}
Both conditions must be met, and since the first one is more important for $\taus>1$ and the second one for $\taus<1$, we can equivalently distinguish the circular and eccentric regimes by comparison of $e$ to the critical eccentricity 
\begin{equation}
e_{\rm crit}=\frac{\sqrt{1+4\taus^2}}{1+\taus^2} \eta. \label{eq:ecrit}
\end{equation}
When $e<e_{\rm crit}$, the eccentricity term has both positive and negative contributions ($\sin\nu_\rmp$ and $\cos\nu_\rmp$ change sign) and can be considered as averaging out to negligibly small values. When $e\gg e_{\rm crit}$, the eccentricity term dominates so that both positive and negative values of $\sin\nu_\rmp$ and $\cos\nu_\rmp$ lead to an increase in the accretion probability. For this case, the mean accretion rate can be written
\begin{equation}
<{\cal R}_{\rm 2D,geo,ecc}>=2\Rp \xi_e e v_\rmK,
\end{equation}
where we have defined 
\begin{equation}
\begin{split}
\xi_e=\int_0^{2\pi}  {1\over e}\bigg\{ &{\left( e\sin \nu_\rmp \right)^2\over 1+e\cos\nu_\rmp} + \\
 &\left[\sqrt{1+e\cos\nu_\rmp}-1\right]^2 \bigg\}^{1/2} d{\cal M}.
\end{split}
\label{eq:xie}
\end{equation}
The value of $\xi_e$ can be calculated by solving Kepler's equation and is only a function of the eccentricity of the object considered. It is shown in Fig.~\ref{fig:xie}. For small eccentricities $e\wig{<}0.2$ we obtain $\xi_e\approx \xi_e(e=0)=0.743$, a value that we adopt from now on. 

Combining the low-eccentricity and high-eccentricity limits, the average encounter velocity can be calculated from the quadratic mean of eqs.~(\ref{eq:Deltavr}) and (\ref{eq:Deltavtheta}) (and assuming $\tausnu\ll 1$)
\begin{equation}
\Delta v \approx\eta v_\rmK {\sqrt{1+4\taus^2}\over 1+\taus^2}f_e,\label{eq:Deltav_e}
\end{equation}
where the eccentricity factor $f_e$ is calculated as:
\begin{equation}
f_e= \sqrt{1+\left[\xi_e \frac{1+\taus^2}{\sqrt{1+4\taus^2}} {e\over \eta}\right]^2}.\label{eq:fe}
\end{equation}

Dropping the mean terms, we thus write the mean accretion rate 
\begin{equation}
{\cal R} _{\rm 2D,geo}= {\cal R} _{\rm 2D,geo,circ} f_e, \label{eq:R2De}
\end{equation}
and the average collision probability as
\begin{equation}
{\cal P} _{\rm 2D,geo}= {\cal P} _{\rm 2D,geo,circ} f_e. \label{eq:P2De}
\end{equation}

\begin{figure}
\includegraphics[width=\hsize,angle=0]{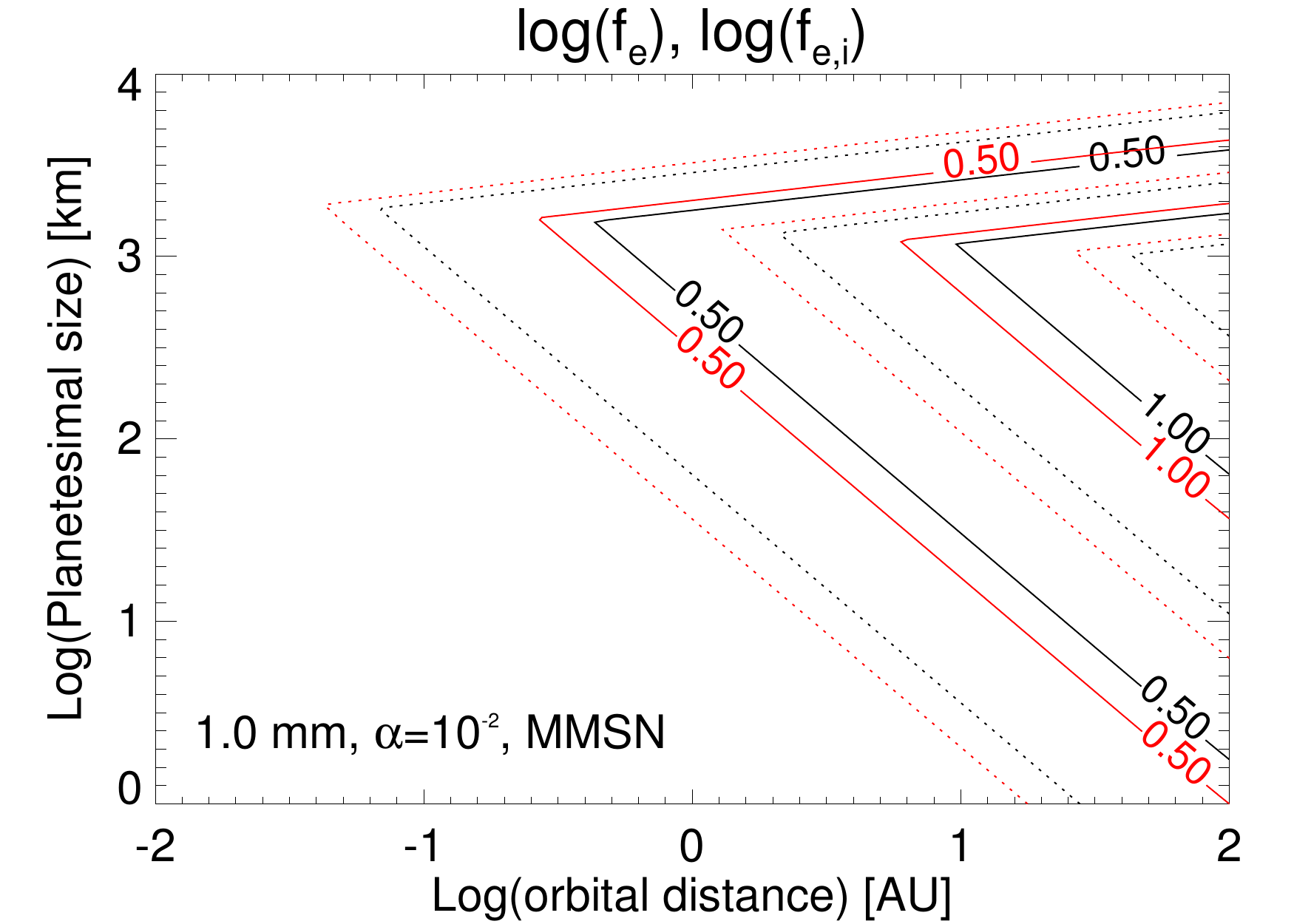}
\caption{Contour plot of the decimal logarithm of the eccentricity focusing factors $f_e$ [see eq.~\protect\eqref{eq:fe}] (in black) and $f_{e,i}$ [see eq.~\protect\eqref{eq:fei}] (in red) as a function of orbital distance in a MMSN disk for dust of $1$\,mm in size. The eccentricity is the same as in Fig.~\protect\ref{fig:ecc_I08}.}
\label{fig:colprob_orbdist_fe}
\end{figure}

The behavior of $f_e$ as a function of dust size is shown in Fig.~\ref{fig:colprob_orbdist_fe}. The diagram of course resembles the contour plot obtained for eccentricities (see Fig.~\ref{fig:ecc_I08}). The eccentricity factor will become important for large planetesimals and at large orbital distances, i.e., when the gas drag is reduced. Quantitatively, when a planetesimal eccentricity $e=0.01$ we obtain $f_e\approx 4.25$ for small dust particles. When plotted as a function of $\taus$ (not shown here), the eccentricity factor is found to be small and flat for small values of $\taus$. A small drop occurs for $\taus=1/\sqrt{2}$, but the eccentricity factor then rapidly increases and becomes proportional to $\taus$ because for large pebbles and boulders, the relative motions between particles and planetesimals due to gas drag become small. Eccentricity effects then play a dominant role in these encounters. 

The difference between these the two low- and high-eccentricity regimes can be understood by calculating the ratio $\Gamma$ of the time for the dust to drift from the apocenter to the pericenter divided by half of the orbital period of the planetesimal: 
\begin{equation}
\Gamma={2ea\over v_r}{\Omega_\rmK\over \pi}.
\end{equation}
It is easy to show using eq.~(\ref{eq:vrnu}) that
\begin{equation}
\Gamma={2\over \pi}{1+\taus^2\over 2(\taus+\tausnu)}{e\over \eta}.
\label{eq:Gamma}
\end{equation}
Thus for $\taus\gg 1$, $e=e_{\rm crit}$ correspond to $\Gamma={2/\pi}$. When $\Gamma<1$, dust grains generally have only one possible encounter with the planetesimal on its orbit and thus the collision probability defined by eq.~(\ref{eq:P2De}) is very close to that obtained for a planetesimal on a circular orbit. Conversely, when $\Gamma\gg 1$, the drifting dust has many possibilities of colliding with the eccentric planetesimal, thus increasing the collision probability proportionally to the number of encounters. For $\taus\ll 1$ however, the multiple encounters are balanced by a collision geometry which is less favorable for collisions so that a higher eccentricity is required for the collision probability to become proportional to eccentricity. 


\subsection{Including inclinations}
We have seen in the 2D case that eccentricity may help collisions. However, as discussed in section~\ref{sec:geometry}, both eccentricities and inclinations are expected to be excited, with $i\sim e/2$. It is thus important to also consider inclined orbits in the calculation of collision probabilities.

Adapting eqs.~\eqref{eq:Deltavr} and \eqref{eq:Deltavtheta} to the 3D case, and neglecting variations of the {\rrev gas} and dust velocity with height in the disk \citep[see][]{TakeuchiLin2002}, we can express the approach velocities between an inclined planetesimal and dust as
\begin{subequations}
\begin{align}
{\Delta v_r\over v_\rmK} &=-{2(\taus+\tausnu)\over 1+\taus^2}\eta - {e\sin\nu_\rmp\over (1+e\cos\nu_\rmp)^{1/2}} ,\label{eq:3Deltavr}\\
{\Delta v_\theta\over v_\rmK} & = -{1\over 1+\tau_\rms^2} \eta -\left[(1+e\cos\nu_\rmp)^{1/2}\cos i-1\right] , \label{eq:3Deltavtheta}\\
{\Delta v_z\over v_\rmK} & = -(1+e\cos\nu_\rmp)^{1/2}\sin i. \label{eq:3Deltavphi}
\end{align}
\end{subequations}
The approach velocity $\Delta v=\sqrt{(\Delta v_r)^2+(\Delta v_\theta)^2+(\Delta v_z)^2}$ can be approximated by retaining only the leading squared terms: 
\begin{equation}
\begin{split}
\left({\Delta v\over v_\rmK}\right)^2\approx &{4\taus^2+1\over (1+\taus^2)^2}\eta^2+\\
 &e^2\left[{\sin^2\nup\over (1+e\cos\nup)^2}+{1\over 4}\cos^2\nup\right]+2(1-\cos i).
\end{split}
\end{equation}
By averaging approximately over the mean anomalies we then obtain  
\begin{equation}
\Delta v \approx\eta v_\rmK {\sqrt{1+4\taus^2}\over 1+\taus^2}\fei,\label{eq:Deltav_ei}
\end{equation}
where the eccentricity-inclination factor $\fei$ is calculated as
\begin{equation}
\fei= \sqrt{1+\frac{(1+\taus^2)^2}{{1+4\taus^2}}\frac{1}{\eta^2}\left(\xi_e^2 e^2+ \sin^2 i\right)}.\label{eq:fei}
\end{equation}
These relations thus replace eqs.~\eqref{eq:Deltav_e} and \eqref{eq:fe}. Given that $i\sim e/2$, and $\xi_e\approx 0.743$, the inclusion of inclination effects thus leads to approach velocities which are about $50\%$ higher than when considering eccentricities alone (see Fig.~\ref{fig:colprob_orbdist_fe}).

\subsection{Gravitational focusing}\label{sec:focusing}

For large-enough planetesimals, gravitational focusing must be taken into account. Given the complex behavior of the three-body problem and the added complexity of gas drag, this can become a challenging problem requiring detailed numerical integrations. Here, we follow \cite{OrmelKlahr2010} in deriving simplified expressions for the focusing factor $f_{\rm focus}$ in three regimes: (1) in the settling regime, gas drag is the dominant mechanism controlling the trajectories of small particles around sufficiently large planetesimals;  (2) in the Safronov regime, gravitational effects dominate over gas drag and lead to the classical gravitational focusing; (3) in the three-body regime, the interaction of (large) particles and (large) planetesimals must include the global geometry of the problem and the presence of the central star and leads to much more complex effects. 

{\rev For simplicity, we will not account for eccentric and/or inclined orbits when calculating the enhancement of the collision probability due to gravitational focusing. First, this is a complex problem, beyond the scope of the present paper. Second, the increased cross section of eccentric planetesimals is generally matched with a larger encounter velocity so that the collision probability is often close to that of a planetesimal on a circular orbit. Last, in any case, planetesimal orbits are expected to have a range of eccentricities from circular to the maximum eccentricity so that the effect of eccentricity and gravitational focusing may be decoupled.}

Because we will consider gravitational effects, it is useful to define the ratio of the planetesimal to the stellar mass
\begin{equation}
\mup\equiv M_\rmp/M_*
\end{equation}
and the Hill radius
\begin{equation}
R_{\rm H}\equiv r (\mup/3)^{1/3}.
\label{eq:RH}
\end{equation} 
The Bondi radius will become important in the settling regime. {\rev Following \cite{LambrechtsJohansen2012}, we define it} as a function of the dust-planetesimal encounter speed $\Delta v$,
\begin{equation}
R_{\rm B}\equiv {G M_\rmp\over \Delta v^2}.
\label{eq:RB}
\end{equation}
For small enough dust particles with $\taus\wig{<}1$, eq.~\eqref{eq:Deltav_e} $\Delta v$ becomes independent of the dust size and $R_{\rm B}$ depends only on the planetesimal properties\footnote{{\rev A definition of the Bondi radius often used involves the isothermal sound speed $c_\rmg$ rather than $\Delta v$ which is useful when estimating the radius at which a protoplanet begins developing an atmosphere. At 1\,AU, $c_\rmg/\Delta v\approx 20$, hence, the capture of an atmosphere occurs for a radius of $\sim 2000\,$km, assuming a physical protoplanet density of $1\,\gcc$. However, as shown by eq.~\eqref{eq:Rp0}, already at $\Rp\sim 50\,$km, the gravity of the planetesimal is sufficient to affect the trajectory of incoming dust and affect the collision probability.}}. 

Using $\Delta v\approx \eta v_\rmK$, one can show that the minimum size for a planetesimal to have a Bondi radius larger than the planetesimal size is
\begin{equation}
R_{\rm p}^{R_{\rm B}=R_{\rm p}} \approx \eta \left(\rhop\over \rhostar\right)^{-1/2} \left(r\over R_*\right)^{-1/2} R_*,
\label{eq:rb=rp}
\end{equation}
where $\rhostar$ and $R_*$ are the central star's mean density and radius, respectively. 
For $\rhop=1\,\gcc$ and our prescriptions for an MMSN-like disk, $R_{\rm p}^{R_{\rm B}=R_{\rm p}} \approx 100\,$km independently of the orbital distance. 

For larger planetesimals, an important limit is the size at which the Bondi radius equals the Hill radius. This occurs for
\begin{equation}
R_{\rm p}^{R_{\rm B}=R_{\rm H}} \approx {\eta\over 3^{1/6}} \left(\rhop\over \rhostar\right)^{-1/3} R_*,
\label{eq:rb=rh}
\end{equation}
implying $R_{\rm p}^{R_{\rm B}=R_{\rm H}} \approx 1200\,{\rm km}\ \rau^{1/2}$ with the same fiducial planetesimal density.

\subsubsection{The settling regime}\label{sec:settling}
\begin{figure}[htb]
\includegraphics[width=\hsize,angle=0]{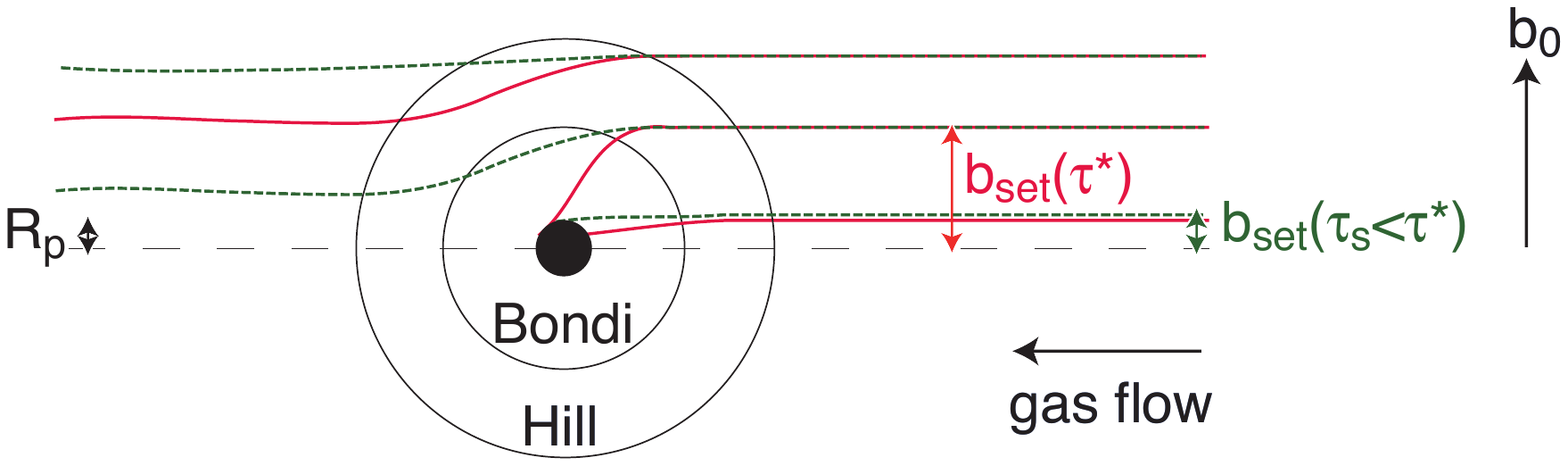}
\caption{Illustration of dust capture in the settling regime. Planetesimals with a radius $\Rp$ between about 100\,km and 1000\,km have a Bondi radius larger than $\Rp$ but smaller than the Hill radius (see text). For the critical stopping time $\taus^*$ corresponding to a stopping time equal to the time necessary for the unperturbed dust particle to cross the Bondi sphere, the effective cross section for capture is $b_{\rm set}\sim R_{\rm B}$. The approximate trajectory of the dust for $\taus=\taus^*$ is indicated by red lines. For smaller dust particles with $\taus<\taus^*$ (dashed green lines), the deflection by the gravitational pull of the planetesimal is smaller, hence a smaller effective capture radius.}
\label{fig:settling-regime}
\end{figure}

For small particles (with $\taus\wig{<}1$) the effect of gas drag can combine with the gravitational pull of the planetesimal to increase the collision cross section, as illustrated in Fig.~\ref{fig:settling-regime}. We consider the case where planetesimal and dust approach each other with an initial velocity $\Delta v_0$. Because we only consider small particles with $\taus\wig{<}1$, we approximate the dust to planetesimal velocity difference as $\eta v_\rmK$. However, because of the increase in the capture cross section we also have to account for the Keplerian shear. The planetesimal being at an orbital distance $r$ and the dust particle being initially at $r+b_0$, by assuming $b_0/r\ll 1$ we obtain
\begin{equation}
\Delta v_0=\left(\eta +{3\over 2}{b_0\over r}\right)v_\rmK,
\label{eq:Deltav0}
\end{equation}
where $b_0$ is the initial impact parameter.   The change in velocity $\delta v$ of the dust particle in the reference frame of the planetesimal can be estimated by multiplying the gravitational acceleration from the planetesimal to the stopping time of the particle,
\begin{equation}
\delta v={\taus\over \Omega_\rmK}{G M_\rmp\over b_0^2}.
\label{eq:delta v}
\end{equation}
We expect the capture cross section in the settling regime to correspond to the impact parameter for which $\delta v$ is on the order of the initial approach velocity $\Delta v_0$. {\rev In the limit of $\taus\ll 1$ it can be shown that the condition for settling is $\delta v\sim \Delta v_0/4$, which \cite{OrmelKlahr2010} then adopted for all $\taus$.}

By combining these equations, we derive a cubic equation that defines the impact parameter for collisions $b_0$:
\begin{equation}
3b_0^3 + 2 \eta rb_0^2 -8\taus \mup r^3=0.
\label{eq:cubical}
\end{equation}
This equation, which has one real positive root, is the same as that of \cite{OrmelKlahr2010}. 

For large values of $\taus$ we obtain $b_0\sim (8\taus)^{1/3}R_\rmH$. This equation hence that the impact parameter should increase indefinitely with $\taus$, simply because eq.~\eqref{eq:delta v} predicts a velocity change that also increases with $\taus$. In reality, this remains true only up to the point when the stopping time and interaction time $b_0/\Delta v_0$ become comparable. As noted by \cite{LambrechtsJohansen2012}, this corresponds to the point when the impact parameter equals the Bondi radius. Particles with larger stopping times tend to be more easily gravitationally scattered and the effective planetesimal cross section to capture them decreases accordingly. This critical stopping time $\taus^*$ and corresponding impact radius $b_0^*$ are linked by
\begin{equation}
\taus^*/\Omega_\rmK=b_0^*/\Delta v.
\label{eq:taus_om}
\end{equation} 
Using eq.~\eqref{eq:delta v} and $\delta v\approx \eta v_\rmK/4$ (neglecting shear this time), we obtain for the critical stopping time:
\begin{equation}
\taus^*=4 \frac{\mu_\rmp}{\eta^3}. 
\label{eq:taus*}
\end{equation}
This may be also written in terms of the Hill and Bondi radius of the planetesimal:
\begin{equation}
\taus^*=\frac{4}{3^{1/2}} \left(R_{\rm B}\over R_{\rm H}\right)^{3/2}.
\end{equation}
{\rev Using eqs.~\eqref{eq:taus_om}, \eqref{eq:taus*}, and \eqref{eq:RB}, one can show that} $b_0^*=4R_{\rm B}$. 

\begin{figure}
\includegraphics[width=\hsize,angle=0]{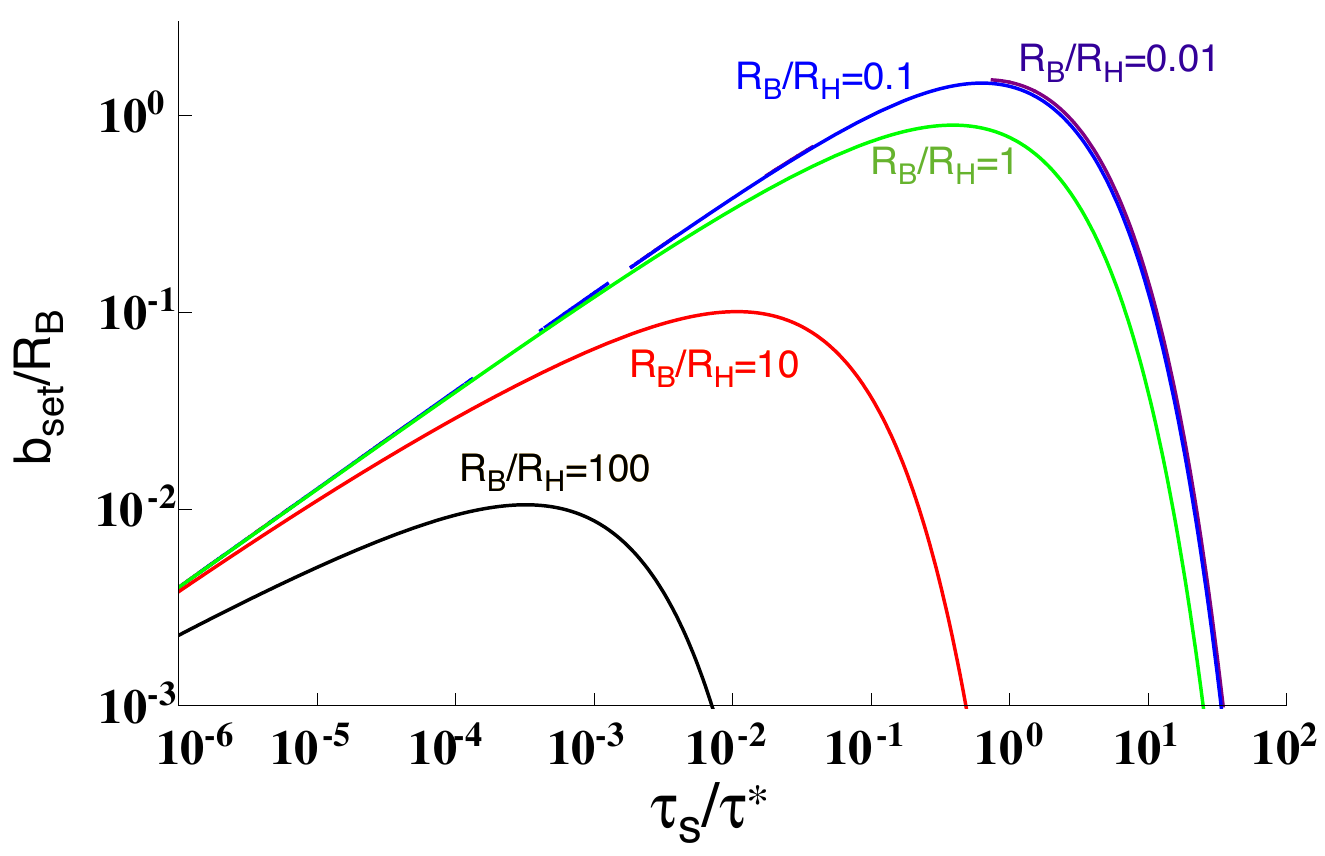}
\caption{Value of the effective capture radius $b_{\rm set}$ in units of the Bondi radius $r_{\rm B}$ as a function of the ratio of the stopping time of the dust particle to the critical stopping time  $\taus/\taus^*$. The curves correspond to different planetesimal sizes expressed in term of the Bondi radius to the Hill radius, $R_{\rm B}/R_{\rm H}=0.01$, 0.1, 1, 10, and 100, as labeled.}
\label{fig:cubical_ok10}
\end{figure}

In order to link the weak and strong coupling regimes and reproduce numerical results, \cite{OrmelKobayashi2012} propose the fit
\begin{equation}
b_{\rm set}=b_0 e^{- (\taus/\min(2,\taus^*))^{0.65}}.
\label{eq:bset}
\end{equation}
Figure~\ref{fig:cubical_ok10} shows how the effective cross section in the settling regime $b_{\rm set}$ changes as a function of the stopping time and of the size of the planetesimal (measured in terms of the ratio $R_{\rm B}/R_{\rm H}\propto \Rp^2$). The results for small planetesimals ($R_{\rm B}/R_{\rm H}\wig{<}1$) are very similar to \cite{LambrechtsJohansen2012} (see their Fig.~4) and are almost independent of planetesimal size: The maximum settling cross section is obtained for $\taus\sim \taus^*$ and is equal to the Bondi radius of the planetesimal for the prescribed encounter velocity. For larger planetesimals with a Bondi radius larger than the Hill radius, the increase in cross section is suppressed {\rev because of the Keplerian shear}. 
(We note that the minimum for $\taus^*=2$ in eq.~\eqref{eq:bset} corresponds to $R_{\rm B}=0.9 R_{\rm H}$.) {\rev A detailed study of the consequences of eqs.~\eqref{eq:cubical} and \eqref{eq:bset} on the form of the solutions is presented in Appendix~\ref{sec:analytical}.}

The encounter velocity for the settling regime is obtained directly from eqs.~\eqref{eq:bset} and \eqref{eq:Deltav0}: 
\begin{equation}
\Delta v_{\rm set}=\left( \eta + {3\over 2}{b_{\rm set}\over r}\right) v_\rmK. 
\end{equation}
The focusing factor for the settling regime is then obtained by dividing the new collision rate $2b_{\rm set}\Delta v_{\rm set}$ by the geometrical rate $2R_\rmp \Delta v$: 
\begin{equation}
f_{\rm set}={b_{\rm set}\over R_\rmp} \left(1+{3\over 2}{b_{\rm set}\over r}{1\over \eta}\right).
\end{equation}

\subsubsection{The Safronov regime}

{\rev
When planetesimals are small enough so that their gravitational reach is not affected by stellar tides, and the settling effect from Sect.~\ref{sec:settling} is not important, the impact {\rrev parameter is given by the well-known \cite{Safronov1972} relation:}
\begin{equation}
b_{\rm Saf}^0=R_\rmp \sqrt{1+\frac{2 G M_\rmp}{R_\rmp \Delta v_{\rm circ}^2}},
\label{eq:bSaf0}
\end{equation}
where $\Delta v_{\rm circ}$ is defined by eq.~\eqref{eq:vcirc}. 
{\rrev This relation, however, fails in the limit when gas drag becomes important, both in the settling regime described previously and in the hydro regime to be explained next. This occurs when the stopping time $\taus$ becomes smaller than the time required to cross the planetesimal Hill sphere, $\tau_{\rm H}=\Omega_\rmK R_{\rm H}/\Delta v_{\rm circ}$ .} Taking into account such an effect for small $\taus$, we set the focusing factor in the Safronov regime as:
\begin{equation}
f_{\rm Saf}={1\over 1+\tau_{\rm H}/\taus}\sqrt{1+2\mup \frac{r}{R_\rmp} \left(\frac{v_\rmK}{\Delta v_{\rm circ}}\right)^2}.
\label{eq:fSaf}
\end{equation}
The corresponding impact parameter is $b_{\rm Saf}=b_{\rm Saf}^0 /( 1+\tau_{\rm H}/\taus)$.  The added factor ensures that the focusing factor has the usual form when $\taus\gg\tau_{\rm H}$, while it becomes smaller than one when gas drag becomes significant ($\taus\ll\tau_{\rm H}$) outside of the validity range of eq.~\eqref{eq:bSaf0}. In this case the impact parameters are given by the expressions for the geometric, settling, or hydro regimes instead\footnote{\rrev{Equation~\eqref{eq:fSaf} may be obtained by adding a drag term $E_{\rm drag}=(1/2)v_{\rm Saf}^2\tau_{\rm H}/\taus$ to the usual energy conservation equation and by assuming angular momentum conservation. This is very approximate because a proper treatment would require including both the physics of the settling and hydro regimes. However, it acts to suppress the value of $f_{\rm Saf}$ outside its validity range and has a negligible effect inside of it.}}.
}

\subsubsection{The three-body regime}\label{sec:3body}
For large particles ($\taus >1$) around large planetesimals, both the gravity of the planetesimals and of the central star must be taken into account. This leads to a much more complex behavior including the presence of horseshoe orbits and the fact that only particles with specific impact parameters can enter the planetesimals' Hill sphere. 

The approach velocity derived from three-body numerical integrations without gas drag is \citep{OrmelKlahr2010}
\begin{equation}
\Delta v_{\rm 3b}\approx 3.2 {R_\rmH \over r}v_\rmK,
\end{equation}
whereas the impact rate is:
\begin{equation}
b_{\rm 3b}=\left(1.7\sqrt{R_\rmp R_\rmH}+R_\rmH/\taus\right)e^{-\left[0.7(\eta/\taus)(r/R_\rmH)\right]^5}.
\end{equation}
This second equation is based on eq.~(31) from \cite{OrmelKlahr2010}, {\rrev but multiplied by an exponential factor to account for gas drag and better reproduce the results of numerical models \citep{OrmelKobayashi2012}.}
Using eqs.~\eqref{eq:R}, \eqref{eq:vcirc}, \eqref{eq:Rgeocirc}, {\rrev and \eqref{eq:RH}}, we write the focusing factor in the three-body regime as
\begin{equation}
\begin{split}
f_{\rm 3b}=&{1\over \eta}{1+4\taus^2\over \sqrt{1+4\taus^2}}\times \\
&3.2\left[\left(\mup \Rp r\right)^{1/2}+{0.5\over \taus}\mup^{2/3} r\right]e^{-(\eta \mup^{-1/3}\taus^{-1})^{5}}.
\end{split}
\end{equation}

\subsubsection{The gravitational focusing factor}

Following \cite{OrmelKobayashi2012}, we write 
\begin{equation}
f_{\rm focus}= \max\left(1,f_{\rm set},f_{\rm Saf},f_{\rm 3b}\right).
\label{eq:ffocus}
\end{equation}
The value of $f_{\rm focus}$ thus determines the regime that determines gravitational focusing, with $1$ indicating a geometrical regime. Based on this determination, we assess the impact radius and the encounter velocity:
\begin{equation}
\begin{cases}
\bcol=b_{\rm geo}, \vcol=\Delta v & \mbox{in the geometrical regime,}\\
\bcol=b_{\rm set}, \vcol=\Delta v_{\rm set} & \mbox{in the settling regime,}\\
\bcol=b_{\rm Saf}, \vcol=\Delta v_{\rm Saf} & \mbox{in the Safronov regime,}\\
\bcol=b_{\rm 3b}, \vcol=\Delta v_{\rm 3b} & \mbox{in the three-body regime.}\\
\end{cases} 
\label{eq:focusing}
\end{equation}

\begin{figure}
\includegraphics[width=\hsize,angle=0]{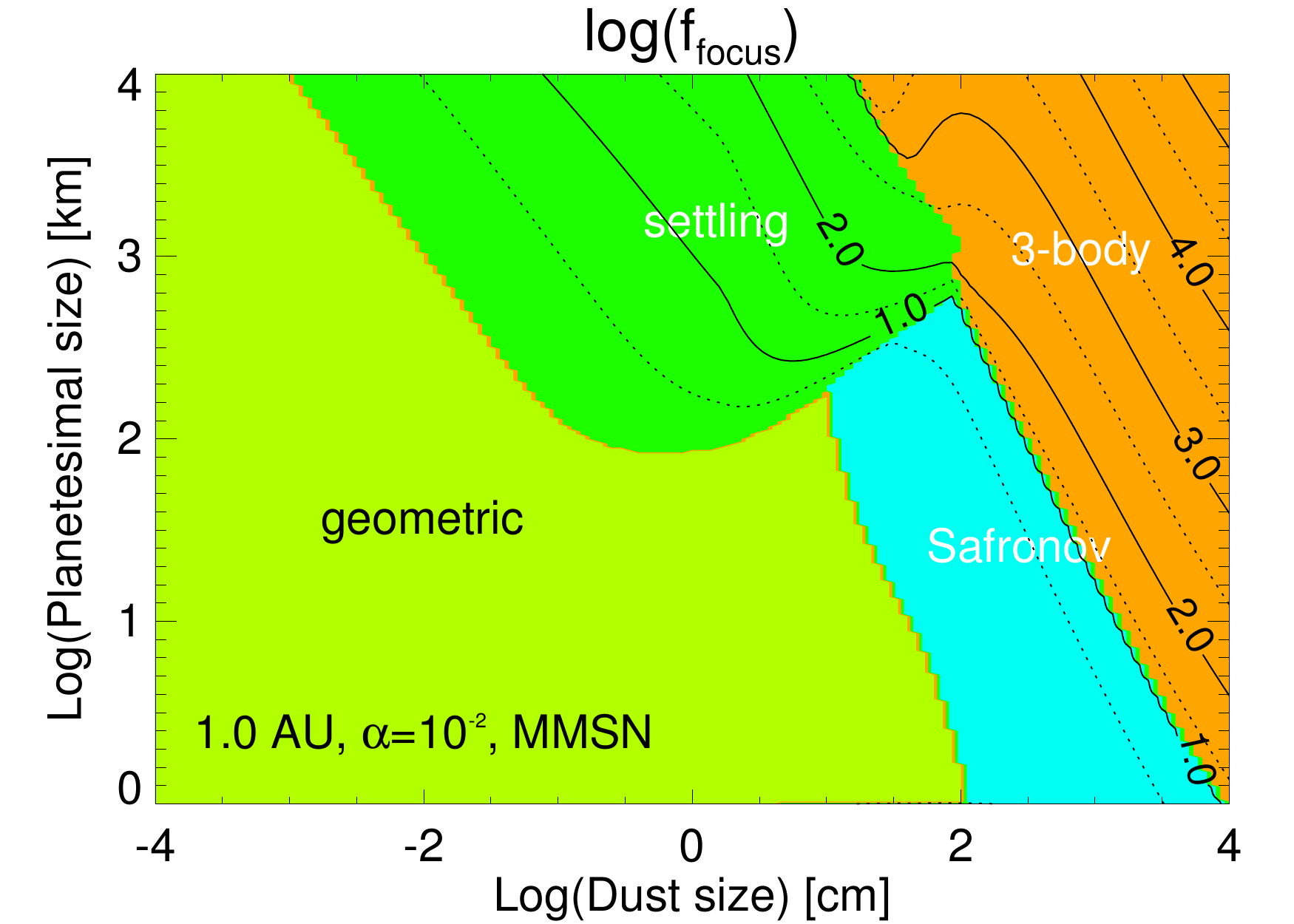}
\caption{Contours of the logarithm of the gravitational focusing factor $\log_{10}(f_{\rm focus})$ [see eq.~(\protect\ref{eq:ffocus})] as a function of dust size (in cm) and planetesimal size (in km), at 1\,AU in the MMSN. The physical density of dust and planetesimals is set to $1$ and $3\,\gcc$, respectively. The different physical regimes (geometric, Safronov, settling, three-body) are identified by different colors. This figure assumes that planetesimals are on circular orbits.}
\label{fig:ffocus_circ}
\end{figure}

Figure~\ref{fig:ffocus_circ} shows how values of $\ffocus$ change as a function of dust and planetesimal sizes, assuming planetesimals on circular orbits. For small planetesimals ($\wig{<}100\,$km) and dust sizes such that $\taus\wig{<}1$ (corresponding to about $1\,$m at 1\,AU), the focusing factor is very close to one, meaning that the accretion probability is defined by the geometric probability. For larger sizes, gravitational effects begin to dominate and lead to very high values of $\ffocus$. In this diagram, most of the area is dominated by collisions {\rev in the geometric and} Safronov regimes. For planetesimals above 100km, the encounters with dust {\rev mostly} take place in the settling regime, in which the gravitational pull of the planetesimals, which tends to increase the collision probability, is strongly suppressed by the gas drag. By comparison, three-body collisions play a role only when considering large particles and they will thus only play a limited role here. 



\subsection{Hydrodynamical effects}\label{sec:hydro}

\subsubsection{The hydrodynamical flow}

\begin{figure}[htb]
\includegraphics[width=\hsize,angle=0]{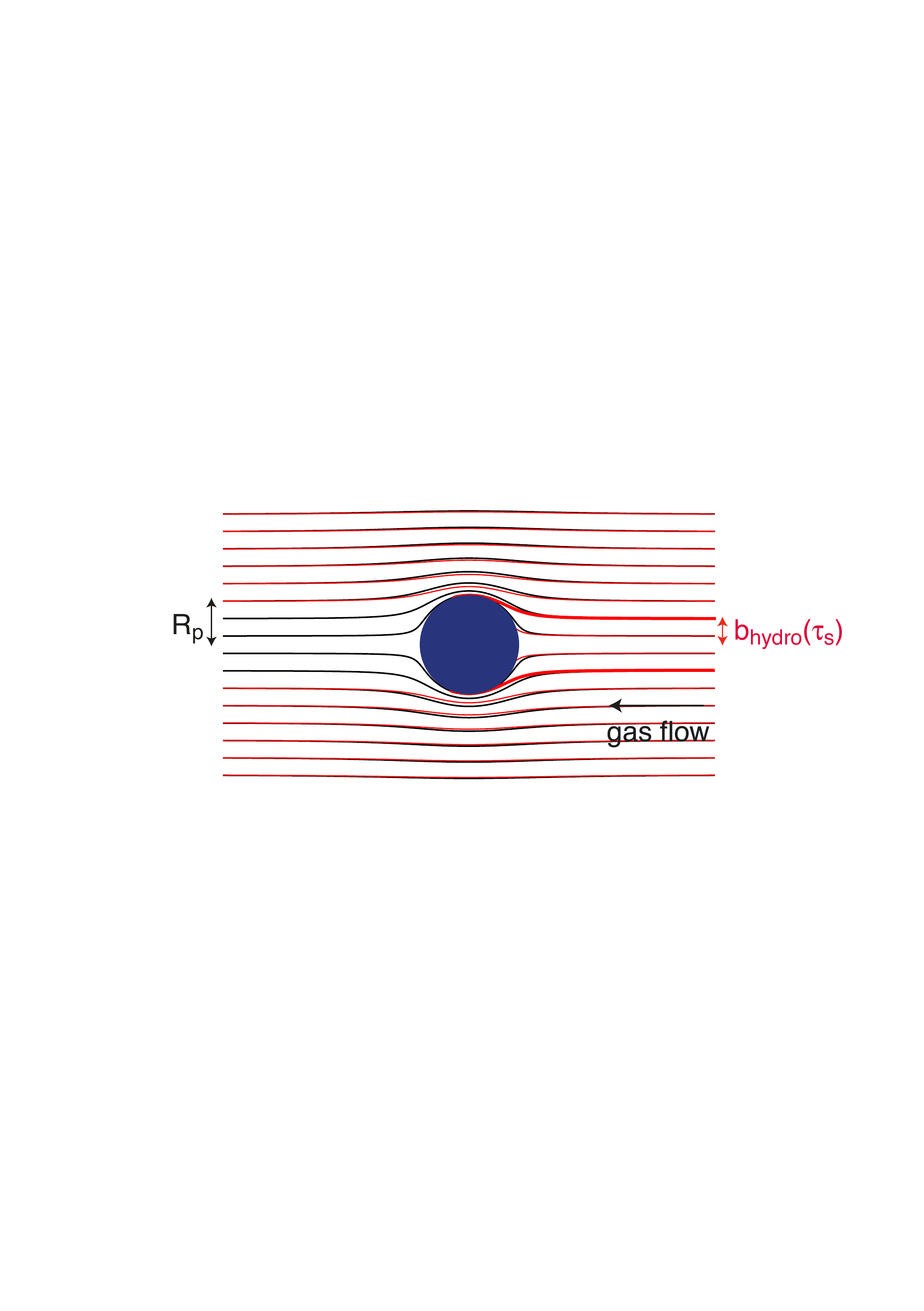}
\caption{Illustration showing how gas streamlines (in black) are deflected around the planetesimal (at the center). The red streamlines correspond to dust particles with a stopping time that is on the same order as the interaction timescale. The thick red lines correspond to the most extreme streamlines leading to a collision, which define $b_{\rm hydro}(\taus)$.}
\label{fig:flow}
\end{figure}

It is important to notice at this point that the collision probability thus derived is valid in the limit that the hydrodynamic flow of gas around the planetesimal can be neglected. As shown by \cite{SekiyaTakeda2003} and illustrated in Fig.~\ref{fig:flow}, this is not the case for small dust particles. More specifically, these authors show that a dust particle will be carried by the flow around the planetesimal if the dust particle stopping time is on the same order as or smaller than the planetesimal crossing time. This problem is well-known in the geophysical context where it affects, for example, the ability of rain drops and snow flakes to collect aerosols \citep[e.g.,][]{Feng2009}. 

We define a non-dimensional friction parameter $\tau_{\rm f}$ which is the ratio of these two timescales,
{\rev
\begin{equation}  
\tau_{\rm f}=t_{\rm s}\Delta v /R_{\rm p},
\label{eq:tauf}
\end{equation}
where $\Delta v$ is the velocity of the flow during the encounter as determined by eq.~\eqref{eq:Deltav_e} (taking possible eccentric orbits into account).
}
When $\tau_{\rm f}\ll 1$, dust particles have a stopping time that is much shorter than the gas interaction time; they rapidly adjust their velocity vector to follow the gas and we thus expect accretion to be inefficient. Conversely, when $\tau_{\rm f}\gg 1$, the particles' stopping time is too long and we expect hydrodynamical effects to be negligible for the calculation of the collision probability. 
The numerical simulations by \cite{SekiyaTakeda2003} \citep[see also][]{Feng2009,Sellentin+2013} show that the collision probability drops when $\tau_{\rm f}\wig{<}1$, and it actually becomes consistent with zero in the simulations they present when $\tau_{\rm f}\wig{<}0.3$. 

\begin{figure}[htb]
\includegraphics[width=\hsize,angle=0]{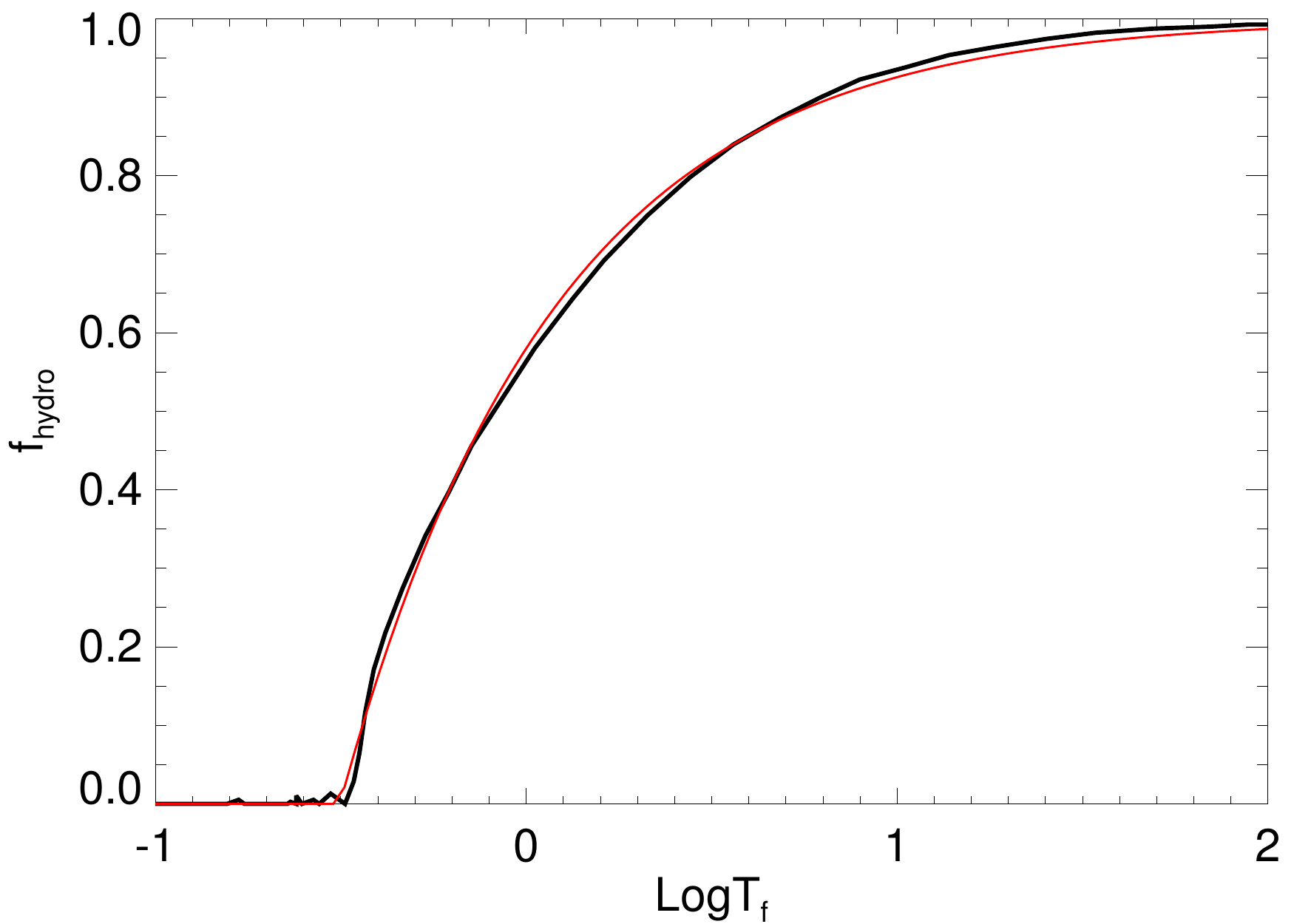}
\caption{Value of $f_{\rm hydro}$ as a function of the dimensionless friction time $\tauf$ as obtained by \protect\cite{SekiyaTakeda2003} (their Fig.~8) (black line) and our fit (red line and eq.~(\ref{eq:fhydro})).}
\label{fig:hydro}
\end{figure}

We write the reduction factor on the collision probability due to this factor $f_{\rm flow}$. From fitting the numerical simulations of \cite{SekiyaTakeda2003} as shown in Fig.~\ref{fig:hydro}, we obtain
\begin{equation}
f_{\rm flow}=1-0.42\tauf^{-0.75}.
\label{eq:fflow}
\end{equation}
This factor tends asymptotically to $1$ for large values of $\tauf$ and it drops to zero when $\tau_{\rm f}\wig{<}0.3$. (The fact that it becomes negative is not important as it will be taken care of in the next section.) 

\subsubsection{The boundary layer}\label{sec:boundary}

When the flow factor becomes too small, one must account for the boundary flow and for the fact that a small fraction of the particles with the right impact parameter will hit. The size of the boundary layer can be written \citep{Wurm+2004}
\begin{equation}
b_{\rm boundary}=R_\rmp\sqrt{q_{\rm boundary}/\Delta v},
\label{eq:bboundary}
\end{equation}
where $q_{\rm boundary}$ can be seen as either the flow velocity inside a porous planetesimal \citep{Wurm+2004} or as the flow velocity difference between the dust particles and the gas \citep{SekiyaTakeda2005}. We use the latter approach to calculate $q_{\rm boundary}$ based on the kinematic theory of gases, which yields
\begin{equation}
{q_{\rm boundary}\over \vcol}={3\over 8}{C_{\rm D}\over 1-\phi}{s\over R_\rmp}{\vcol\over c_\rmg},
\end{equation}
where $C_{\rm D}$ is the drag coefficient from eq.~\eqref{eq:CD} and $\phi$ is the porosity of the planetesimal. It can thus be seen that even non-porous bodies with $\phi=0$ will have non-zero $q_{\rm boundary}$ flow. (We will assume $\phi=0$, but retain it in the equations for reference).  

Because this concerns small particles, it is useful to consider the behavior of this boundary layer in the Epstein regime, in which case eq.~(\ref{eq:bboundary}) takes a simple form
\begin{equation}
\left({b_{\rm boundary}\over R_\rmp}\right)_{\rm Epstein}= \left({3\over 8}{C_{\rm D}\over 1-\phi}{\rho_\rmg\over \rhos}\tau_\rmf\right)^{1/2}.
\end{equation}
Given that $C_{\rm D}$ is on the order of unity for the relevant Reynolds numbers and that $\rho_\rmg/\rhos$ is tiny, the boundary layer is generally very small. The reduction factor to be applied to particles with $\tau_\rmf\wig{<}0.3$ is thus on the order of $10^{-6}$ in most cases. 

\subsubsection{The hydrodynamical factor}

{\rev This suppression of accretion is expected to occur in the hydrodynamical regime but not in the settling regime
because of the slow encounters. In the settling regime, the particle is dragged towards the planetesimal. Once this process starts, it is irreversible because the gravitational pull of the gravitating body steeply increases. The outcome does not depend on the physical size (cross section) of the body as it does in the Safronov regime and it should not depend on the precise flow pattern in the vicinity of the body. As an illustration, the recent calculations by \cite{Ormel2013} show that a protoplanet (or big planetesimal) strongly affects the gas flow around it, yet can still accrete particles at large cross sections when the interactions take place in the settling regime. This implies a steep jump in the accretion rate as the particle transitions from the Safronov/hydrodanamic regime to the settling regime. Further work is needed to understand the details of this transition, which we omit here for simplicity.} 

In summary, if the settling regime does not apply, the hydrodynamical correction factor is given by
\begin{equation}
f_{\rm hydro}=\max\left[\left({3\over 8}{C_{\rm D}\over 1-\phi}{s\over R_\rmp}{\vcol\over c_\rmg}\right)^{1/2},1-0.42\tauf^{-0.75}\right].
\label{eq:fhydro}
\end{equation}
{\rev Conversely, if the settling regime applies (i.e., when $b_{\rm set}>R_\rmp$), we set $f_{\rm hydro}=1$.}

\begin{figure}
\includegraphics[width=\hsize,angle=0]{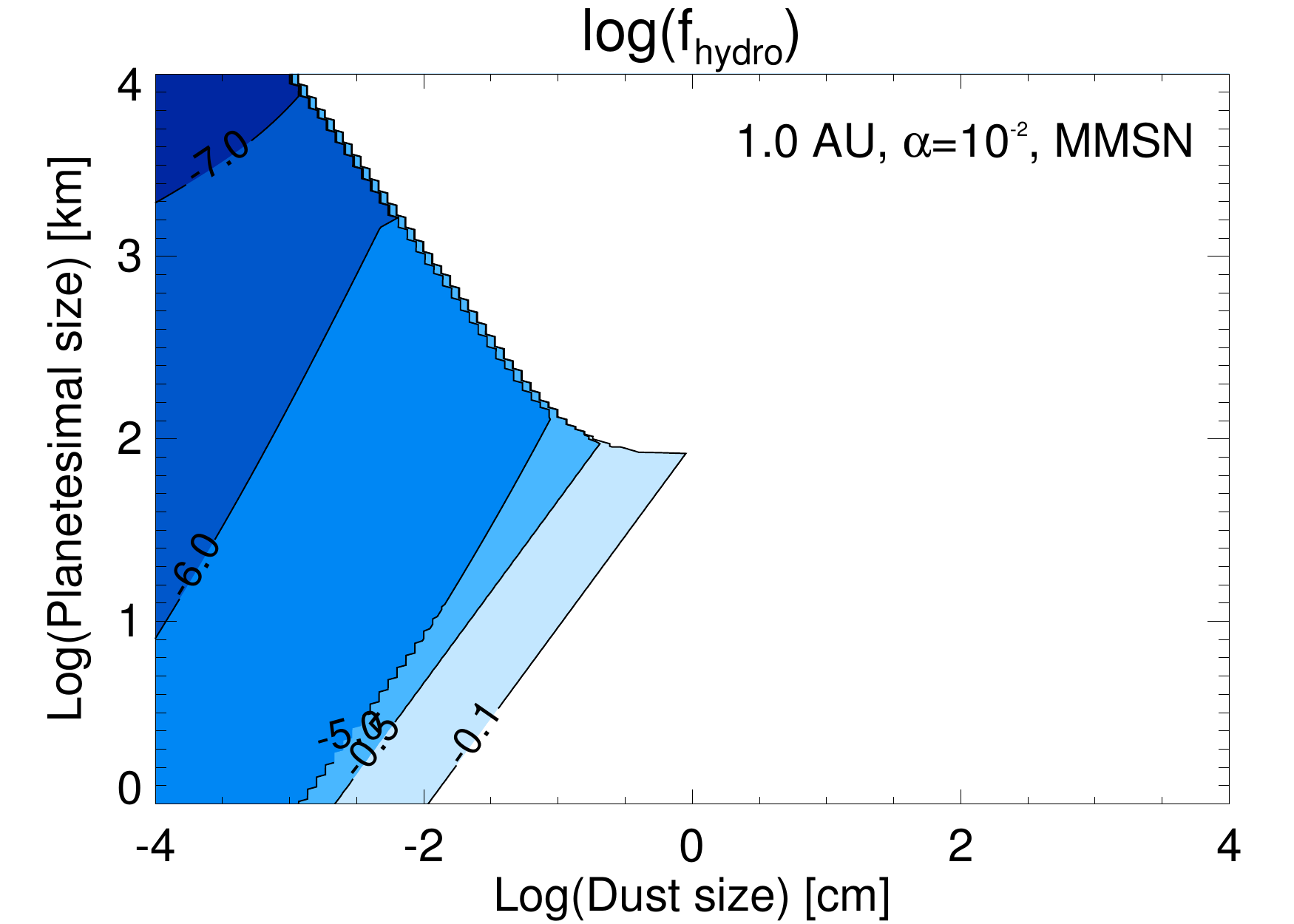}
\caption{Contours of the logarithm of the hydrodynamical factor $\log_{\rm 10}(f_{\rm hydro})$ [see eq.~(\protect\ref{eq:fhydro})] as a function of dust size (in cm) and planetesimal size (in km), at 1\,AU in the MMSN. The physical density of dust and planetesimals is set to $1$ and $3\,\gcc$, respectively. The contours are for $f_{\rm hydro}=10^{-7}$, $10^{-6}$, $10^{-5}$, $10^{-0.5}$, $10^{-0.1}$. {\rev The transition from the hydrodynamic flow regime to the boundary regime (dark to light blue) takes place over a narrow range of dust sizes for planetesimals smaller than 100\,km in radius. For larger planetesimal sizes, the transition between the hydrodynamic and settling regimes is discontinuous by construction. A proper modeling of this transition will require more work.}}
\label{fig:fhydro}
\end{figure}

Figure~\ref{fig:fhydro} shows the variations of $f_{\rm hydro}$ as a function of dust size and planetesimal size at 1\,AU in the MMSN. The hydrodynamic flow prevents the accretion of very small dust particles (centimeter-sized or less) especially by large planetesimals {\rev (except when these are large enough to grab particles in the settling regime)}. The high sensitivity on $\tauf$ (and the large range of possibilities for this factor) implies that this is an almost binary process: either $\tauf$ for the collisions is smaller than one and the collision probabilities should be expected to be very small, or it is larger and the hydrodynamical flow may be neglected. The strong sensitivity of $f_{\rm hydro}$ with planetesimal size is important when accounting for a size distribution of planetesimals. 

\subsection{Effect of turbulence}

\subsubsection{Amplitude of velocity fluctuations}

Turbulence in the disk generates density and pressure fluctuations. For example, using direct numerical simulations in the shearing box approximation, \cite{HeinemannPapaloizou2009} find that spiral density waves excited by magneto-rotational instabilities (or another suitable mechanism) yield relative density fluctuations of about $15\%$ of the mean values on length scale smaller than the pressure scale height. This would thus lead $\eta$, as calculated from eq.~\eqref{eq:eta} to vary significantly (even changing sign) over both space and time. However, this $\eta$ is not a proper measure of the headwind that the particles feel because the gas associated with these density fluctuations is still moving azimuthally at a velocity close to the average gas velocity. Actually, the velocity dispersion of the particles is linked to that of the gas and may thus be written \citep{YoudinLithwick2007}
\begin{equation}
\delta v_{\rm t} \sim {\sqrt{\alpha}\over {\rm max}(1,\tau_\rms)} v_{\rm th}.
\label{eq:deltavt}
\end{equation}
In the simulations of \cite{HeinemannPapaloizou2009}, $\delta v_{\rm t}\sim 0.15  v_{\rm th}$ {\rev (i.e., $\sim 100\rm\,m\,s^{-1}$ at 1\,AU)} for $\alpha\sim 5\times 10^{-3}$ for the gas, which agrees well with eq.~\eqref{eq:deltavt}. (We note that $v_{\rm th}\sim 1\,\rm km\,s^{-1}$ at 1\,AU in the MMSN). \cite{Flock+2013} also obtain similar values of $\alpha$ and slightly lower turbulent velocities in the range $10-100\rm\,m\,s^{-1}$ in the mid-plane, but an order of magnitude higher in the disk atmosphere heated both by stellar irradiation and MRI dissipation.  

{\rev These velocity fluctuations obtained in MRI-active regions are of the same order of magnitude as the mean headwind felt by planetesimals $\eta v_\rmK\sim 50\rm\,m\,s^{-1}$. In these regions, turbulence can have an important role. In dead zones and generally in regions with much smaller values of $\alpha\sim 10^{-4}$ it can perhaps be neglected. We examine below two consequences of turbulence on the collision probability.}

{\rev
\subsubsection{Consequence for the hydro mode}

A first consequence of turbulence is that the flow around the planetesimal may not be adequately described as laminar as assumed in Sect.~\ref{sec:hydro} and in simulations by \cite{SekiyaTakeda2003} and \cite{Sellentin+2013}. Recently, \cite{Mitra+2013} simulated the accretion of dust by boulders (less than a km in size) in a turbulent flow and obtained results that strongly differ from the laminar case. In their simulations, dust (with sizes in the range $10-200\,\rm\mu m$) is efficiently accreted by boulders of about $6$ to $200\,$m. The speeds and accretion trajectories seem to differ from the laminar case. Unfortunately, in the absence of a description of the assumed turbulent flow and without a comparison to a purely laminar case, it is not possible to quantify the consequences of these results. 

Qualitatively, we can tell that if the smallest scales of turbulence remain large compared to the size of the object considered, the laminar prescription defined by eq.~\eqref{eq:fhydro} should remain valid although modified to account for a mean flow velocity that differs from the laminar headwind velocity. This modification should lead to the accretion of grains of slightly smaller sizes without fundamentally affecting the picture. If, however, the turbulent cascade extends to scales as small as the size of the planetesimal and lower, it will modify the picture considerably and probably lead to a drastic reduction in the extent of the parameter regime affected by the hydro mode.

\subsubsection{Turbulence factor and random walk}

Beyond its effect on the hydrodynamical regime, turbulence also affects the collision probability. This can be estimated with the same approach as for the eccentric planetesimal 
}
in Sect.~\ref{sec:eccentricity}: the fluctuations of the velocity field along the planetesimal's trajectory can be considered to average out to zero if they are smaller than the mean collision velocity, {\rev and to lead to an increased collision probability otherwise.} We thus estimate the factor by which the final collision probability is increased over the geometrical limit due to turbulence by
\begin{equation}
f_{\rm t}=\sqrt{1+\left[{1+\taus^2\over \sqrt{1+4\taus^2}}{1\over \max (1,\taus)}{\sqrt{\alpha}\over \eta}{v_{\rm th}\over v_\rmK}\right]^2}.
\label{eq:ft}
\end{equation}
{\rev In the other regimes, we do not expect turbulence to lead to a significant increase of the collision probability and hence define the collision probability as }
${\cal P} _{\rm 2D,geo,circ} f_{\rm hydro} f_{\rm t}$ only if $f_{\rm t}>\max(f_e,f_{\rm focus})$. 

\begin{figure}
\includegraphics[width=\hsize,angle=0]{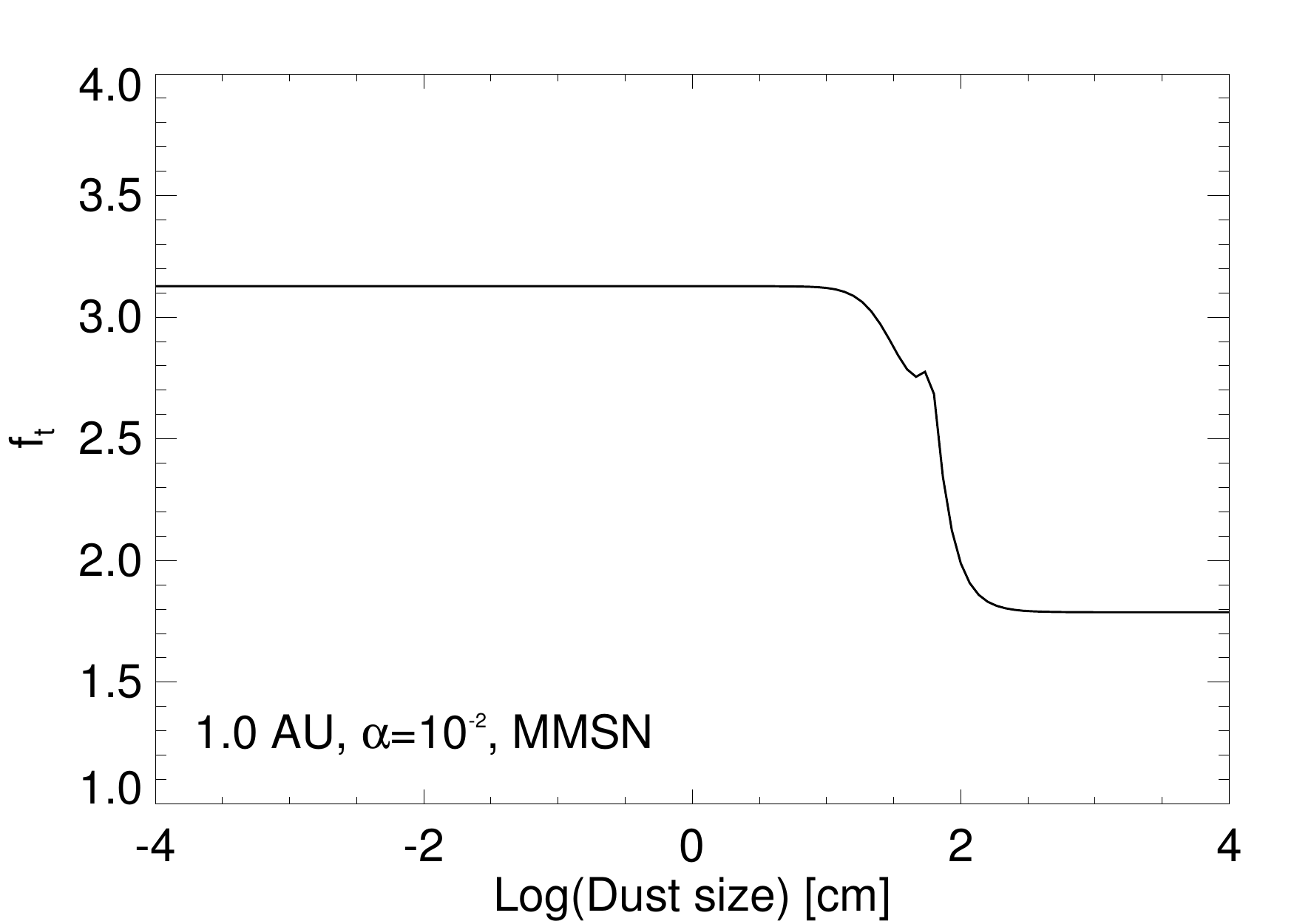}
\caption{Value of the turbulence factor $f_{\rm t}$ [see eq.~(\protect\ref{eq:ft})] as a function of dust size (in cm), at 1\,AU in the MMSN for a turbulence viscosity parameter $\alpha=10^{-2}$. We note that contrary to the other plots, the linear value of the factor is plotted instead of its log.}
\label{fig:ft}
\end{figure}

As shown in Fig.~\ref{fig:ft}, {\rev turbulence is expected to increase the mean collision probability in the geometrical limit only by a modest amount}. When considering for example dust of $1$\,cm in size, the value of $f_{\rm t}$  for $\alpha=10^{-3}$ varies from a maximum of $1.9$ at $0.1$\,AU to $1.03$ at $100$\,AU.  These values scale with $\sqrt{1+\alpha/10^{-3}}$.

\subsection{Additional effects relevant for massive embryos}

{\rev 

The effects discussed previously apply to a wide range of planetesimal sizes. When considering planetary embryos with large masses, two effects should be included but are not taken into account in the present calculation. 

\subsubsection{Presence of an atmosphere}

Big planetesimals (or protoplanets) with masses above $0.1\,\rm M_\oplus$ (equivalently $\Rp\wig{>}5000\,$km for $\rhop=1\,\gcc$) start to acquire an atmosphere. \cite{InabaIkoma2003} showed that the presence of a dense atmosphere increases the cross section of planetesimals and they derived expressions to quantify this effect. However, \cite{OrmelKobayashi2012} showed that for small particles interacting in the settling regime the presence of an atmosphere does not yield an increase in the cross section. The presence of an atmosphere becomes important only when considering the three-body interactions between planetary embryos and very large dust particles, namely boulders or planetesimals.

\subsubsection{Gap opening}

Tidal interactions between massive protoplanets lead to the opening of gaps in circumstellar disks \citep[e.g.,][]{LinPapaloizou1985, LinPapaloizou1993}. These gaps suppress the flow of material between the outer disk and the inner disk and could be the reason behind the existence of the so-called transitional disks, i.e., disks with accretion onto the central star and a central cavity \citep[e.g.,][]{Rice+2006}. Actually, gaps are opened more easily (i.e., by smaller-mass protoplanets) in the dust and/or planetesimal disk \citep[e.g.,][]{TanakaIda1999,PaardekooperMellema2004} than in the gas itself, implying a potentially high degree of filtering of material from the outer system. This effect cannot be accounted for in the expressions derived here.  We merely note that the presence of massive protoplanets with masses larger than $10\,\mea$ that can open up gaps in gas disks will certainly be accompanied by a drastic reduction of the metallicity of the gas flowing into the inner regions and accreted by the central star. 
}

\section{Collision probabilities}

\subsection{The 2D collision probability}

We can thus calculate the final collision probability of a planetesimal of radius $R_\rmp$, physical density $\rhop$, orbital semi-major distance $r$, eccentricity $e$, with a dust grain of size $s$ and physical density $\rhos$  inside a disk as follows. We first need to define the characteristics of the gas disk at that orbital distance, in particular the disk temperature $T$, mid-plane density $\rho_\rmg$, scale height $h_\rmg$, Keplerian velocity $v_\rmK$ and $\eta$ parameter (Sect.~\ref{sec:geometry}). We then need both the molecular and turbulent viscosities $\nu_\rmg$ and $\nu$, respectively, to calculate the drag forces, stopping time $\tau_\rms$ and $\tausnu$  and thus the relative velocities $\Delta v_r$ and $\Delta v_\theta$ between our planetesimal and dust particle (Sect.~\ref{sec:velocities}). 

Armed with these quantities, we can easily calculate with eq.~\eqref{eq:P2Dcirc} the 2D geometrical probability in the circular case ${\cal P}_{\rm 2D,geo,circ}$. For eccentric and/or inclined planetesimals, we estimate the probability increase by averaging the planetesimal's cross section on its trajectory. The expression for the eccentricity-inclination factor $f_{e,i}$ is found in eq.~\eqref{eq:fei} (this factor is always larger than unity and is to be used instead of $f_e$ defined by eq.~\eqref{eq:fe} to account for a non-zero inclination). The inclusion of gravitational focusing is calculated analytically in three regimes, the Safronov, settling, and three-body regimes (see Sect.~\ref{sec:focusing}). The corresponding focusing factor $f_{\rm focus}$ is defined in eq.~\eqref{eq:ffocus}. The planetesimal linear cross section $b$ (equal to $R_\rmp$ in the geometrical limit) and encounter velocity $\vcol$ (equal to $\eta v_\rmK f_e$ in the geometrical limit and assuming a small $\taus$; see eq.~\eqref{eq:Deltav_e}) are defined in eq.~\eqref{eq:focusing} (see Sect.~\ref{sec:focusing}). The value of $f_{\rm focus}$ is always larger than unity. 

We account for a decrease in the collision probability due to the hydrodynamical flow around the planetesimal through the factor $f_{\rm hydro}$, defined in eq.~\eqref{eq:fhydro}. This factor is always smaller than unity, and in fact becomes very small when the dust particle stopping time becomes smaller than the planetesimal crossing time, i.e., when the friction parameter $\tau_{\rm f}$ defined by eq.~\eqref{eq:tauf} becomes smaller than unity. We also account for the velocity dispersion of dust particles due to turbulent fluctuations in the gas disk through the factor $f_{\rm t}$ defined by eq.~\eqref{eq:ft}. This factor is larger than unity, i.e., turbulence always increases the probability of a collision. 

The final 2D collision probability is written
\begin{equation}
 {\cal P} _{\rm 2D}= {\cal P} _{\rm 2D,geo,circ}\,f_{\rm hydro}\, \max(f_{\rm t},f_{e,i}, f_{\rm focus}),
\label{eq:P2D}
\end{equation}
{\rev where $f_{\rm hydro}\le 1$, whereas other factors are always equal to or larger than unity. We note that values of ${\cal P}_{\rm 2D}$ above unity correspond to multiple collisions.} 
 
\begin{figure}
\includegraphics[width=\hsize,angle=0]{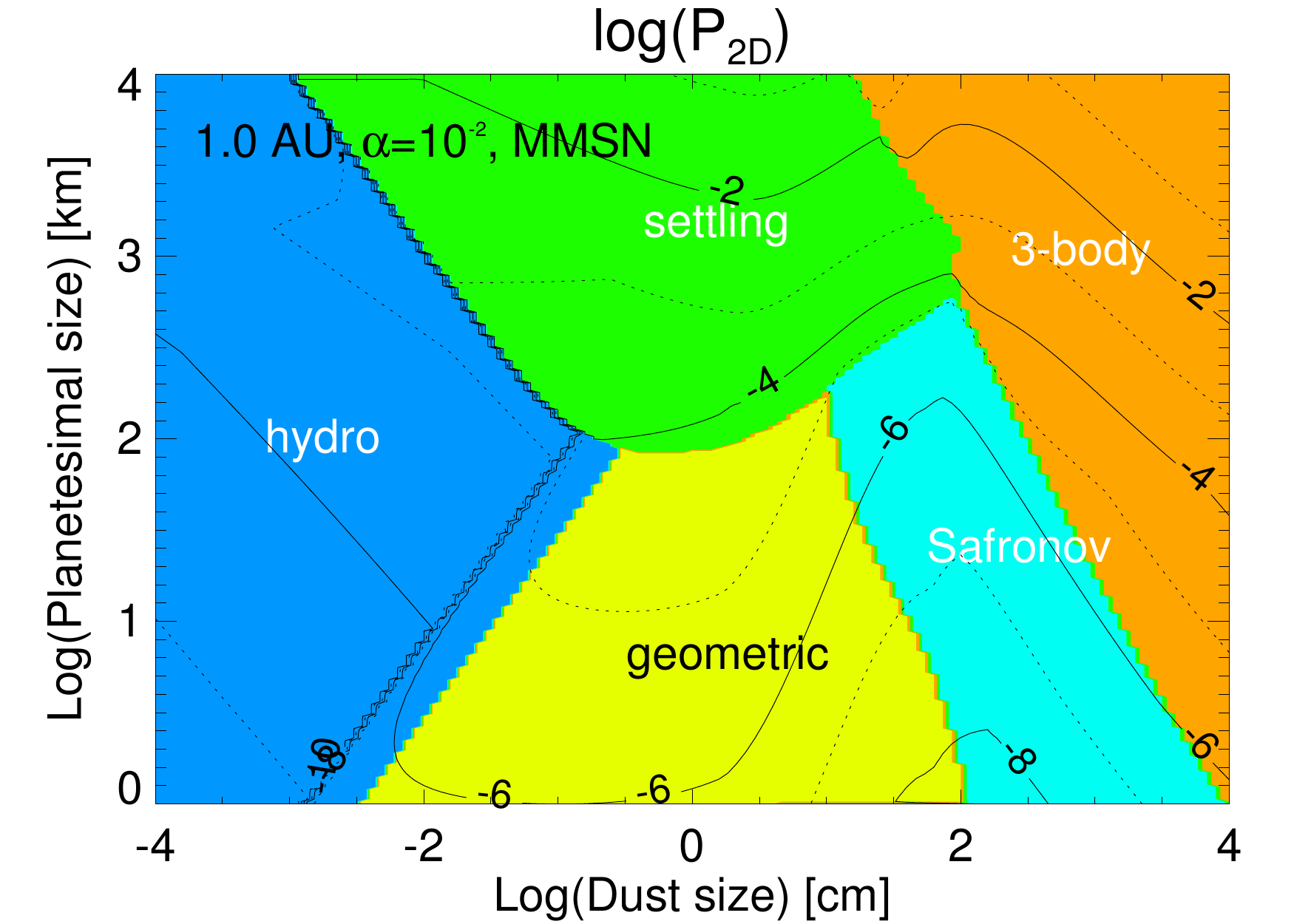}
\caption{2D collision probability ${\cal P}_{\rm 2D}$ between a dust particle of radius $s$ (x-axis, in centimeters) and a planetesimal of radius $R_\rmp$ (y-axis, in kilometers), calculated for the MMSN at 1AU, physical densities $\rhos=\rhop=1\,\gcc$, a turbulence parameter $\alpha=10^{-2}$, and the eccentricity distribution shown in Fig.~\protect\ref{fig:ecc_I08}. The values of ${\cal P}_{\rm 2D}$ range from $10^{-12}$ (inefficient) to close to $1$ (perfect). {\rev We note that given the assumed $\rhop$ value, a $0.1\rm\,M_\oplus$ (resp. $1$) embryo is equivalent to a planetesimal with $\Rp=5225$\,km (resp. $11250\,$km).}}
\label{fig:p2d}
\end{figure}

Figure~\ref{fig:p2d} shows the 2D probabilities that result from including all these processes at 1AU in the MMSN case (see Appendix~\ref{sec:MMSN}) for a variety of combinations of dust and planetesimal sizes. At this location, particles with $s\sim 1$\,m have $\tau_\rms\sim 1$ and thus are the ones that have the largest relative velocities and smallest accretion probability. The probability decrease due to the hydrodynamic flow around planetesimals (for $\tau_\rmf\wig{<}1$) is evident in the upper left hand corner of the figure, but it is limited to relatively large planetesimals (above 1km for micron-sized dust): this is because the interaction time for the gas around small planetesimals is too short compared to the stopping time of even small dust particles.  

\subsection{The 3D collision probability}\label{sec:p3d}

\begin{figure}
\includegraphics[width=\hsize,angle=0]{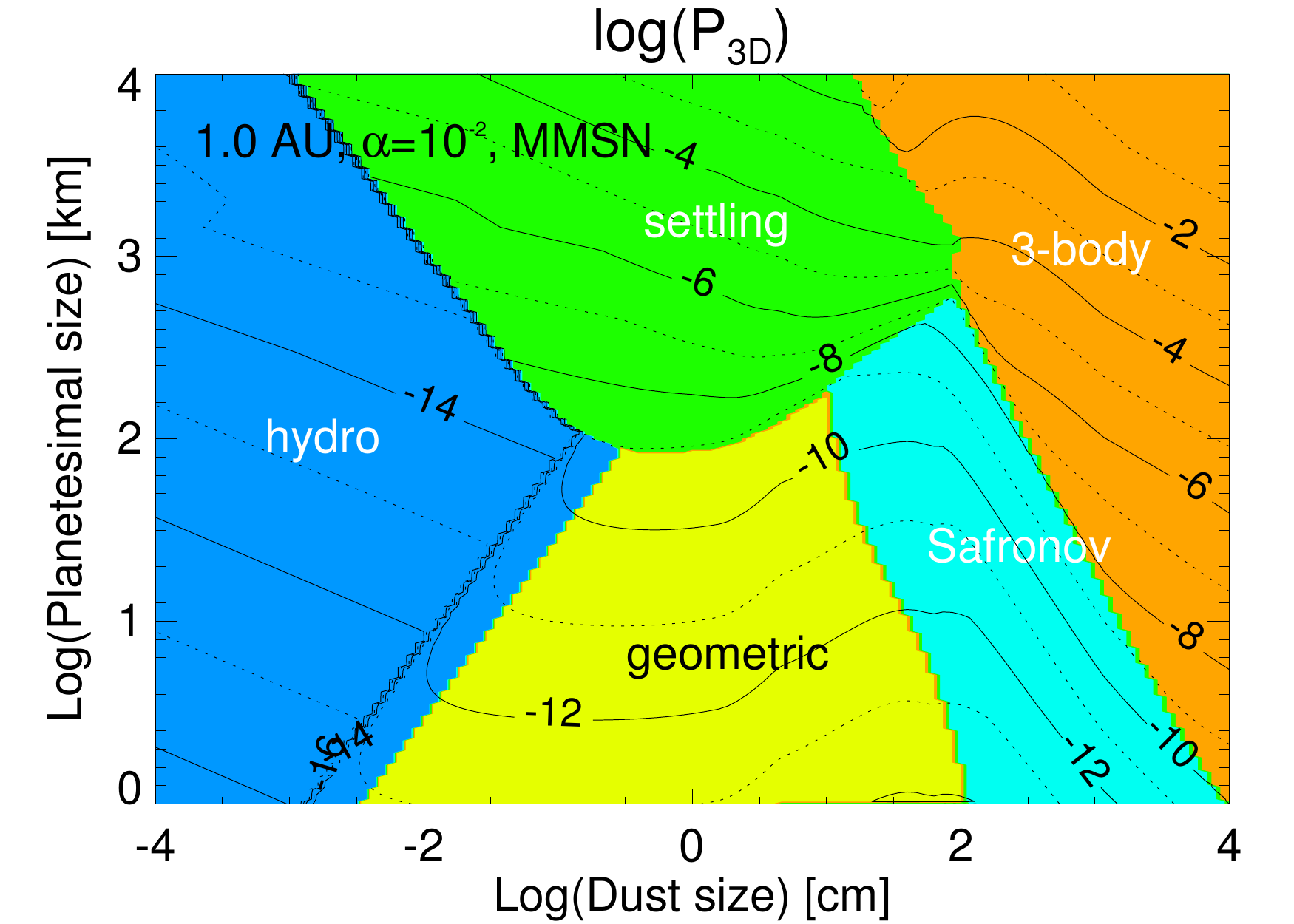}
\caption{Collision probability ${\cal P}_{\rm 3D}$ including 3D effects between a dust particle of radius $s$ (x-axis, in centimeters) and a planetesimal of radius $R_\rmp$ (y-axis, in kilometers), at 1AU with the same conditions as Fig.~\ref{fig:p2d}. The values of ${\cal P}$ range from $10^{-18}$ (very inefficient accretion) to almost $1$ (perfect accretion). The probabilities are much lower than in Fig.~\protect\ref{fig:p2d} primarily because of vertical (turbulent) diffusion of dust particles.}
\label{fig:p3d}
\end{figure}

{\rev The final collision probability accounting for 3D effects, ${\cal P}$, is obtained from ${\cal P}_{\rm 2D}$ and eq.~\eqref{eq:P2D} and eq.~\eqref{eq:P3D}.}
Figure~\ref{fig:p3d} shows the values of ${\cal P}$ at $1\,$AU as a function of dust and planetesimal sizes. {\rev Except when it reaches very large sizes for the planetesimals (i.e., in the planet-regime), the probability is several orders of magnitude smaller than ${\cal P}_{\rm 2D}$ because the dust layer is much thicker than the planetesimal cross section.} This reduction is more pronounced for small grains, because they are spread over a thicker disk and for small planetesimals because of their smaller cross section. Otherwise, we can see that planetesimals with sizes below $\sim 100$\,km have collision probabilities that are close to the geometrical ones [see eq.~\eqref{eq:P3Dgeocirc}] and scale with the square of their size. These collisions take place in the Safronov regime, except for those with small dust grains which take place in the hydrodynamical regime and become much less likely. For larger planetesimals, collisions can take place in the settling regime or, for large grains, in the three-body regime. Because of a significant focusing factor, these collisions become more dependent on planetesimal size (and mass). 

{\rev Importantly, Fig.~\ref{fig:p3d} shows that for a wide variety of planetesimals dust sizes, the collision probability remains significantly smaller than unity. This means that embryos less massive than an Earth mass can capture only a small fraction of the grains flowing through the disk. As shown in Appendix~\ref{sec:analytical}, the situation is more favorable for smaller values of $\alpha$ since ${\cal P}\propto\alpha^{-1/2}$, but this is not sufficient for a large fraction of the dust sizes to be considered.}

\begin{figure}
\includegraphics[width=\hsize,angle=0]{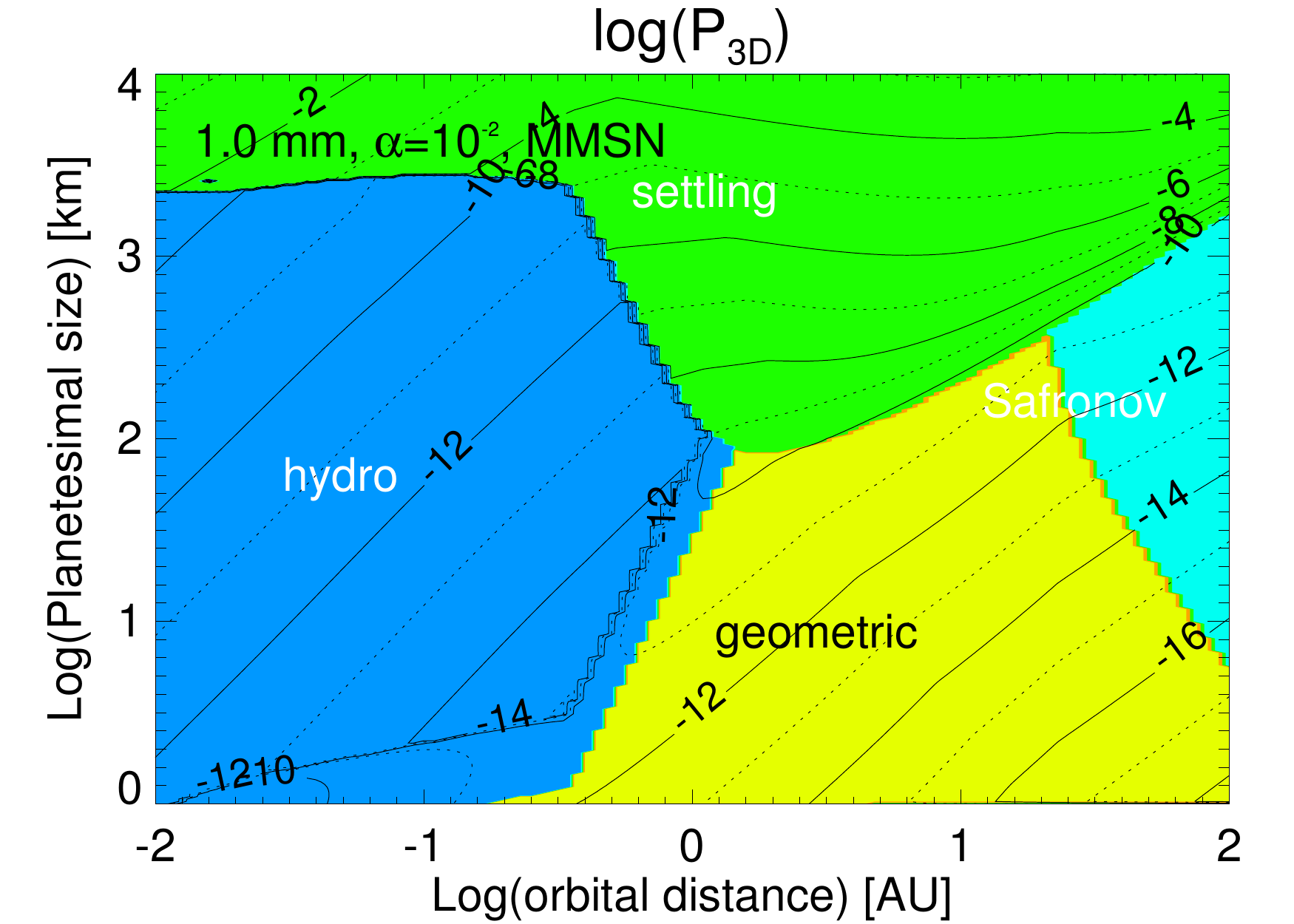}
\caption{Same as Fig.~\protect\ref{fig:p3d}, but with the collision probability ${\cal P}_{\rm 3D}$ for 1\,mm dust grains in a MMSN disk shown as a function of orbital distance.}
\label{fig:p3d_orbdist}
\end{figure}

Figure~\ref{fig:p3d_orbdist} shows the same probability ${\cal P}$, but for 1\,mm dust grains as a function of orbital distance. {\rev Most collisions in this figure fall into a hydro, geometric, or settling regime. The hydro regime is widespread and extends to up to $\sim 1$\,AU for a variety of planetesimal sizes, indicating that grains smaller than millimeter size must be either captured at larger orbital distances by small planetesimals ($\wig{<} 3\,$km in radius), or through same size collisions. In the geometrical regime, ${\cal P}$ depends strongly on planetesimal size and inversely on orbital distance. As demonstrated in Appendix~\ref{sec:analytical}, in that regime ${\cal P}\propto \Rp^2 r^{-9/4}$,  because the geometrical probability scales as $1/(h_\rmg r)$ and $h_\rmg \propto r^{5/4}$ in the MMSN. On the other hand, in the settling regime, ${\cal P}$ is even more strongly dependent on planetesimal size (it becomes dependent on its {\em mass}), but very weakly on orbital distance. Thus, as shown in Appendix~\ref{sec:analytical} for the settling/Bondi regime, ${\cal P}\propto \Rp^{3}r^{-3/4}$ for a fixed $\taus$. The dependence of $\taus$ and particle size on orbital distance implies that the probability can increase with orbital distance, as is the case in some regions of Fig.~\ref{fig:p3d_orbdist}.}

\section{Are planetesimals efficient dust filters?}\label{sec:filtering}

\subsection{Filtering efficiency at 1\,AU}

{\rev We now revisit the filtering efficiency as defined in Sect.~\ref{sec:xfilter} for the geometrical circular limit, but this time applying the full treatment including eccentricities, gravitational focusing, hydrodynamical effects and turbulence. We employ the parameter $x_{\rm filter}$ as defined by eq.~\eqref{eq:xfilter} and a value of the surface density of planetesimals $\Sigma_\rmp(r)$ equal to the MMSN disk of solids. A value of $x_{\rm filter,MMSN}(s)$ close to unity or larger thus implies an efficient filtering of dust grains of size $s$ by a mature disk of planetesimals. A lower value implies an inefficient filtering of the dust particles by this disk.}

\begin{figure}[tb]
\includegraphics[width=\hsize]{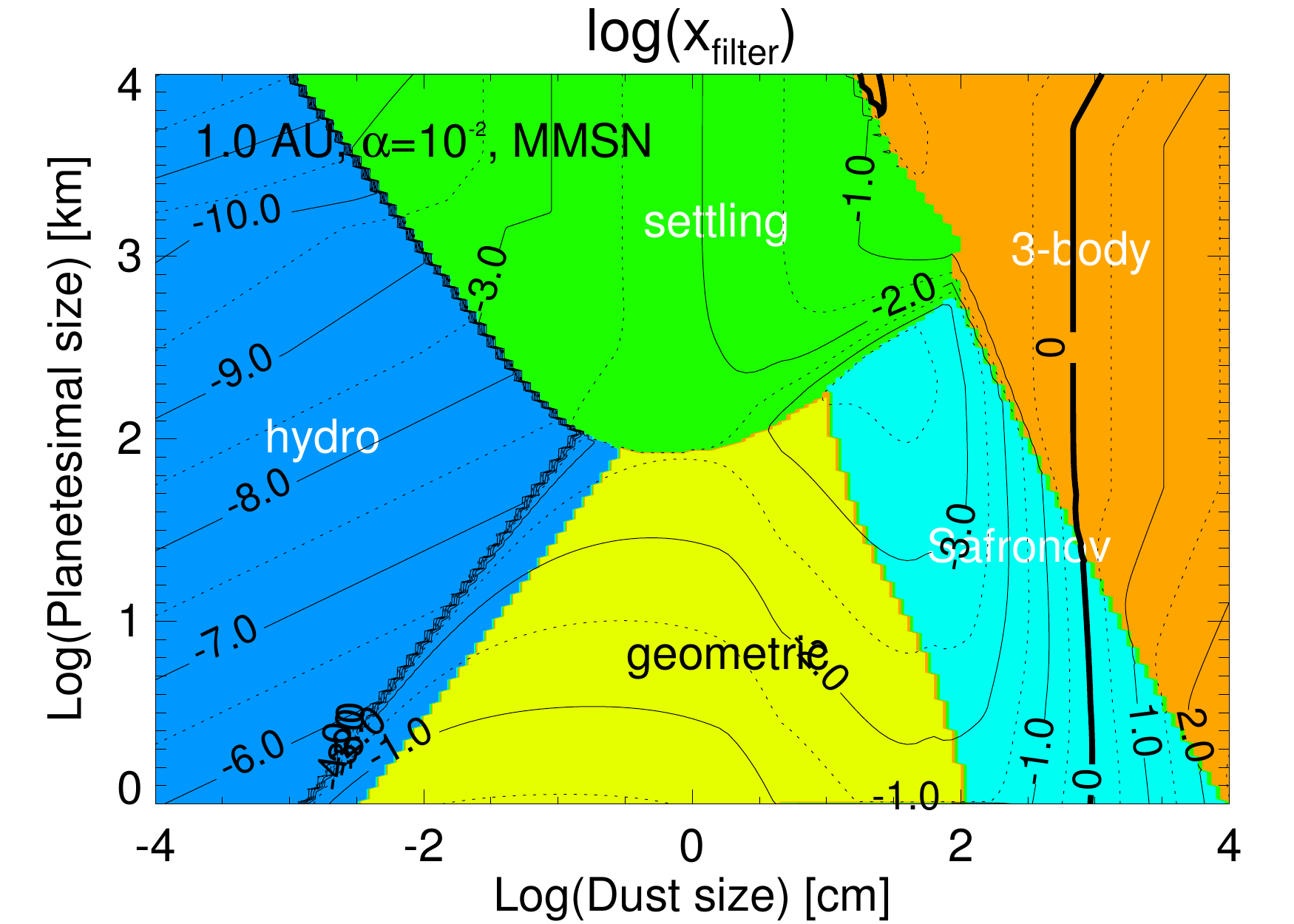}
\caption{Contours of the filtering efficiency by a MMSN planetesimal disk at 1 AU for $\alpha=10^{-2}$, assuming monodisperse size distributions for planetesimals and dust grains. The contour indicating perfect filtering efficiency for $\log_{10}x_{\rm filter}=0$ is shown in bold.}
\label{fig:xeffic_1AU}
\end{figure}

Figure~\ref{fig:xeffic_1AU} shows the values of $x_{\rm filter,MMSN}(s)$ at 1\,AU as a function of dust and planetesimal size (both are assumed to have a single size distribution). {\rev Dust of millimeter size or less is generally transported by the flow around the planetesimals and can only be accreted by small planetesimals in the geometrical regime or by large ones in the settling regime.}  Its filtering in the inner solar system appears to be inefficient. In the geometrical limit, filtering is relatively efficient for small planetesimals but decreases inversely with the planetesimal size until it reaches the settling regime. There, for planetesimals in the range $100-1000$\,km and beyond, a higher efficiency results from the Bondi-accretion regime. For large dust particles with a large stopping time, the slow radial drift has the consequence that dust and planetesimals have more time to interact gravitationally resulting in a more efficient filtering. However, only particles of 10 meters or more appear to be efficiently captured by planetesimals of any size. (This does not necessarily imply high accretion rates, however, because these relations implicitly assume an infinite time for the interaction.)

\subsection{Filtering efficiency vs. orbital distance}

\begin{figure}
\includegraphics[width=\hsize,angle=0]{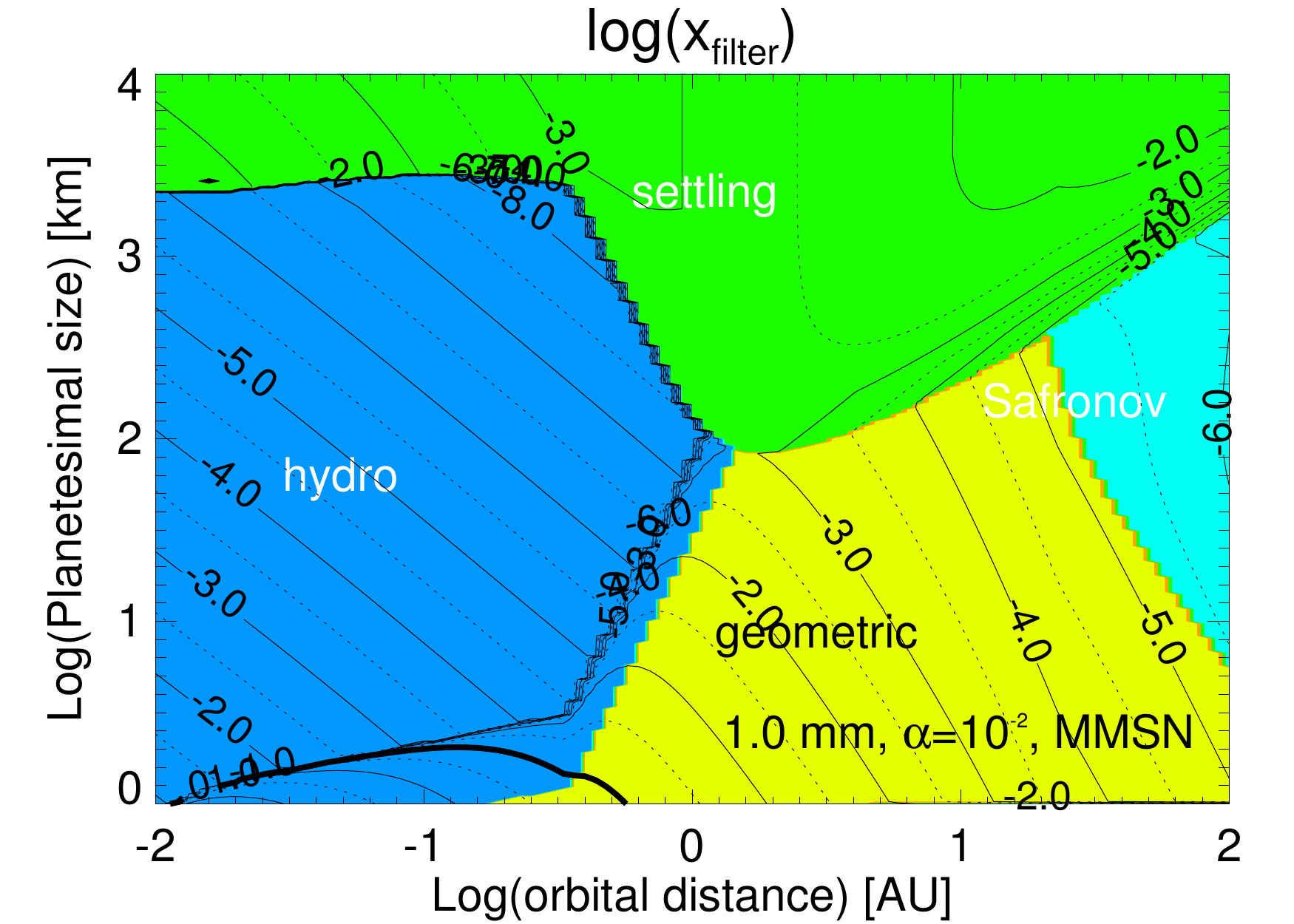}
\caption{Contours of the filtering efficiency by a MMSN planetesimal disk for 1 mm dust for $\alpha=10^{-2}$, assuming a monodisperse size distribution for planetesimals. The contour indicating perfect filtering efficiency for $\log_{10}x_{\rm filter}=0$ is shown in bold.}
\label{fig:xeffic_1mm}
\end{figure}
 
Figure~\ref{fig:xeffic_1mm} shows the filtering efficiency for millimeter particles as a function of orbital distance. Strikingly, these particles are very inefficiently filtered inside of 1\,AU, or only by {\rev either very small planetesimals or relatively large planetary embryos. The orbital dependence of filtering depends strongly on the collision regime: for small planetesimals, for which collisions with mm-sized grains take place in the geometrical regime, filtering is highly dependent on orbital distance. As shown in Appendix~\ref{sec:analytical}, $x_{\rm filter}\propto r^{-7/4}$, which would favor the growth of the very first planetesimals very close to the central star. On the other hand, Fig.~\ref{fig:xeffic_1mm} shows that in the settling regime, filtering is more weakly dependent on orbital distance. Appendix~\ref{sec:analytical} shows that $x_{\rm filter}\propto r^{-1/4}$ in the settling/Bondi regime and $x_{\rm filter}\propto r^{1/4}$ in the settling/Hill regime. Farther from the central star, planetary embryos can perturb the motion of dust particles to much larger distances. 

Thus, while small planetesimals can filter dust only if they are close to the central star, the filtering properties of planetary embryos are relatively independent of orbital distance. This could help to understand why the inner solar system appears to have a gradation  of composition whereas the outer solar system seems more uniform.}

\subsection{Filtering as a function of the mass of gas in the disk}

\begin{figure}
\includegraphics[width=\hsize,angle=0]{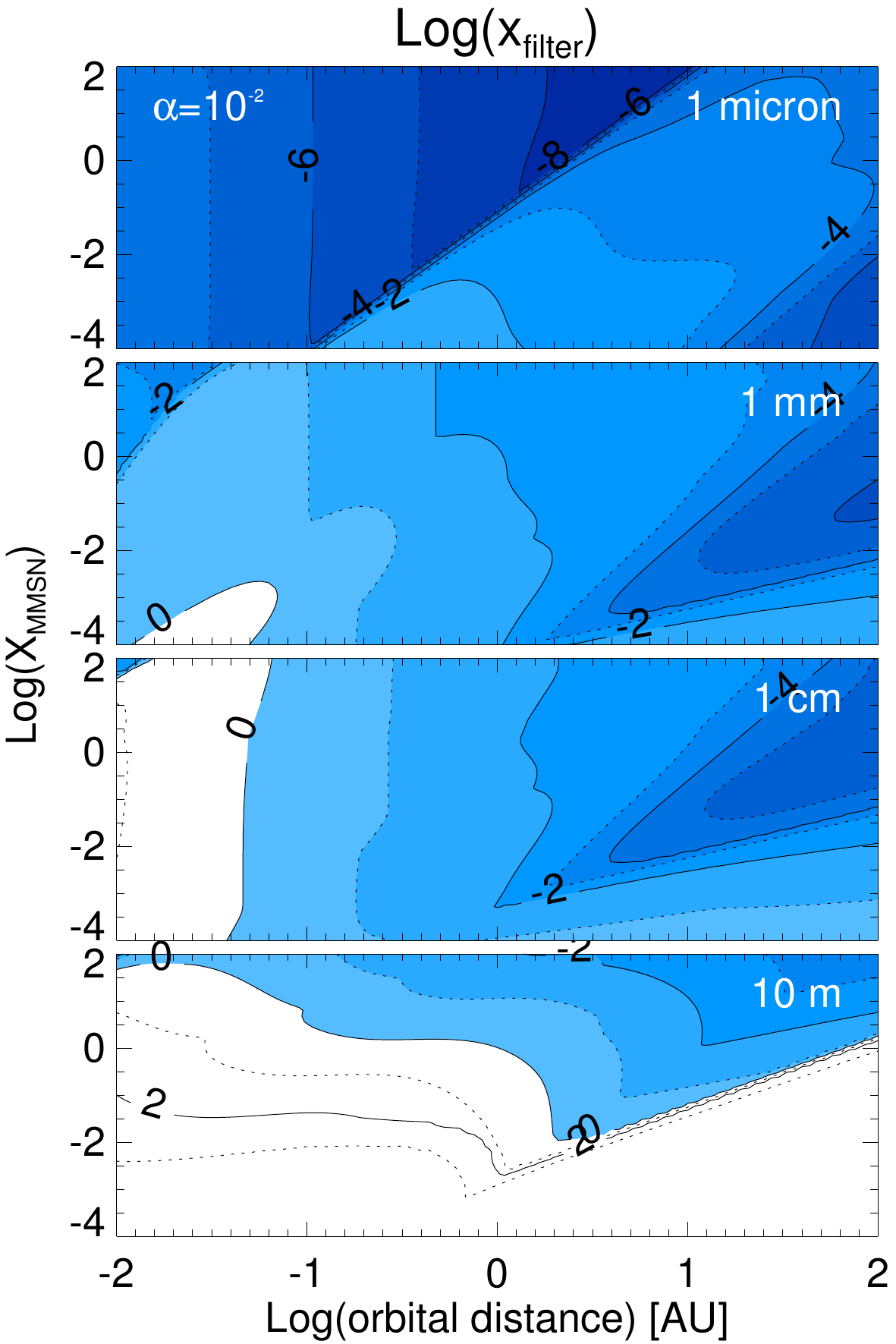}
\caption{Filtering efficiency of a swarm of planetesimals with radii between 1\,km and 1000\,km as a function of orbital distance and {\rrev $\xmmsn$, the mass of the gas disk in MMSN units,} for $\alpha=10^{-2}$. The panels corresponds to various dust sizes ranging from 1 micron (top) to 10 meters (bottom). Within each panel, disk evolution proceeds from top (large $\xmmsn$) to bottom (small $\xmmsn$). A negative value of Log$(x_{\rm filter})$ indicates inefficient filtering. Efficient filtering is shown as white areas. }
\label{fig:filtering_orbdist_mmsn}
\end{figure}

Figure~\ref{fig:filtering_orbdist_mmsn} shows how the filtering efficiency varies with dust size and as a function of the $\xmmsn$ factor, i.e., as the gas disk evolves and progressively becomes less dense, for a turbulent viscosity set by $\alpha=10^{-2}$. We have assumed in all cases that a collision cascade maintains a population of planetesimals with radii between $1$\,km and $1000$\,km with a mass exponent ${q}=3.5$ and a surface density equal to the minimum mass solar nebula in solids. 

Clearly, filtering is most ineffective for small, micron-sized particles, mostly because they are in a hydrodynamical regime and thus avoid planetesimals {\rev and/or because turbulence lifts them up}. This occurs at all orbital distances and for all values of $\xmmsn$. Larger particles are progressively easier for the planetesimal swarm to collect, with a filtering that is more efficient in the inner regions than in the outer ones. Millimeter-sized particles begin to be more efficiently collected at orbital distances shorter than 0.1\,AU and only for $\xmmsn<10^{-2}$, i.e., when the gas density has become small enough to reduce the inward drift of dust significantly. Centimeter-sized particles may also be collected more efficiently regardless of $\xmmsn$, but only inside 0.1\,AU. Particles of 10 meters or more are efficiently collected as soon as the gas density becomes of the same order as or lower than that of the MMSN. This is a direct consequence of the slower gas drag and hence slower inward drift of these particles at later ages. 

{\rev Apart from small dust particles in the hydrodynamical regime and large boulders, the filtering properties of planetesimals remain remarkably stable and independent of the evolution of the gas disk inside about 3\,AU. It is interesting to see that the filtering efficiency remains relatively large (between $0.1$ and $0.01$) in a zone between 0.1\,AU and 1\,AU, but that it drops and becomes dependent on the evolution of the disk beyond about 3\,AU. }

Even when $x_{\rm filter}\approx 1$, filtering can be considered to be moderately efficient because this corresponds to a mass in planetesimals equals to that of the present solar system. For a smaller mass, as would presumably occur in a young planetesimal disk, the filtering efficiency would be reduced by a ratio equal to that between the mass of the planetesimal disk to the MMSN. Figure~\ref{fig:filtering_orbdist_mmsn} thus shows that in order to efficiently capture small (centimeter-sized or less) dust particles, they have to be first assembled into larger planetesimals of at least 10 meters or more by a different mechanism such as orderly growth (by collision of grains of the same-size) or a streaming instability. Alternatively, filtering could be done more efficiently by planetesimals smaller than 1\,km which will also drift and have thus not been considered in this study.

{\rrev
\subsection{Filtering by an extended belt of planetesimals}

We now consider the filtering of dust by a belt of planetesimals assumed to extend between 0.1\,AU and 35\,AU. We calculate $X_{\rm filter}$ as defined by eq.~\eqref{eq:Xfilter_def}, but assuming that a monodisperse size distribution of planetesimals. We neglect any variations of dust size (e.g., due to growth or vaporization at the ice line) in the integration over orbital distances. A value of $X_{\rm filter}$ larger than unity indicates an efficient filtering of dust by the planetesimal belt. It also implies that the star accretes gas that is poor in dust of that size. 

Figure~\ref{fig:xfiltering_alpha1d-2} shows that for our fiducial value of $\alpha=10^{-2}$ the filtering by a belt of planetesimals is efficient only for large dust sizes beyond about 10 meters. Two islands of higher filtering are for planet embryos ($\Rp\wig{>}1000$\,km) and pebbles of centimeter to meter size, or when planetesimals are very small ($\Rp\wig{<}10\,$km) for dust larger than millimeter size. These correspond to collisions in the settling and geometric regimes, respectively. 

\begin{figure}
\includegraphics[width=\hsize,angle=0]{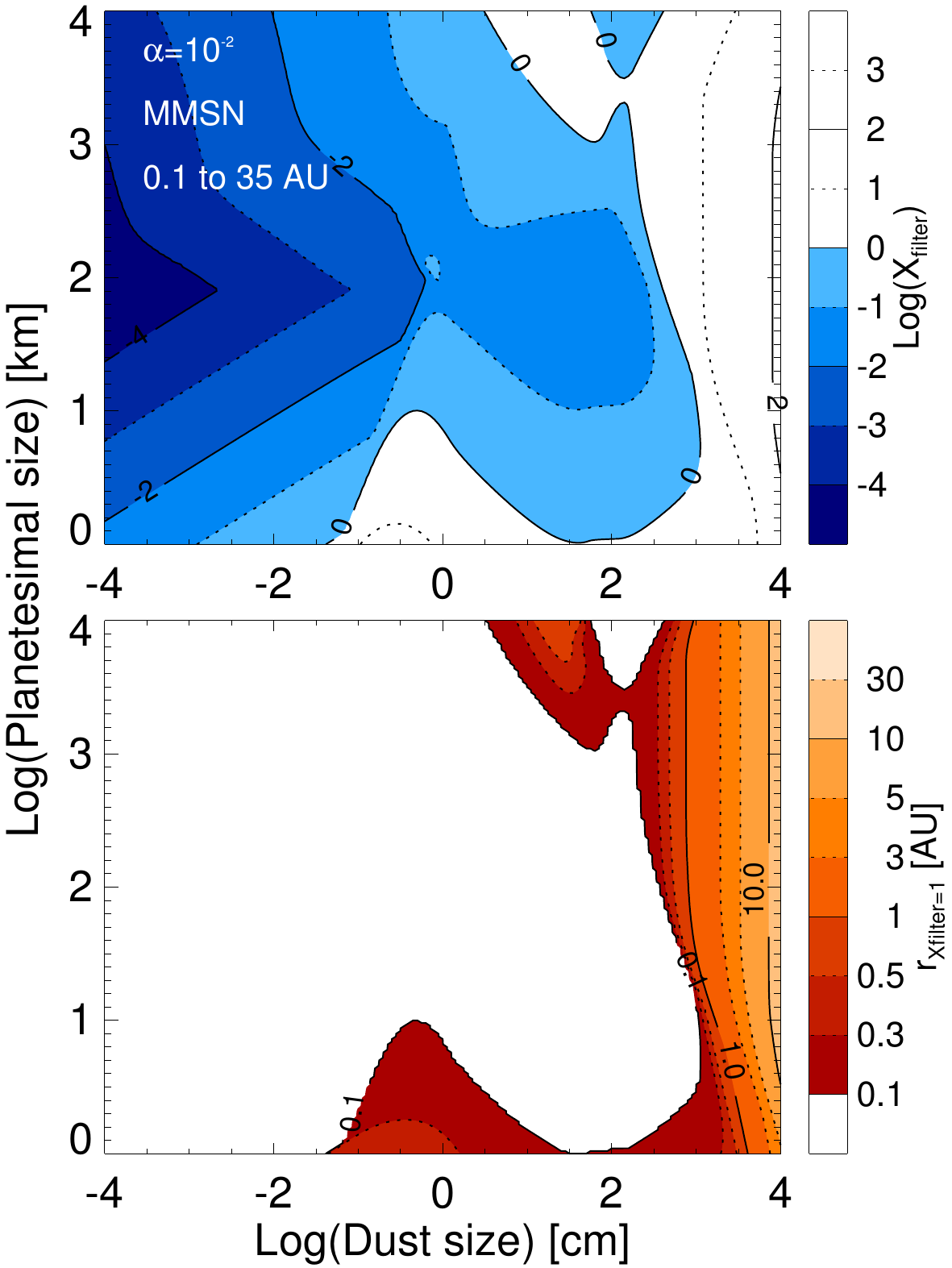}
\caption{Filtering efficiency of dust of 1 micron to 100 meters by a MMSN planetesimal belt with planetesimals of 1 to 10,000\,km in radius extending from 0.1 to 35 AU for a turbulent parameter $\alpha=10^{-2}$ and assuming monodisperse size distributions for planetesimals and dust grains. {\em Top panel:} Contours of the disk-integrated filtering factor $X_{\rm filter}$. {\em Bottom panel:} Orbital distance at which $X_{\rm filter}(r)=1$ for dust particles drifting in from beyond 35\,AU. (The integration thus starts from 35 AU and goes inwards.)}
\label{fig:xfiltering_alpha1d-2}
\end{figure}

Even in these islands, dust is able to penetrate deep inside the planetesimal belt, most of it reaching orbital distances between 0.1\,AU to 1\,AU. The clear gradient in the abundance of ices measured in the asteroid belt and its scarcity in asteroids that are closer than 2\,AU to the Sun implies that either the gas disk was much less turbulent or another physical mechanism led to the present composition of the solar system.

\subsection{Filtering in a weakly turbulent disk}

We now turn to the case of a weakly turbulent disk (or equivalently, a disk with an extended dead-zone) by considering the case of a turbulent parameter $\alpha=10^{-4}$. This case is more favorable because dust settles closer to the mid-plane and the mean gas flow is also slower. Appendix~\ref{sec:analytical} demonstrates that for most grain sizes, $x_{\rm filter}\propto\alpha^{-1/2}$. Appendix~\ref{sec:weak-turbulence} provides the same figures as in the previous sections, but for the weak-turbulence case instead of the fiducial $\alpha=10^{-2}$. In all cases, filtering is found to be much more efficient. 

\begin{figure}
\includegraphics[width=\hsize,angle=0]{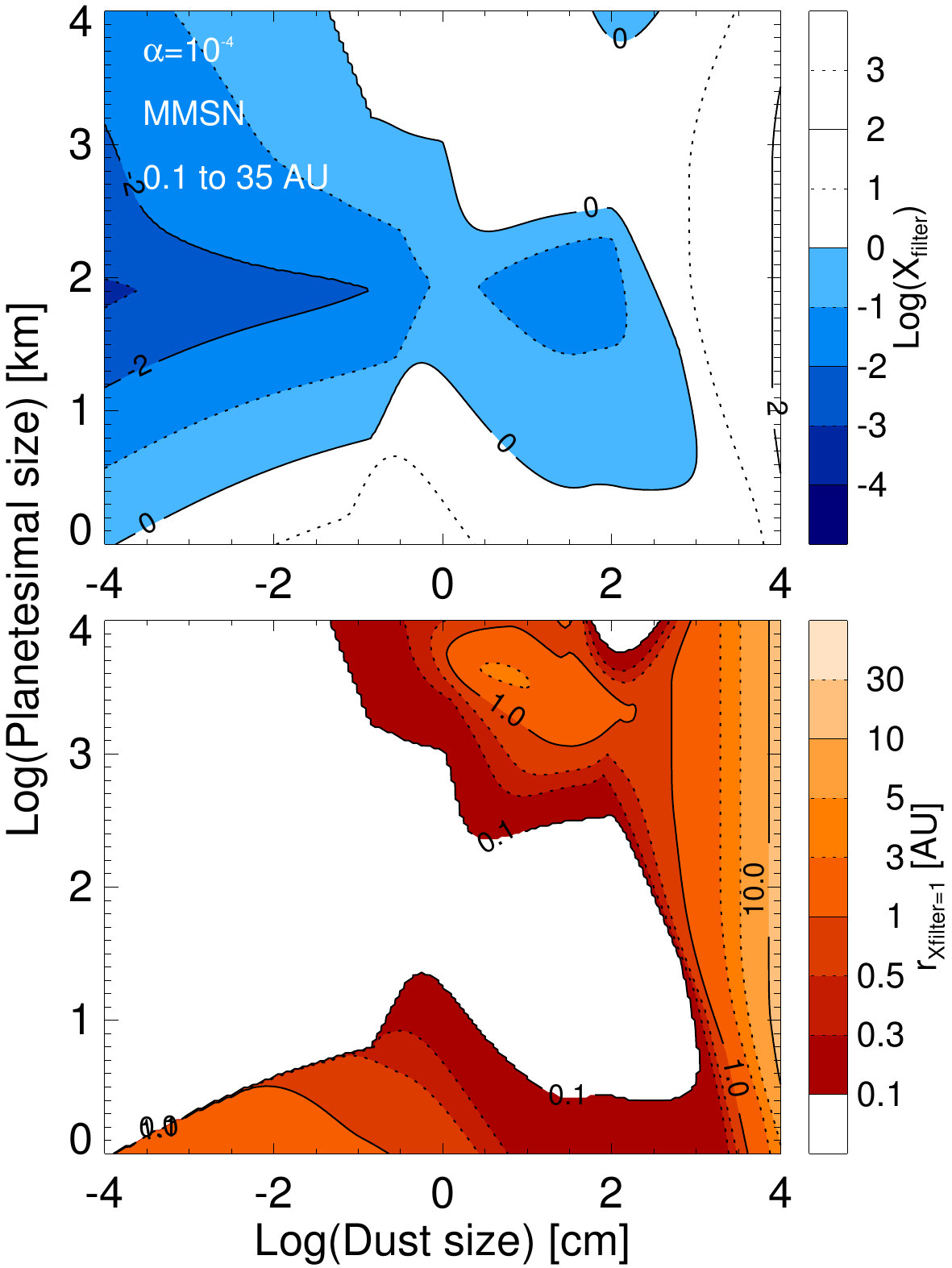}
\caption{Same as Fig.~\protect\ref{fig:xfiltering_alpha1d-2}, but for a turbulent parameter $\alpha=10^{-4}$.}
\label{fig:xfiltering_alpha1d-4}
\end{figure}

This is exemplified by Fig.~\ref{fig:xfiltering_alpha1d-4} which shows the integrated filtering coefficient as a function of planetesimal and dust sizes for $\alpha=10^{-4}$. The two islands of efficient filtering previously observed in Fig.~\ref{fig:xfiltering_alpha1d-2} are now much wider and more pronounced. Large embryos ($\Rp\wig{>}1000\,$km) can now efficiently filter dust between millimeter and meter size in the settling regime. This is directly linked to the efficient accretion of pebbles by protoplanetary cores discussed by \cite{OrmelKlahr2010} and \cite{LambrechtsJohansen2012}. (It should be noted that this requires a small value of the planetesimal scale height $\hp$, as discussed in Appendix~\ref{sec:hp}.)  Small planetesimals are able to efficiently filter small dust in the geometric regime and millimeter-sized dust particles can now be captured beyond 1\,AU, in better agreement with the meteoritic record. 

It is thus possible to imagine that the relative lack of water in the inner solar system may have been due to filtering of ice grains by planetesimals and/or planet embryos beyond 2 to 3 AU in a weakly turbulent disk and/or by its inefficient capture around 1\,AU. Conversely, the presence of abundant chondrules of micron to millimeter-size in meteorites may be due both to their inefficient filtering by planetary embryos and their more efficient capture by small size planetesimals. At this point, without a proper model to predict the distribution of sizes of planetesimals and its evolution, these are only conjectures. However, we may affirm that the complex dependency of filtering with planetesimal and dust size must have played a crucial role in shaping our solar system.
}

{\rev
\subsection{Compact vs. porous grains and planetesimals}

We have so far assumed a canonical $1\,\gcc$ physical density both for planetesimals and grains. We know that small particles, from dust grains to asteroids are porous and may therefore be characterized by small densities \citep[e.g.,][]{ConsolmagnoBritt1998,DominikTielens1997,Kataoka+2013}. On the other hand, large planetesimals and protoplanets are affected by compression effects and will generally have higher densities. Interestingly both tend to lead to a higher filtering efficiency than for our baseline density. For small planetesimals in the geometrical limit, the filtering efficiency derived by eq.~\eqref{eq:xfilter_geo_MMSN} is $\propto 1/\rhop$ because, for a given total mass of planetesimals (that defined by the MMSN) and a given size, the number of planetesimals is proportional to the inverse of their physical density. Porous planetesimals will therefore be more efficient filters in the geometric regime. In the settling regime, the filtering efficiency directly depends on the mass of the planetesimal, not on its density. 
}

\subsection{Filtering by protoplanets}

Even though our parameter range extended to radii of $10^4$\,km (corresponding to a mass of $0.7\,\rm M_\oplus$ for $\rhop=1\,\gcc$), our study has focused on smaller planetesimals. For protoplanets, two effects that have not been taken into account become gradually more important. 

First, protoplanets with masses over $\sim0.1\,\rm M_\oplus$ (corresponding to radii of about 5000\,km for $\rhop=1\,\gcc$) begin building a hydrogen-helium atmosphere which progressively enhances their cross section \citep{InabaIkoma2003, TanigawaOhtsuki2010}. When modeling this effect with the approach of \cite{OrmelKobayashi2012}, we find that the effect is significant at the high end of the radius range considered here, but without affecting our conclusions: the filtering efficiency remains low even for these Earth-mass protoplanets when in the settling regime, and extremely low in the hydrodynamical regime. 

For even larger masses (outside the range of this study), we can expect atmospheric capture to continue to increase, but it is progressively balanced by the fact that at masses $\sim 30\,\rm M_\oplus$ ($\sim 0.1\rm\,M_J$) protoplanets start opening gaps in the dust disk, thereby strongly suppressing or even halting their capture \citep{PaardekooperMellema2004, MorbidelliNesvorny2012}. The consequence is that the growth of giant planet cores should lead to a parking of dust outside of the orbit of the core farthest from the central star. This effect is most significant for particles with $\taus\sim 1$. These particles would neither be captured efficiently nor be accreted by the central star, which is a form of filtering that differs from the one discussed thus far. Presumably, these should eventually lead to the formation of a belt of planetesimals such as the Edgeworth-Kuiper belt in the solar system.

\subsection{Consequence for planet growth}

{\rrev 
As discussed previously, the island of higher filtering efficiency in the settling regime (for $\Rp\wig{>}1000\,$km) seen in figs.~\ref{fig:xfiltering_alpha1d-4} (see also Fig.~\ref{fig:xeffic_1AU_1d-4})} is directly linked to the efficient accretion of pebbles by protoplanetary cores found by \cite{LambrechtsJohansen2012}.
The increase in the filtering efficiency with planetesimal radius in the settling regime results from the very fast increase in the Bondi radius and hence planetesimal cross section (for dust of the right size), which more than compensates for the fewer objects for a given mass of planetesimals. This effect begins for sizes around 100\,km when the Bondi surface becomes larger than the planetesimal itself, and continues to the point where the Bondi and Hill radii are equal (around 1200\,km for $\rhop=1\,\gcc$ planetesimals at 1\,AU; see Appendix~\ref{sec:MMSN}). For larger sizes the effective cross section is not set by the Bondi surface and decreases rapidly (see Fig.~\ref{fig:cubical_ok10}). 

Although the filtering {\rrev factors $x_{\rm filter}$ and $X_{\rm filter}$} approach or exceed 1, we stress that this occurs in a relatively small region of the parameter space in terms of planetesimal and dust sizes. Given that planetesimals must have had a wide range of sizes, {\rrev it is not clear that filtering could be complete, even for weakly turbulent disks. We notice that} {\rev the collision probabilities obtained for large embryos are generally significantly smaller than unity (see Fig.~\ref{fig:p3d} and Appendix~\ref{sec:analytical}). This seems to imply that the rapid growth of giant planets cores by pebble accretion \citep{LambrechtsJohansen2012} is mostly driven by the fast inward flow of pebbles and requires a very large mass of solids in the disk, several times larger than the embryos to be created ($\sim 10\,\rm M_\oplus$ or more for giant planet cores). However, {\rrrev a recent work by \cite{Lambrechts+Johansen2014} shows that pebbles could grow to the right sizes (cm to mm) to be effectively captured by large embryos beyond an Earth mass. We note that the high filtering efficiency bands seen in the settling regime in figs.~\ref{fig:xfiltering_alpha1d-2} and \ref{fig:xfiltering_alpha1d-4} for $s\sim1-100$\,cm and $\Rp\wig{>}1000\,$km extend to high $\Rp$ values (large embryo masses) beyond the limit of the graphs. Thus, assuming that large embryos (Mars to Earth mass) may be formed rapidly by another mechanism, this opens the possibility of an efficient conversion of dust to planets, as obtained by Lambrechts \& Johansen. 
Also,} because of the ability of small planetesimals to filter dust grains (and pebbles), a convoluted environment of interacting planetesimals, fragments, and pebbles, like that studied by \cite{Chambers2014}, could, under the right conditions, limit the loss of solids onto the central star. 

More work is required, however, especially since collisions do not necessarily result in accretion \citep[e.g.,][]{SekiyaTakeda2003,Blum+Wurm2008,Johansen+2014} as we have implicitly assumed for this discussion. 
}

\section{Conclusions}

{\rev 
Abundant micron- to centimeter-sized dust is detected in protoplanetary disks in spite of theoretical predictions of a fast inward drift of these particles due to gas drag. We have studied whether this dust may be efficiently filtered by planetesimals and thus contribute to their growth, the formation of planets and the final composition of stars with planets.  

In most cases, the planetesimal disk can be considered vertically thinner than the dust subdisk which is itself thinner than the gas disk. Dust particles drift inward as a result of gas drag but also because of the advection of gas [eq.~\eqref{eq:vrnu}]. For all of these grains, turbulence and epicyclic motions lead to frequent crossings of the mid-plane. We showed that the {\em filtering length}, i.e., the distance over which we can expect most particles to have crossed the mid-plane during their inward drift, is on the order of} $\eta r\sim 10^{-3}r$ where $r$ is the orbital distance [eq.~\eqref{eq:lambda_f}]. {\rev This low value implies that particles may be filtered by the planetesimals even if they are initially high up in the disk. 

We thus derived the collision probability between a non-drifting planetesimal and a drifting dust particle in the geometrical limit [eq.\eqref{eq:P3Dgeocirc_chi}]. This probability depends on the ratio of the surface of the planetesimal $\Rp^2$ to the vertical surface of the disk $\hg r$ multiplied by a factor that depends on the values of $\alpha$ and $\taus$ which control both the characteristic vertical extent of the dust subdisk and the inward drift speed of the particles. Assuming a size distribution of planetesimals set by a collisional cascade, we derived the filtering efficiency [eq.~\eqref{eq:Xfilter_def}], i.e., the efficiency at which particles would be trapped by the planetesimal swarm. Small planetesimals dominate this regime because of their large collective surface area. The global filtering efficiency for an MMSN disk of 1\,km radius planetesimals at 1\,AU in a quiescent disk with $\alpha=10^{-4}$ was found to be close to unity for millimeter-sized grains, and between 0.1 and 10 for all grain sizes considered. It is inversely proportional to the planetesimal density and radius. It also scales with $\alpha^{-1/2}$ for all except the smallest grain sizes, implying a better filtering in quiescent parts of the disk (dead zones). Last but not least, it has a strong dependence on the orbital distance, $r^{-7/4}$, implying that filtering in the geometric regime is more efficient close to the central star [see eq.~\eqref{eq:X_filter_geo} and Fig.~\ref{fig:xfilter_geo}]. 

We then considered additional mechanisms. We showed that given the small expected values of the eccentricities and inclinations of the planetesimals in these young gas disks, the increase of the collision probabilities caused by these non-circular orbits is expected to be small [Fig.~\ref{fig:colprob_orbdist_fe}]. Gravitational focusing was found to be significant for planetesimals larger than about 100\,km and/or dust (or rather boulders) beyond meter size [Fig.~\ref{fig:ffocus_circ}]. However, small grains with a stopping time shorter than the planetesimal crossing time [eq.~\eqref{eq:tauf}] are carried by the hydrodynamical flow around the planetesimals, yielding a strongly reduced collision probability [eq.~\eqref{eq:fhydro} and Fig.~\ref{fig:fhydro}]. Turbulence in the disk was estimated to have a limited effect [eq.~\eqref{eq:ft} and Fig.~\ref{fig:ft}], but confirming this result would require direct simulations. 

On the basis of these calculations, we examined whether planetesimals are efficient filters of dust in protoplanetary disks. Large ``dust'' particles over 10 meters drift inward so slowly that they can be captured efficiently by planetesimals of any size [Fig.~\ref{fig:xeffic_1AU}]. Small dust particles (below millimeter size) generally have a low collision probability with planetesimals because of the hydrodynamical effect, at least at orbital distances of an AU or less. In between, millimeter dust to pebbles of tens of centimeters are captured in the geometrical regime by planetesimals smaller than 100\,km, but in the settling regime by larger planetesimals [figs.~\ref{fig:xeffic_1AU} and \ref{fig:xeffic_1mm}]. The capture law for the latter is very different than in the geometrical regime: it is independent of the planetesimal density $\rhop$ and is proportional to $r^{-1/4}$ in the settling/Bondi regime and $r^{1/4}$ in the settling/Hill regime [eq.~\eqref{eq:xfilter_mmsn}]. This implies that filtering in the settling regime is more uniform than in the geometrical regime. We can therefore expect that early, small planetesimals grew rapidly by capturing dust in the geometrical regime, creating an inside-out planetesimal formation front. Larger planetesimals would have been able to capture dust out to larger orbital distances. 

{\rrev Locally, the filtering efficiencies defined by $x_{\rm filter}$}  are generally low: they reach unity only at orbital distances below about 0.05 AU and for grains of millimeter size and larger for a turbulence parameter $\alpha=10^{-2}$. At 1\,AU, grains from millimeter to meter size are captured with only a $\sim 1\%$ efficiency for an MMSN disk of planetesimals for $\alpha=10^{-2}$, and $\sim 10\%$ for $\alpha=10^{-4}$ [figs.~\ref{fig:xeffic_1AU} to \ref{fig:xeffic_1mm_1d-4}]. Less turbulence increases the filtering efficiency, but not enough for an efficient capture of the solids in the disk. Separately, the fact that the gas disk progressively becomes less massive does not change the filtering efficiencies, except for very small grains (which are more easily accreted as time goes by), and beyond a few AUs (where filtering becomes less efficient before increasing again for very tenuous disks) [figs~\ref{fig:filtering_orbdist_mmsn} and \ref{fig:filtering_orbdist_mmsn_1d-4}]. Accounting for the fact that small planetesimals may be more porous and that large planetesimals should be more dense than our canonical $1\,\gcc$ density would help to increase the filtering efficiency close to 100\,\% at around 1\,AU from the central star. 
}

{\rrev Once a belt of planetesimals forms and grows over a large range of orbital distances, it may be able to filter the incoming dust more effectively, especially in weakly turbulent disks. Values of the integrated filtering efficiency $X_{\rm filter}$ defined by eq.~\eqref{eq:Xfilter_def} [figs~\ref{fig:xfiltering_alpha1d-2} and \ref{fig:xfiltering_alpha1d-4}] show the existence of essentially three regimes: planetesimals smaller than about 10\,km are very efficient at filtering dust, but because of the strong radial dependency, this must occur close to the central star. Planetesimals between 10\,km and 1000\,km in size are inefficient at filtering because of their smaller surface-to-mass ratio and absence of gravitational focusing. Planet embryos larger than 1000\,km are more efficient at filtering dust, and contrary to small-size planetesimals, this occurs over a wide range of orbital distances. 

Accounting for the size distribution of individual particles assembled in meteorites is a complex task that would require a full model accounting for the evolution of the size distribution of planetesimals and dust particles everywhere in an evolving protoplanetary disk. We simply mention that the absence of grains larger than about $\sim 1$\,cm in meteorites appears to be consistent with our finding that these would be captured preferentially in large embryos. Conversely, two features can explain the size distribution of chondrules in meteorites \citep[see, in a different context][]{Cuzzi+2001}: grains smaller than centimeter-size are captured preferentially in small-planetesimals, and moreover, grains of smaller sizes are increasingly more difficult to capture, as a consequence of the hydrodynamical flow around these planetesimals.} 

{\rev
Obviously, the question of whether dust can be efficiently filtered by planetesimals is a complex one that touches on many physical problems ranging from the unclear structure of circumstellar disks, angular momentum transport processes, efficiency of turbulence, growth processes, etc. We hope that our study can lay the ground for further studies. For example, direct simulations of planetesimals and dust at the interface between the hydrodynamical and settling regimes would be needed. The influence of turbulence has been approximated here and appears to be limited, but this would need to be confirmed on the basis of realistic simulations. We have also used mean approaches, but rare statistical events could matter and change the picture. Last but not least, we implicitly assumed that collisions result in sticking. In reality, this will strongly depend on the velocities involved, leading to a wide range of possibilities. Understanding the evolution of young stars, their composition, and that of planetary systems requires these dedicated studies. 
}

------------

\begin{acknowledgements}
T.G. warmly thanks the staff at the Tokyo Institute of Technology for their hospitality and support during his stay. This work has also benefited from multiple discussions with Taku Takeuchi, Hidekazu Tanaka, Holger Homann, Satoshi Okuzumi, Hiroshi Kobayashi, Guy Libourel, Alessandro Morbidelli, {\rrev Hal Levison and Michiel Lambrechts. We thank the referee, John Chambers, for remarks that helped to clarify the manuscript.} Part of this work was supported by the French \emph{Programme National de Plan\'etologie} and by the \emph{Agence Nationale de la Recherche} through the MOJO project (ANR-13-BS05-0003-01). C.W.O. acknowledges support for this work by NASA through Hubble Fellowship grant No. HST-HF-51294.01-A awarded by the Space Telescope Science Institute, which is operated by the Association of Universities for Research in Astronomy, Inc., for NASA, under contract NAS 5-26555.
\end{acknowledgements}

\bibliography{filtering}

\Online
\onecolumn
\begin{appendix}
\section{The minimum mass solar nebula scaling}\label{sec:MMSN}

We provide in Table~\ref{tab:MMSN} the main quantities that are used in the article and present their scaling in the so-called minimum mass solar nebula formalism \citep{Hayashi1981,Nakagawa+1986}.

\def\Z{\vphantom{$\left(R_{\rm p}^{R_{\rm B}}\right)^2_1$}}
\begin{table*}[hp]
\caption{Expressions for the minimum mass nebula scaling used in this article.}
\label{tab:MMSN}
\begin{tabular}{@{}llll}
\hline
\Z Quantity & Description & Equation & Value \\
\hline
\Z$m_\rmg$ & Gas molecular mass & & $3.9\times 10^{-24}\,\rm g$ \\
\Z$\sigma_\rmg$ & Gas collisional cross section (H$_2$) & & $2\times 10^{-15}\,\rm cm^2$\\
\Z$\mu_\rmg$ & Gas dynamic viscosity (H$_2$) & & $\sim 1.06\times 10^{-4}\ r_\AU^{-1/4}\rm\, g\,cm^{-1}\,s^{-1}$ \\
\Z$M_*$ & Stellar mass & & $1.989\times 10^{33}\,\rm g$\\
\Z$\Sigma_\rmg$ & Gas surface density & &$1.7\times 10^3\, \xmmsn r_\AU^{-3/2}\ \gcms$\\
\Z$T$ & Disk temperature & & $280\ r_\AU^{-1/2}\ {\rm K}$\\
\Z$\Omega_\rmK$ & Keplerian frequency & $\sqrt{GM_*/ r^3}$ & $1.99\times 10^{-7}\, r_\AU^{-3/2}\ \rm s^{-1}$\\
\Z$v_\rmK$ & Keplerian velocity & $\sqrt{GM_*/ r}$  & $2.98\times 10^6\, r_\AU^{-1/2}\ \rm cm\,s^{-1}$\\
\Z$c_\rmg$ & Isothermal sound speed & $\sqrt{k_{\rm B}T/m_\rmg}$ & $1.0\times 10^5\,r_\AU^{-1/4}\,\rm cm\,s^{-1}$\\
\Z$h_\rmg$ & Disk scale height & $c_\rmg/\Omega_\rmK$ & $0.033\ r_\AU^{5/4}\,\AU$\\
\Z$\rho_\rmg$ & Mid-plane gas density & $\Sigma_\rmg/(\sqrt{2\pi}h_\rmg)$ & $1.4\times 10^{-9}\, \xmmsn r_\AU^{-11/4}\,\gcc$\\
\Z$\nu_\rmg$ & Gas kinematic viscosity (H$_2$) & $\mu_\rmg/\rho_\rmg$ & $\sim 7.6\times 10^{4}\ \xmmsn^{-1} r_\AU^{5/2}\rm\, cm^{2}\,s^{-1}$ \\
\Z$\nu_t$ & Gas turbulent viscosity & $\alpha c_\rmg h_\rmg$ & $\alpha\times 4.9\times 10^{16}\ r_\AU\rm\, cm^{2}\,s^{-1}$ \\
\Z$\eta$ & Sub-Keplerian factor & eq.~(\ref{eq:eta}) & $1.8\times 10^{-3}\, r_\AU^{1/2}$ \\
\Z$\lambda_\rmg$ & Gas mean free path & $m_{\rm g}/\sigma_{\rm g}\rho_{\rm g}$ & $1.44\ \xmmsn^{-1} r_\AU^{11/4}\,\rm cm$ \\
\Z$s_{\rm lim}$ & Epstein to Stokes limiting size & $9/4\,\lambda_\rmg$ & $ 3.23\ \xmmsn^{-1} r_\AU^{11/4}\,\rm cm$ \\
\Z$s_{\rm max}$ & Stokes to quadratic limiting size & $27 \lambda_{\rm g} c_{\rm g}/ (2\eta r\Omega_{\rm K})$ & $357\ \xmmsn^{-1} r_\AU^{5/2}\,\rm cm$\\
\Z$\tau_\rms$ & Dimensionless stopping time & eq.~(\ref{eq:taus}) & $4.8\times 10^{-3}\, \xmmsn^{-2}\rho_\rms r_\AU^{17/4} {\rm min}\left( \delta_{\rm lim},\delta_{\rm lim}^2\right) $ \\
\Z$t_\rms$ & Stopping time & $\tau_\rms /\Omega_\rmK$ & $24000\ \xmmsn^{-2}\rho_\rms r_\AU^{17/4} {\rm min}\left( \delta_{\rm lim},\delta_{\rm lim}^2\right) \,\rm s$ \\
\Z$\left(r_{\rm lim}\right)_{\tau_\rms=1}$ & Stokes to Epstein orbital distance for $\tau_\rms=1$ & & $3.5\ \xmmsn^{8/17}\rho_\rms^{-4/17}\,\AU$ \\
\Z$\tau_{\rm f}$ & Dimensionless friction time & eq.~(\ref{eq:tauf}) & $1.3\times 10^{8}\ \xmmsn^{-2}\rho_\rms R_\rmp^{-1} r_\AU^{23/4} {\rm max}\left( \delta_{\rm lim}, \delta_{\rm lim}^2 \right)$ \\
\Z$R_{\rm B}$ & Bondi radius & eq.~\eqref{eq:RB} & $95.6 \rhop R_{\rm p,100km}^3$\ [km]\\
\Z$R_{\rm H}$ & Hill radius & eq.~\eqref{eq:RH} & $1.33\times 10^5 \rhop^{1/3} R_{\rm p,1000km} \rau$\ [km] \\
\Z$R_{\rm p}^{R_{\rm B}=R_{\rm p}}$ & geometrical-Bondi limit & eq.~\eqref{eq:rb=rp} & $102 \rhop^{-1/2}$\ [km]\\
\Z$R_{\rm p}^{R_{\rm B}=R_{\rm H}}$ & Bondi-Hill limit & eq.~\eqref{eq:rb=rh} & $1180 \rhop^{-1/3}\rau^{1/2}$\ [km]\\
\hline\\
\end{tabular}
Notes: $r_\AU\equiv r/\AU$; $\delta_{\rm lim}\equiv s/s_{\rm lim}$; $x_{\rm lim}\equiv r/r_{\rm lim}$.\\
$\mu_\rmg =8.76\times10^{-5} 365.85/(T+72\,\rmK) (T/293.85\,\rmK)^{3/2}$ (Sutherland's formula).

\end{table*}

\section{Geometric collision probability with shear}\label{sec:shear}

We derive the expression for the geometric collision probability and show that it is identical to that derived by \cite{OrmelKlahr2010}. From eqs.~(\ref{eq:vrnu}) and (\ref{eq:vthetanu}) and accounting for both the (assumed inward) radial velocity of the gas due turbulent viscosity $v_\nu$ and the shear in the disk,
\begin{subequations}
\begin{align}
\Delta v_r&=-{2 (\taus+\tausnu)\over 1+\taus^2}\eta v_{\rmK,0},\label{eq:A_vx}\\
\Delta v_\theta&=-{1\over 1+\taus^2}\eta v_{\rmK,0}-{3\over 2}x\Omega_{\rmK,0}, \label{eq:A_vy}
\end{align}
\end{subequations}
where, following \cite{OrmelKlahr2010}, we have used an expansion around the location of the planetesimal ($\Omega_{\rmK,0}\equiv \Omega_\rmK(x=0)$), and $x$ is the radial distance from the planetesimal at which the collision with the dust particle takes place. For simplicity, we assume that on average $x=R_\rmp/2$. This yields the impact rate in the geometrical limit as:
\begin{equation}
{\cal R}_{\rm 2D,geo,circ}=2R_\rmp\eta v_\rmK {2(\taus+\tausnu)\over 1+\taus^2}
\sqrt{1+\left[{4\eta +3(R_\rmp/ r)(1+\tau_\rms^2)\over 8\eta(\taus+\tausnu)}\right]^2},
\label{eq:A_Rgeocirc}
\end{equation}
where we have dropped the 0 labels. When $v_\nu=0$, this expression is equivalent to eq.~(22) of \cite{OrmelKlahr2010} who derive this impact rate by an analysis of the trajectory of dust particles. In most cases, we will consider $\eta\tau_\rms\ll R_\rmp/r$ so that shear may be neglected and ${\cal R}\sim 6 R_\rmp \eta v_\rmK \tau_\rms/(1+\tau_\rms^2)$: the collision rate is the product of the cross section $2 R_\rmp$ and the encounter velocity $3 \eta v_\rmK \tau_\rms/(1+\tau_\rms^2)$. 

The 2D probability that a planetesimal will accrete a given dust grain is then given by eq.~\eqref{eq:P2Dbasic}: 
\begin{equation}
{\cal P}_{\rm 2D,geo,circ}={R_\rmp\over \pi r}\sqrt{1+\left[{4\eta +3(R_\rmp/ r)(1+\tau_\rms^2)\over 8\eta(\taus+\tausnu)}\right]^2}.
\label{eq:A_P2Dcirc}
\end{equation}
This equation may be approximated in the different regimes (from small to large particles):
\begin{equation}
{\cal P}_{\rm 2D,geo,circ}\approx {R_\rmp\over \pi r}\max\left[{1\over 2(\taus+\tausnu)},1,{3\over 8}{R_\rmp\over r}{\taus\over \eta}\right].  
\label{eq:A_P2Dcirc_approx}
\end{equation}
The lowest capture probability $R_\rmp/\pi r$ corresponds to dust particles with $\taus\sim 1$ which drift in with a speed that is comparable to the headwind felt by the planetesimal. One can expect that these particles will be the most difficult to filter. For smaller $\taus$ values, the slower drift leads to a higher probability, limited by the gas drift. For large $\taus$ values, the drift rate also decreases and the effect of the Keplerian shear across the planetesimal becomes dominant. The capture probability then becomes a very steep function of the planetesimal size ($\propto R_\rmp^3$ in the Epstein regime and $\propto R_\rmp^4$ in the Stokes regime). {\rev However, shear becomes important only for very large planetesimals ($\Rp\wig{>}7200{\,\rm km}\taus^{-1}\rau^{3/2}$ for the MMSN). It only concerns cases for which the collisions take place in the settling or three-body regimes. Shear can thus be neglected for the geometrical regime.}

{\rev
\section{Analytical estimates for small grains (geometric, hydro and settling regimes)}\label{sec:analytical}

We provide here analytical estimates for the collision probabilities and filtering efficiencies for the geometric and settling regimes. In order to provide tractable relations, we make the following simplifications: 
\begin{itemize}\itemsep0.2em
\item We study only small particles such that $\taus<1$, corresponding generally to $s\wig{<}1$\,m.
\item Given the above assumption, we use $\Delta v=\eta v_\rmK$ for the dust-planetesimal encounter velocity.
\item We approximate $\chi_{\alpha,\taus}$ defined by eq.~\eqref{eq:chialpha} as follows:
\begin{equation}
\chialpha\approx{\rm min}(1.7/\alpha,1/(2\sqrt{\alpha\taus})).
\label{eq:chialpha_approx}
\end{equation}
We thus separate small particles with $\taus\le\tausnu\sim 0.3^2 \alpha$ which have not settled from larger particles for which $\hdust(\taus)<h_\rmg$.
\item We consider cases for which the planetesimal capture radius is smaller than the dust disk scale height and thus eq.~\eqref{eq:P3D} can be simplified as:
\begin{equation}
{\cal P}={\sqrt{\pi}\over2}{b\over \hdust}{\cal P}_{\rm 2D}.
\label{eq:P3D_simplified}
\end{equation}
\item We do not consider the Safronov and three-body regimes. 
\end{itemize}

\begin{figure*}[bht]
\includegraphics[width=\hsize,angle=0]{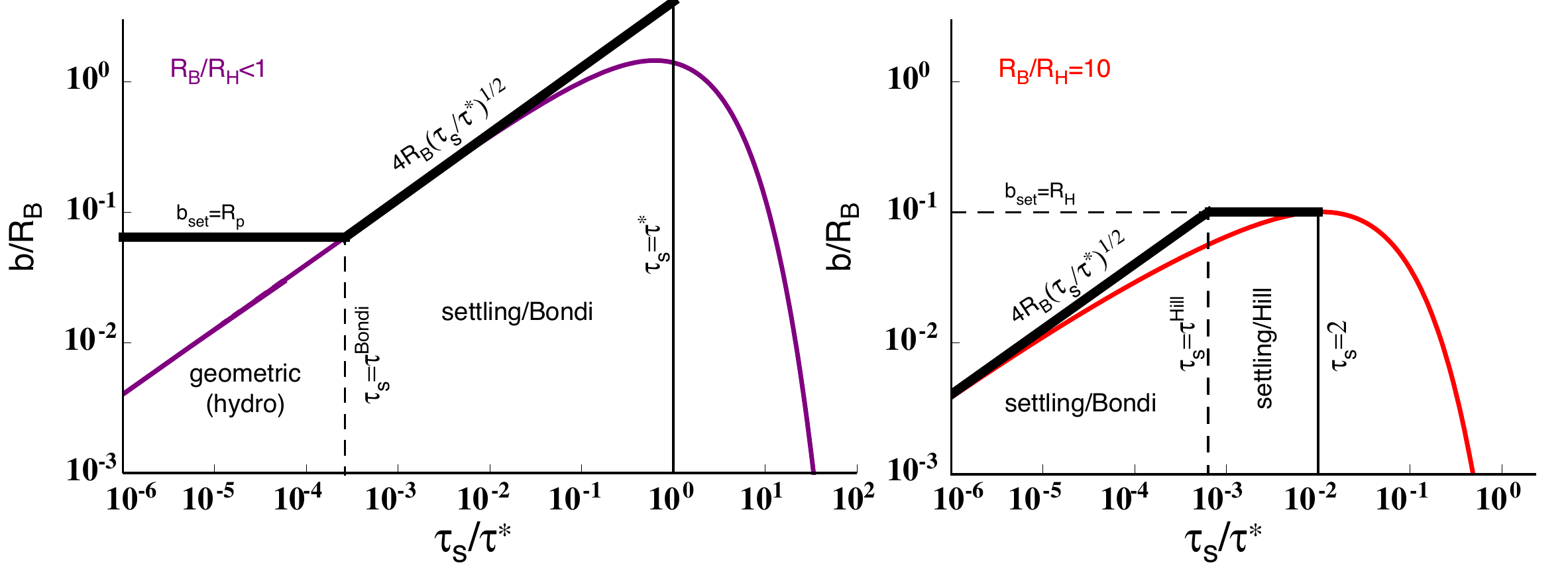}
\caption{Value of the effective capture radius $b_{\rm set}$ in units of the Bondi radius $r_{\rm B}$ from eq.~\eqref{eq:bset} (colored curves) together with the approximation used in eq.~\eqref{eq:b-fit} (thick black lines). 
{\em Left panel}: Solution when the Bondi radius is smaller than the Hill radius. 
{\em Right panel}: Solution when the Bondi radius is larger than the Hill radius (specifically, the solution of eq.~\eqref{eq:bset} is shown for $R_{\rm B}/R_{\rm H}=10$). }
\label{fig:cubical_ok10_panels}
\end{figure*}

Given a known encounter velocity, the calculation of the collision probability only requires that of the planetesimal effective capture radius. We know that it is extremely small in the hydro regime and equal to $\Rp$ in the geometric regime. In the settling regime, it is given by eq.~\ref{eq:bset}. As shown in Fig.~\ref{fig:cubical_ok10}, the settling regime may be subdivided into a Bondi regime in which the capture radius is proportional to $\taus^{1/2}$, and a Hill regime in which the capture radius is approximatively independent of $\taus$ and equal to the Hill radius. We thus approximate the effective capture radius of planetesimals/embryos in the different regimes as
\begin{equation}
b(\taus)\approx 
\begin{cases}
0 & \mbox{in the hydro regime,}\\
\Rp &\mbox{in the geometric regime,}\\
4 R_{\rm B}(\taus/\taus^*)^{1/2}&\mbox{in the Bondi regime,}\\
R_{\rm H}&\mbox{in the Hill regime,}
\end{cases}
\label{eq:b-fit}
\end{equation}
and $\taus^*$ is defined by eq.~\eqref{eq:taus*}. The different regimes are defined by the following relations:
\begin{equation}
\begin{cases}
\taus\le\taus^{\rm hydro}{\rm\ and\ }\taus\le\taus^{\rm Bondi}& \mbox{for the hydro regime,}\\
\taus>\taus^{\rm hydro}{\rm\ or\ }\taus >\taus^*& \mbox{for the geometrical regime,}\\
\taus >\taus^{\rm Bondi}{\rm\ and\ }\taus\le {\rm min}(\taus^*,\taus^{\rm Hill}) & \mbox{for the settling/Bondi regime,}\\
\taus >\taus^{\rm Hill}{\rm\ and\ }\taus\le 2 & \mbox{for the settling/Hill regime.}
\end{cases}
\end{equation}
The dimensionless stopping times for the different regimes are defined by eq.~\eqref{eq:taus*} for $\taus^*$, eq.~\eqref{eq:tauf} and $\tau_{\rm f}=1$ for $\taus^{\rm hydro}$, $4R_{\rm B}(\taus^{\rm Bondi}/\taus^*)^{1/2}=\Rp$ for $\taus^{\rm Bondi}$, and {\rrrev $4R_{\rm B}(\taus^{\rm Hill}/\taus^*)^{1/2}=R_{\rm H}$ for $\tau^{\rm Hill}$}. These and their MMSN approximations are thus respectively: 
\begin{alignat}{3}
&\taus^{\rm hydro}&&={1\over\eta}{\Rp\over r}\quad&&\approx 3.68\times 10^{-4}\Rpcent\raudef^{-3/2},\label{eq:taus_hydro}\\
&\taus^*&&=4{\mu_\rmp\over \eta^3}\quad&&\approx 1.41\times 10^{-3} \rhopdef \Rpcent^3 \raudef^{-3/2},\label{eq:taus_star}\\
&\taus^{\rm Bondi}&&={1\over 4}{\eta\over\mu_\rmp}{\Rp^2\over r^2}\quad&&\approx 3.85\times 10^{-4}\rhopdef^{-1}\Rpcent^{-1}\raudef^{-3/2},\label{eq:taus_Bondi}\\
&\taus^{\rm Hill}&&={1\over 24^{2/3}}{\eta\over\mu_\rmp^{1/3}}\quad&&\approx 0.681 \rhopdef^{-1/3}\Rpmille^{-1}\raudef^{1/2}. \label{eq:taus_Hill}
\end{alignat}
It can be seen that the value of $\taus^*$ is proportional to $\mup$ and reaches $\taus^*=2$ for the Hill-dominated settling regime ($\Rp\sim 1000\,$km). Importantly, $\taus^*=\taus^{\rm Bondi}=\taus^{\rm hydro}\equiv\taus^0$ for $\Rp\equiv\Rp^0$, with
\begin{alignat}{3}
&\taus^0&&={3^{1/2}\over 4\pi^{1/2}}\left(M_*\over \rhop r^3\right)^{1/2}\quad&&\approx 1.88\times 10^{-4}\rhopdef^{-1/2}\raudef^{-3/2},\label{eq:taus0}\\
&\Rp^0&&=\taus^0\eta r \quad&&\approx 51.1 \rhopdef^{-1/2}\ {\rm km}.\label{eq:Rp0}
\end{alignat}



\subsection{Collision probabilities}

Using eqs.~\eqref{eq:P3D_simplified}, \eqref{eq:R}, \eqref{eq:P2Dbasic}, \eqref{eq:RH}, \eqref{eq:RB}, and \eqref{eq:taus*}, the collision probabilities for the three regimes defined in eq.~\eqref{eq:b-fit} can be written
\begin{equation}
{\cal P}\approx
\begin{cases}
0&\mbox{in the hydro regime,}\\
\disp{1\over 2\sqrt{\pi}}{\Rp^2\over r \hg}\chialpha& \mbox{in the geometric regime,}\\
\disp{2\over \sqrt{\pi}}{r\over \hg}{\mu_\rmp\taus\over\eta}\chialpha& \mbox{in the settling/Bondi regime,}\\
\disp{1\over 24^{2/3}\sqrt{\pi}}{r\over\hg}\mu_\rmp^{2/3}\taus\chialpha& \mbox{in the settling/Hill regime.}
\end{cases}
\end{equation}
When using the approximation from eq.~\eqref{eq:chialpha_approx} and the MMSN scalings from section~\ref{sec:MMSN}, this yields for small particles such that $\taus<0.09\alpha$:
\begin{equation}
{\cal P}_{\rm MMSN}\approx
\begin{cases}
\disp 4.44\times 10^{-14}\alphadef^{-1}\Rpun^2\raudef^{-9/4}& \mbox{in the geometric regime,} \\
\disp 4.15\times 10^{-9}\left(\taus\over 0.09\alpha\right) \rhopdef \Rpcent^3\raudef^{-3/4}& \mbox{in the settling/Bondi regime,}\\ 
\disp 7.86\times 10^{-4}\alphadef^{-1} \rhopdef^{2/3} \Rpmille^2\raudef^{-1/4}& \mbox{in the settling/Hill regime.}\\ 
\end{cases}
\end{equation}
and for medium-sized particles such that $0.09\alpha\le\taus< {\rm min}(\taus^*,2)$:
\begin{equation}
{\cal P}_{\rm MMSN}\approx
\begin{cases}
\disp 4.44\times 10^{-14}\left(\taus\over 0.09\alpha\right)^{-1/2}\alphadef^{-1} \Rpun^2\raudef^{-9/4}& \mbox{in the geometric regime,} \\
\disp 5.19\times 10^{-9} \left(\taus\over\taus^*\right)^{1/2} \alphadef^{-1/2}\rhopdef^{3/2}\Rpcent^{9/2}\raudef^{-3/2}& \mbox{in the settling/Bondi regime,}\\
\disp 5.71\times 10^{-5}\left(\taus\over \taus^{\rm Hill}\right)^{-1/2}\alphadef^{-1/2}\rhopdef^{5/6}\Rpmille^{5/2}\raudef^{-1/2}& \mbox{in the settling/Hill regime.}
\end{cases}
\end{equation}
We note that in these expressions, the scaling of the particle size (stopping time) was adjusted to be centered on the maximum collision probability in each regime. 

\begin{figure*}[tb]
\includegraphics[width=\hsize,angle=0]{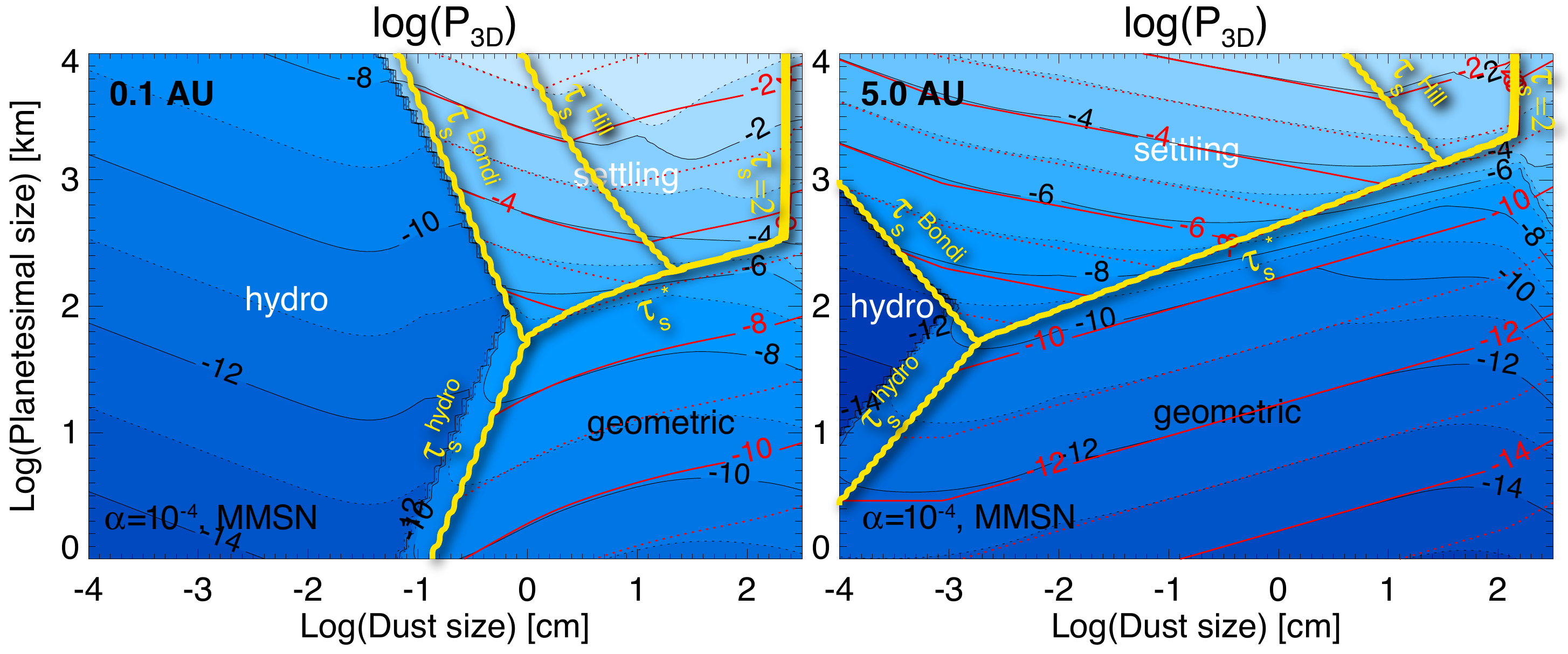}
\caption{Value of the collision probability $\log{\cal P}$ obtained by the full theory (black contours) and by the simplified one (red contours), for the MMSN and $\alpha=10^{-4}$. The two panels are for two orbital distances: 0.1\,AU (left) and 5\,AU (right). The yellow lines are limiting values of the stopping time, $\taus^{\rm hydro}$ [eq.~\eqref{eq:taus_hydro}], $\taus^*$ [eq.~\eqref{eq:taus_star}], $\taus^{\rm Bondi}$ [eq.~\eqref{eq:taus_Bondi}], and $\taus^{\rm Hill}$ [eq.~\eqref{eq:taus_Hill}], respectively.}
\label{fig:colprob_analytical_p3d}
\end{figure*}

Figure~\ref{fig:colprob_analytical_p3d} shows that the simplified solutions are good approximations of the full solutions, except for large grains with $\taus\wig{>}2$ and at the interface between the settling and geometric regimes (along the $\taus^*$ line). In the settling regime, the maximum collision probability is obtained for dust such that $\taus=\taus^{\rm Hill}$, i.e., with a stopping time equal to the Hill sphere crossing time. For a low value of the turbulence parameter $\alpha=10^{-4}$, a $1\,\rm M_\oplus$ embryo (corresponding to $\Rp=11250\,$km for a $\rhop=1\,\gcc$ density) would have ${\cal P}_{\rm MMSN}\approx 0.24/\sqrt{\rau}$ for particles such that $\taus=\taus^{\rm Hill}\approx 0.015\sqrt{\rau}$, i.e., for centimeter-sized pebbles between 1\,AU and 10\,AU.

\subsection{Filtering efficiency}

The filtering efficiency as defined by eq.~\eqref{eq:xfilter} writes for the regimes considered:
\begin{equation}
x_{\rm filter}\approx
\begin{cases}
0& \mbox{in the hydro regime,}\\
\disp{3\over 8\sqrt{\pi}}{r\over \hg}{\Sigma_\rmp\over \rhop\Rp}\chialpha& \mbox{in the geometric regime,} \\
\disp{2\sqrt{\pi}}{r\over \hg}{r^2\Sigma_\rmp\over M_*}{\taus\over\eta}\chialpha& \mbox{in the settling/Bondi regime,}\\
\disp{\sqrt{\pi}\over 24^{2/3}}{r\over\hg}{r^2\Sigma_\rmp\over m_\rmp^{1/3} M_*^{2/3}}\taus\chialpha& \mbox{in the settling/Hill regime.}
\end{cases}
\end{equation}
The MMSN scaling then yields, for small particles such that $\taus<0.09\alpha$:
\begin{equation}
x_{\rm filter,MMSN}\approx
\begin{cases}
\disp 0.127 \alphadef^{-1} \rhopdef^{-1} \Rpun^{-1}\raudef^{-7/4} & \mbox{in the geometric regime,} \\
\disp 0.0119 \left(\taus\over 0.09\alpha\right) \raudef^{-1/4}& \mbox{in the settling/Bondi regime,}\\
\disp 2.24 \alphadef^{-1} \rhopdef^{-1/3} \Rpmille^{-1} \raudef^{1/4}& \mbox{in the settling/Hill regime,}
\end{cases}
\end{equation}
and for medium-sized particles such that $0.09\alpha\le\taus< {\rm min}(\taus^*,2)$:
\begin{equation}
x_{\rm filter,MMSN}\approx
\begin{cases}
\disp 0.127\left(\taus\over 0.09\alpha\right)^{-1/2}\alphadef^{-1} \rhopdef^{-1} \Rpun^{-1}\raudef^{-7/4} & \mbox{in the geometric regime,}\\
\disp 0.0148 \left(\taus\over\taus^*\right)^{1/2} \alphadef^{-1/2} \rhopdef^{1/2} \Rpcent^{3/2}\raudef^{-1}& \mbox{in the settling/Bondi regime,}\\
\disp 0.163 \left(\taus\over \taus^{\rm Hill}\right)^{-1/2} \alphadef^{-1/2} \rhopdef^{-1/6} \Rpmille^{-1/2} & \mbox{in the settling/Hill regime.}
\end{cases}
\label{eq:xfilter_mmsn}
\end{equation}

\begin{figure*}[tb]
\includegraphics[width=\hsize,angle=0]{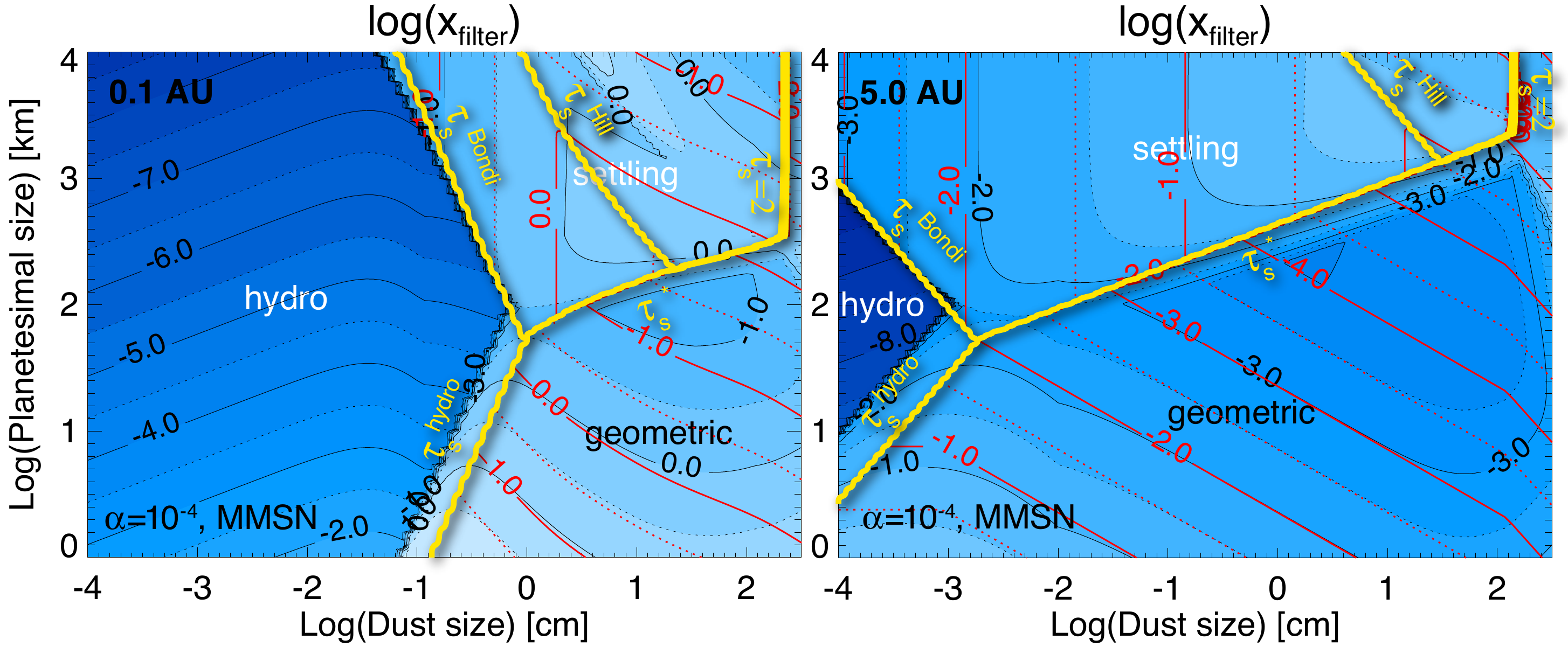}
\caption{Value of the filtering efficiency $\log{x_{\rm filter}}$ obtained by the full theory (black contours) and by the simplified one (red contours), for the MMSN and $\alpha=10^{-4}$. The two panels are for two orbital distances: 0.1\,AU (left) and 5\,AU (right). (See Fig.~\ref{fig:colprob_analytical_p3d} for a full description.)}
\label{fig:colprob_analytical_xfilter}
\end{figure*}

Figure~\ref{fig:colprob_analytical_xfilter} compares the values of $x_{\rm filter}$ obtained with the full theory and the simplified one, again showing good agreement except for large grains beyond meter size and near the $\taus^*$ line. For increasing planetesimal sizes, the maximum filtering efficiency is obtained along $\taus=\taus^{\rm hydro}$, then $\taus=\taus^*$ and finally $\taus=\taus^{\rm Hill}$. 
}

{\rrev
\section{Results for a disk with $\alpha=10^{-4}$} \label{sec:weak-turbulence}

Given the inefficient filtering obtained in the high turbulence ($\alpha=10^{-2}$) case, we now consider the weak turbulence case ($\alpha=10^{-4}$). This is more favorable because dust settles closer to the mid-plane and the mean gas flow is also slower.  

Figures~\ref{fig:xeffic_1AU_1d-4} and \ref{fig:xeffic_1mm_1d-4} show the resulting filtering efficiency, both at 1\,AU and for 1\,mm dust, as a function of orbital distance. At 1\,AU, efficient filtering is achieved for a wider range of dust and planetesimal sizes, basically for dust of $10\,\micron$ to 1\,cm and planetesimals of less than a few kilometers in radius. A small island with $x_{\rm filter}\wig{>}1$ also appears in the settling regime, for meter-sized dust and planetesimals of $\sim 1000\,$km in radius.  Compared to figs.~\ref{fig:xeffic_1AU} and \ref{fig:xeffic_1mm} in the $\alpha=10^{-2}$ case, there is about an order of magnitude increase in the filtering efficiency.

\begin{figure}[tb]
  \includegraphics[width=12cm]{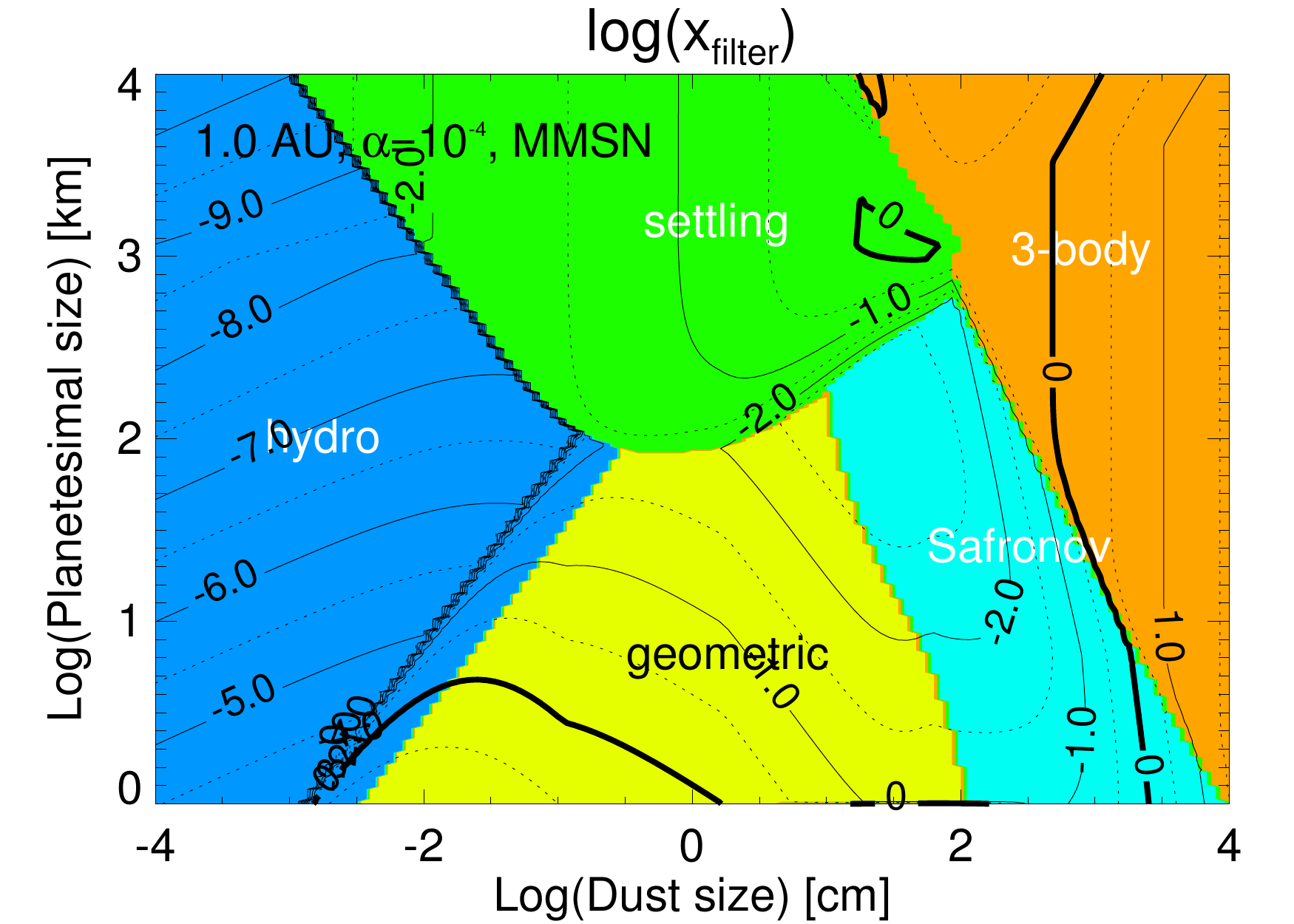}
\caption{Contours of the filtering efficiency by a MMSN planetesimal disk at 1 AU for $\alpha=10^{-4}$, assuming monodisperse size distributions for planetesimals and dust grains. The contour indicating perfect filtering efficiency for $\log_{10}x_{\rm filter}=0$ is shown in bold.}
\label{fig:xeffic_1AU_1d-4}
\end{figure}

\begin{figure}
\includegraphics[width=12cm,angle=0]{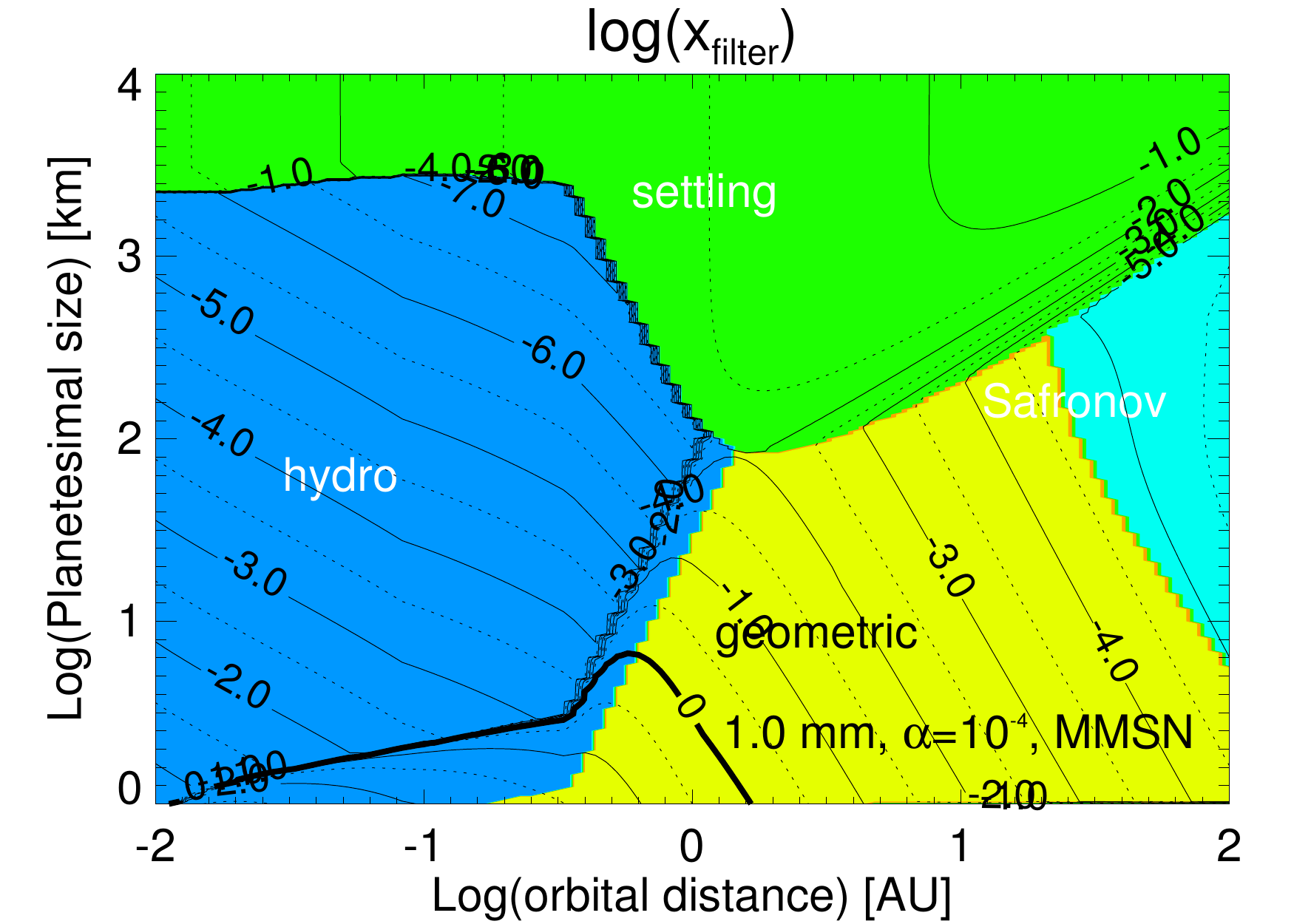}
\caption{Contours of the filtering efficiency by a MMSN planetesimal disk for 1 mm dust for $\alpha=10^{-4}$, assuming a monodisperse size distributions for planetesimals. The contour indicating perfect filtering efficiency for $\log_{10}x_{\rm filter}=0$ is shown in bold.}
\label{fig:xeffic_1mm_1d-4}
\end{figure}

{\rev
Figure~\ref{fig:filtering_orbdist_mmsn_1d-4}, shows how a lower value of the turbulence parameter $\alpha=10^{-4}$ affects filtering by a distribution of planetesimals between 1\,km and 1000\,km. Compared to Fig.~\ref{fig:filtering_orbdist_mmsn} for $\alpha=10^{-2}$, filtering is found to be more efficient of course. For example, dust of 1\,mm in size can now be efficiently captured in an MMSN disk inside about $0.03\,$AU while a value of $x_{\rm filter}=1$ is never reached for the high turbulence case until the gas disk has shrunk to about 1\% of the MMSN value.} Similarly, 1\,cm grains can be captured with an efficiency close to unity inside of 0.1\,AU compared to about 0.05\,AU for the $\alpha=10^{-2}$ case. In the $0.1-1$\,AU range, a wide range of small grains may be captured with an efficiency between $1$ and $0.1$. {\rrev The dependence as a function of $\xmmsn$ remains relatively weak, as in the strong turbulence case. 
}

\begin{figure}
\includegraphics[width=10cm,angle=0]{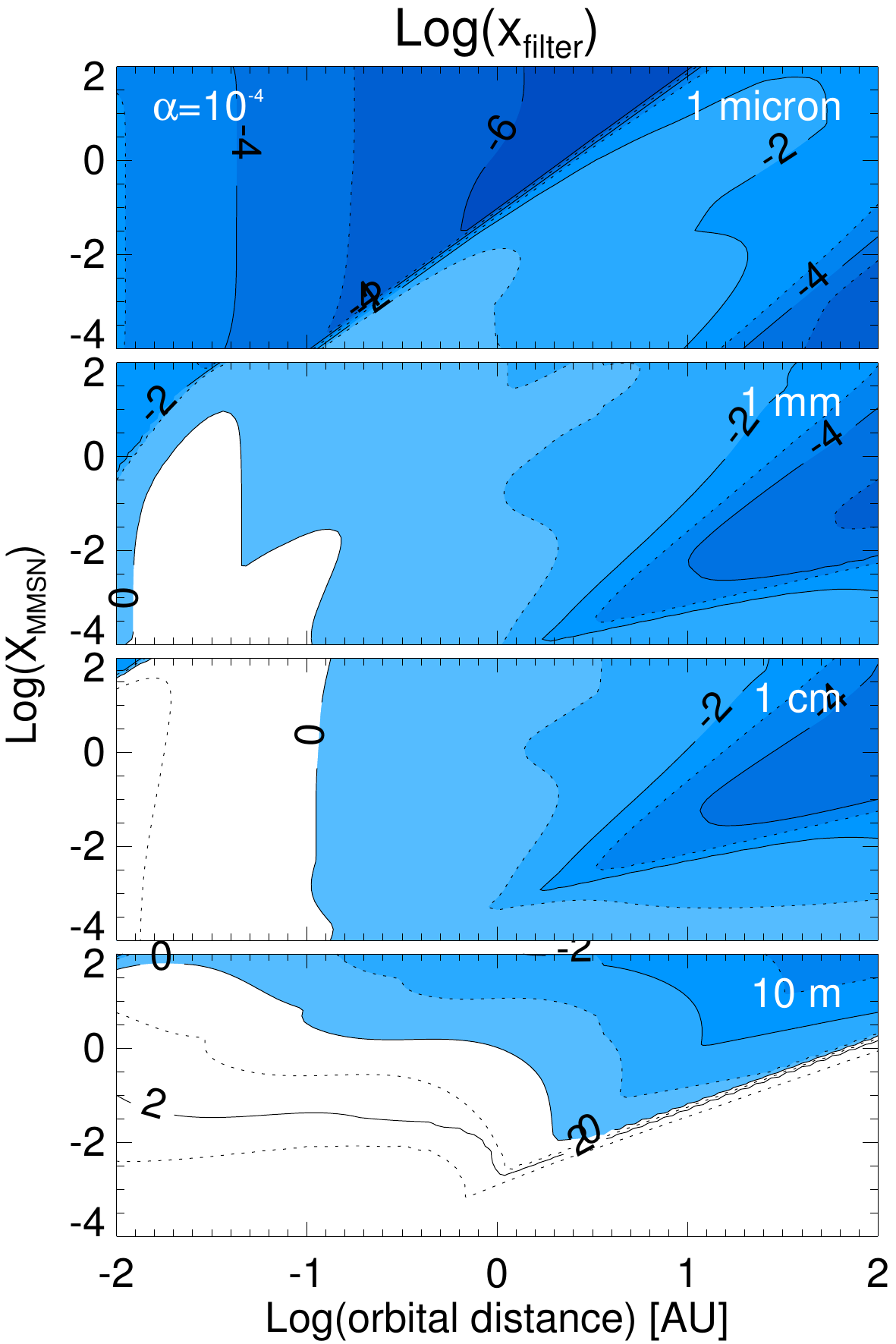}
\caption{Filtering efficiency of a swarm of planetesimals with radii between 1\,km and 1000\,km as a function of orbital distance and mass of the gas disk for $\alpha=10^{-4}$. (See Fig.~\ref{fig:filtering_orbdist_mmsn} for details.) }
\label{fig:filtering_orbdist_mmsn_1d-4}
\end{figure}

\newpage
\section{Dependence on the planetesimal scale height}\label{sec:hp}

All the results presented so far have assumed a rather low value of the planetesimal scale height $\hp=0.01\hg$. This favors the filtering of particles able to settle to the mid-plane in a thin plane. In Fig.~\ref{fig:xfiltering_alpha1d-4_panels} we show how the integrated filtering efficiency $X_{\rm filter}$ depends on the value of $\hp$. We select a low value of the turbulence parameter $\alpha=10^{-4}$ because this is where the differences are most important. With an infinitely thin planetesimal disk ($\hp=0$, left panels), we obtain results that are very similar to our fiducial case ($\hp/\hg=0.01$) shown in Fig.~\ref{fig:xfiltering_alpha1d-4}. 

\begin{figure}
\includegraphics[width=\hsize,angle=0]{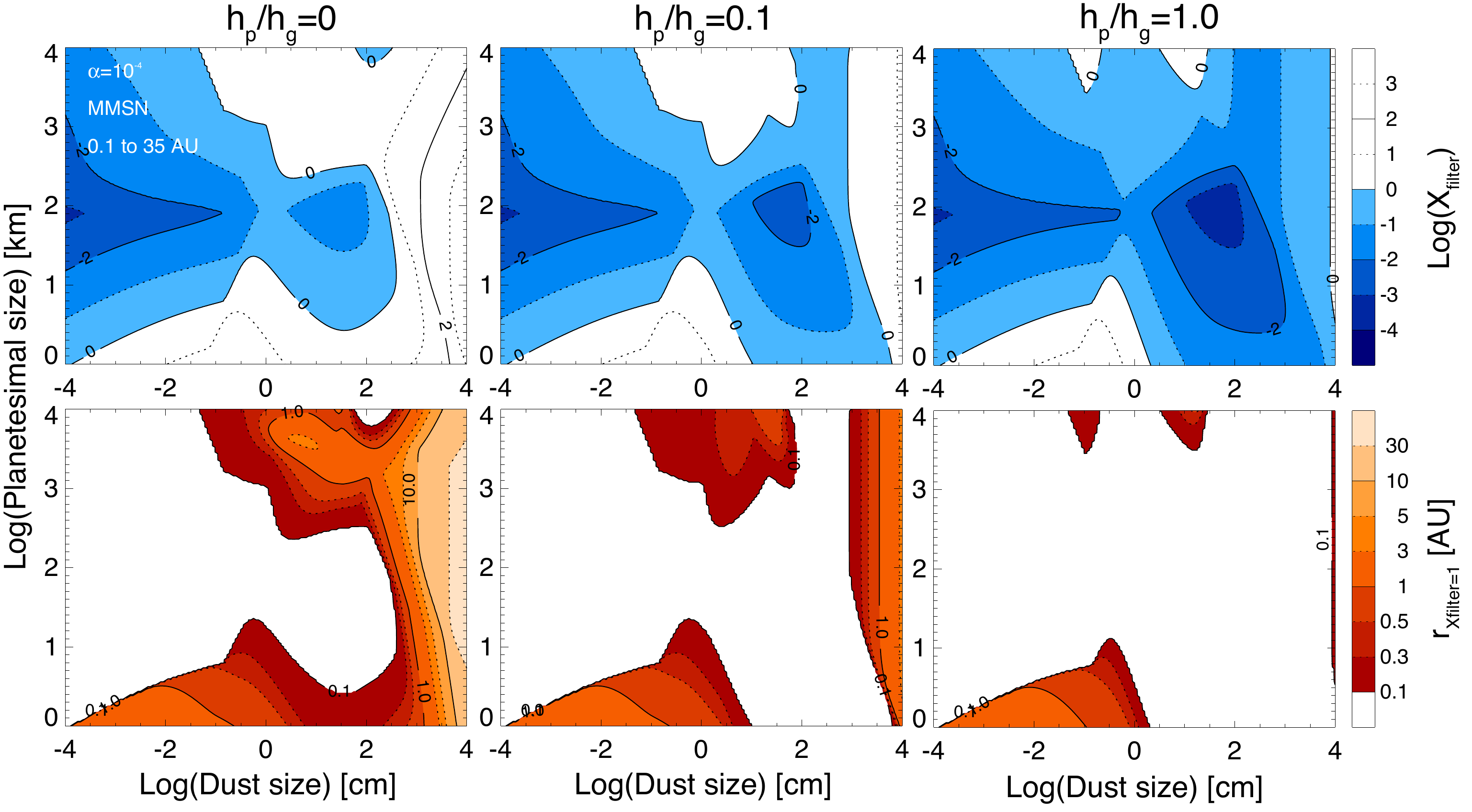}
\caption{Filtering efficiency of dust from 1 micron to 100 meters by a MMSN planetesimal belt with planetesimals of $1-10,000$\,km in radius extending from 0.1 to 35 AU for a turbulent parameter $\alpha=10^{-4}$ and various ratios of the planetesimal to gas scale height, from an infinitely thin planetesimal disk with $\hp/\hg=0$ (left panels), $\hp/\hg=0.1$ (middle panels), and $\hp/\hg=1.0$ (right panels). As in figs.~\ref{fig:xfiltering_alpha1d-2} and \ref{fig:xfiltering_alpha1d-4}, the top panels show the  contours of the disk-integrated filtering factor $X_{\rm filter}$ while the bottom panels show the orbital distance at which $X_{\rm filter}(r)=1$ for dust particles drifting in from beyond 35\,AU.}
\label{fig:xfiltering_alpha1d-4_panels}
\end{figure}

When we increase the planetesimal scale height, the filtering of large particles becomes less efficient, as shown by the middle and right panels of Fig.~\ref{fig:xfiltering_alpha1d-4_panels}. Specifically, compared to the case when $\hp=0$, changes occur for particles larger than about 1 cm when we consider $\hp/\hg=0.1$ and for particles larger than about 0.1\,cm when we consider $\hp/\hg=1.0$. 

We first consider the limit for which eq.~\eqref{eq:P3D} becomes independent of $\hp$ and the problem may be considered 2D. This occurs when the cross section of the embryos becomes so large that $b>(2/\sqrt{\pi})\hp$. By writing $\beta\equiv R_\rmH/b$, one may show that this occurs when
\begin{equation}
\mu_\rmp>{24\over \pi^{3/2}}\left(\beta\hp\over r\right)^3.
\label{eq:mup_condition}
\end{equation}
For the MMSN disk at 1\,AU, this is equivalent to 
\[
\Rp> 41,800 \beta (\hp/\hg) (\rhop/1\,\gcc)^{-1/3}\,\rm km
\]
\[
M_\rmp> 51.2 \beta^3 (\hp/\hg)^3 \ \rm M_\oplus.
\]
For large embryos we expect $\beta\wig{>}1$ with a significant dependence on $\taus$ (see Appendix~\ref{sec:analytical}). In the regime that we considered in this work eq.~\eqref{eq:mup_condition} is not satisfied, but it would be for embryos larger than a few Earth masses which may then accrete more efficiently. 

The decrease of the filtering efficiency with increasing $\hp$ value seen in Fig.~\ref{fig:xfiltering_alpha1d-4_panels} is hence a direct consequence of the fact that when eq.~\eqref{eq:mup_condition} is not satisfied, the 3D collision probability given by eq.~\eqref{eq:P3D} depends on the maximum of $\hdust$ and $\hp$. The probability thus becomes dependent on $\hp$ when $\hp>\hdust$, that is when
\begin{equation}
\taus>\alpha(\hg/\hp)^2.
\label{eq:taus_condition}
\end{equation}
For $\alpha=10^{-4}$, this implies that we expect lower collision probability and filtering efficiency when compared to the $\hp=0$ case when $\taus=10^{-2}$ for $\hp/\hg=0.1$ and $\taus=10^{-4}$ for $\hp/\hg=1$. According to Fig.~\ref{fig:taus} (around 1\,AU), this corresponds to $s\approx 1\,$cm and $1\,$mm, in good agreement with the results of Fig.~\ref{fig:xfiltering_alpha1d-4_panels}.

The question whether large boulders may be efficiently filtered by planetary embryos thus depends crucially on whether these embryos effectively lie close to the mid-plane. In the case of a weakly turbulent disk with $\alpha=10^{-4}$, this is the case when the ratio of the embryo to gas scale height $\hp/\hg=0.01$ or lower, but we notice that an efficient filtering in the settling regime becomes limited to pebbles less than about a meter in size when $\hp/\hg=0.1$ and that it mostly disappears for $\hp=\hg$. A detailed, size-dependent calculation of $\hp$ is therefore critical to determine whether planetary embryos may filter large particles in the disk and grow through that mechanism. 
}

\end{appendix}


\end{document}